\documentclass[10pt]{article}

\usepackage{amsmath}
\usepackage{amssymb}

\usepackage{graphicx}

\usepackage{cite}

\usepackage{color} 

\usepackage[labelfont=bf,labelsep=period,justification=raggedright]{caption}

\makeatletter
\renewcommand{\@biblabel}[1]{\quad#1.}
\makeatother

\date{}

\def\be{\begin{equation}} 
\def\ee{\end{equation}} 
\newcommand \bea {\begin{eqnarray} } 
\newcommand \eea {\end{eqnarray}} 
\newcommand{\nn} {\nonumber} 
\def\s{\sigma}

\def \s{\sigma}      
\def \e{\epsilon}

\begin{document}

\begin{flushleft}
{\Large
\textbf{The role of incoherent microRNA-mediated feedforward loops in noise buffering}
}
\\
Matteo Osella$^{1,2,\dagger,\ast}$, 
Carla Bosia$^{1,2,\dagger}$, 
Davide Cor\'{a}$^{2,3}$,
Michele Caselle$^{1,2}$
\\
\bf{1} Dipartimento di Fisica Teorica and INFN University of Torino, V. Pietro Giuria 1, I-10125 Torino Italy.
\\
\bf{2} Center for Complex Systems in Molecular Biology and Medicine, University of Torino, V. Accademia Albertina 13, I-10100 Torino, Italy.
\\
\bf{3} Systems Biology Lab, Institute for Cancer Research and Treatment (IRCC), School of Medicine, University of Torino, Str. Prov. 142, Km. 3.95, Candiolo I-10060 Torino, Italy
\\
$\dagger$ These authors contributed equally to this work.
\\
$\ast$ E-mail:mosella@to.infn.it
\end{flushleft}

\section*{Abstract}

MicroRNAs are endogenous non-coding RNAs which negatively regulate the expression of protein-coding genes in plants and animals. They are known to play an important role in several biological processes and, together with transcription factors, form a complex and highly interconnected regulatory network. Looking at the structure of this network it is possible to recognize a few overrepresented motifs which are expected to perform important elementary regulatory functions. Among them a special role is played by the microRNA-mediated feedforward loop in which a master transcription factor regulates a microRNA and, together with it, a set of target genes. In this paper we show analytically and through simulations that the incoherent version of this motif can couple the fine-tuning of a target protein level with an efficient noise control, thus conferring precision and stability to the overall gene expression program, especially in the presence of fluctuations in upstream regulators. Among the other results, a nontrivial prediction of our model is that the optimal attenuation of fluctuations coincides with a modest repression of the target expression. This feature is coherent with the expected fine-tuning function and in agreement with experimental observations of the actual impact of a wide class of microRNAs on the protein output of their targets. Finally we describe the impact on noise-buffering efficiency of the cross-talk between microRNA targets that can naturally arise if the microRNA-mediated circuit is not considered as isolated, but embedded in a larger network of regulations.

   
\section*{Author Summary}

The expression of protein-coding genes is controlled by a complex network of regulatory interactions. It is becoming increasingly appreciated that the post-transcriptional repression by microRNAs, a class of small non-coding RNAs, is a key layer of regulation in several biological processes. Since gene expression is a fundamentally stochastic process, the mixed network (comprising transcriptional and microRNA-mediated regulations) has to reliably perform its functions in the presence of noise. In this paper we investigate the function of one of the recurrent architectures of this network, the microRNA-mediated feedforward loops, using a detailed analytical model and simulations. With this approach we show that these regulatory circuits are appropriately designed so as to control noise, giving a rigorous mathematical proof of a previously proposed biological intuition. Moreover the theoretical framework introduced in this paper allows us to make nontrivial predictions that are presently in agreement with observed features of microRNA regulation and that could be more specifically tested experimentally in the future.

\section*{Introduction}

MicroRNAs (miRNAs) are endogenous small non-coding RNAs which negatively regulate the protein production of their targets in metazoans and plants. 
They are expected to target a substantial portion of the human genome \cite{Flynt08} and have  been shown to play 
 key roles in several biological processes ranging from development and metabolism to apoptosis and signaling pathways  \cite{Ambros04,Bartel04,Bushati07,Stefani08,Pillai07}. 
 Moreover their profiles 
 are altered in several human diseases 
\cite{Garcia05,Kerscher06}, making miRNAs a major focus of research in nowadays molecular biology.\\
Recent work, reviewed in \cite{Martinez09}, has shown that the actions of miRNAs and transcription factors (TFs) are often highly coordinated, 
suggesting that the transcriptional and post-transcriptional layers of regulation are strongly correlated and that
miRNA functions can be fully
 understood only by addressing  TF and  miRNA regulatory interactions together in a single ``mixed" network.
 As in  the case of purely transcriptional networks \cite{Milo02}, 
 in this mixed network several recurrent wiring patterns can be detected, called network motifs 
 \cite{Shalgi07,Tsang07,Yu08,Re09}.  The common lore is that network motifs were selected by evolution (and are thus overrepresented in the network) to perform
  elementary  regulatory functions. 
Among these motifs one of the most interesting is the miRNA-mediated feedforward loop (FFL) in which 
a master TF  regulates a miRNA and, together with it, a set of target genes (see Figure 1).
This motif, which shall be the main interest of our paper,  was recently discussed  
 in  \cite{Shalgi07,Tsang07,Re09}. 
In all these papers, despite the fact that the authors used very different computational approaches, 
the FFL was shown to be remarkably overrepresented in the network, thus supporting the idea that 
it should play an important regulatory role. 
Depending on the sign of the transcriptional regulations, FFLs can be divided into two classes: coherent and incoherent  \cite{Hornstein06,Tsang07,Re09}. 
In the coherent FFLs both pathways from the TF to the target have the same effect 
(both repressing or activating target expression), while in the 
incoherent ones the two pathways have opposite effects. Correspondingly one finds different expression 
patterns in the two cases: coexpression of miRNA 
and its target for incoherent FFLs and mutually exclusive expression for the coherent ones (Figure 1). 
This dual picture allows to better understand the complex patterns of correlated expression of 
miRNAs and their targets observed in experiments \cite{Flynt08,Shkumatava09,Tsang07}. 
In many cases the targets show low expression in miRNA-expressing cells, suggesting coherent regulation. On the other hand, 
several other cases present an opposite trend, 
showing that miRNA repression can act in opposition to transcriptional regulation. Indeed, 
different degrees of expression overlap, due to different regulatory 
circuitries, have been related to 
different miRNA functions in several recent papers \cite{Hornstein06,Bartel04,Bartel09,Bushati07,Flynt08}.
 For example, in a coherent FFL as the one in Figure 
1D, the miRNA expression is induced by an upstream TF 
that at the same time represses the target transcription, with the effect of enforcing mutually exclusive domains of expression as the ones observed in the 
fruit fly \cite{Stark05} or for miR-196 and its target Hoxb8 in mouse \cite{Mansfield04} and chicken \cite{Hornstein05}. In this 
cases the miRNA can help the transcriptional 
repression of a target protein that should not be expressed in a particular cell type, acting as a post-transcriptional failsafe control. Instead, an 
incoherent FFL (Figure 1C) can  promote 
high target expression in miRNA-expressing cells,  suggesting that miRNAs may have in this case a fine-tuning function, keeping
the protein level in the correct functional range. A typical example is the regulation of the atrophin gene by the miRNA miR-8 in 
 {\it Drosophila}. It was shown \cite{Karres07} that both a too high and a too low level of expression of the atrophin gene could be  detrimental for the
 organism and that miR-8 is mandatory to keep the expression level exactly in the correct range.\\
 It is by now well understood that gene espression is inherently a stochastic process \cite{Kaern05,Maheshri07,Raj08}. This has 
particularly relevant effects when the number of proteins and/or messenger RNAs (mRNAs) involved is small and stochastic fluctuations may give sizeable deviations from the mean value of
the final protein product.  Thus, the question that naturally arises is how the cell can reconcile 
the fine-tuning function described above with these fluctuations. If there is only a relatively narrow protein level which is optimal, the tuning regulation must also prevent protein fluctuations outside the 
functional range. In fact, it has been 
conjectured that the incoherent FFLs that enable tuning interaction, can also have a role in buffering noise in the target expression \cite{Hornstein06,Tsang07,Wu09}. 

The main goal of our paper is to introduce and solve analytically a stochastic model describing these incoherent FFLs in order to give a proof to this conjecture. 
Our results show that with respect to the simple gene activation by a TF, the introduction of a miRNA-mediated repressing  pathway can significantly dampen fluctuations in the target protein output 
for essentially all the choices of input parameters and initial conditions. As a test of our analysis we also performed extensive numerical simulations which nicely agree with our analytical results. 
It is important to stress (and we shall discuss this issue in detail in the following) that this noise buffering function is actually a precise consequence of the peculiar 
topolgy of the FFL. In fact, in order to fine-tune the level of a target protein one would not necessarily need a FFL topology. The same result 
could well be obtained with an independent miRNA (not under the control of the master TF which activates the target), but this choice would 
lead to strong fluctuations in the target expression. In the same theoretical framework we can show that the construction of an optimal noise filter does not necessarily imply a strong repression, in agreement with the observation that the miRNA down-regulation of a target is often modest \cite{Baek08,Selbach08}.  
  
\section*{Results}

\subsection*{The theoretical framework}
\label{theoretical}

Here we focus on the incoherent FFL in Figure 2A to present our modeling strategy. For each gene in the circuit we take into account the essential 
features of 
transcription, translation, degradation and 
interactions between genes in the regulatory network (detailed scheme in Figure 2A'). Accordingly, the state of the system is described by five variables 
$(w,q,s,r,p)$ representing: $w$ the number of 
mRNAs transcribed from the TF gene, $q$ the number of TF molecules, $s$ the number of miRNAs, 
$r$ the number of mRNAs transcribed from the target gene and $p$ the number of target proteins. We want to explore the mean ($<x_{i}>$) and the standard deviation (
$\s_{x_i}$) of each molecular species $x_i\in
(w,q,s,r,p)$ and we will show that these quantities can be obtained 
analitically at the steady-state. The noise strength of the species $x_i$ will be expressed by the coefficient of variation defined as 
$CV_{x_i} = \s_{x_i} / <x_i>$. As usual in this type of models, transcriptional activation is introduced by choosing the rate of transcription of the regulated gene ($k_{s}(q),k_{r}(q)$ in our case) 
as a nonlinear increasing function of the number 
of TFs ($q$) present in the cell \cite{Alon,Thattai01,Komorowski09,Bintu05}:

\bea
k_r(q) & =& \frac {k_r q^c} {h_{r}^c +q^c}\nn \\
k_s(q) & =& \frac {k_s q^c} {h_{s}^c +q^c}, 
\label{activ-hill}
\eea

where $h_r$ and $h_s$ are dissociation constants, specifying the amount of TFs at which the transcription rate is half of its maximum value ($k_{r}$ and $k_{s}$ 
respectively). $c$ is the Hill coefficient and fixes the steepness of the activation curve.\\
 The miRNA action can direct translational repression or destabilization of target mRNAs \cite{Sanchez06}, i.e. it decreases the rate of translation or increases 
  the rate of degradation of target mRNAs. We choose to model the effect of miRNA regulation  by taking the translation rate of the target 
  ($k_{p}(s)$) to be a repressive Hill function of the number of miRNAs ($s$):

\be
k_p(s) =\frac {k_p} {1 + (\frac{s}{h})^c}.
\label{repr-hill}
\ee

The parameter  $h$  specifies the quantity of miRNAs that determines a rate of translation $k_p/2$, and $c$ is again the Hill coefficient. 
For simplicity we use the same Hill coefficient $c$ for each Hill function, but the analysis can be straigthforwardly generalized to the case of different steepnesses.\\
 The alternative choice of a degradation rate of mRNAs as a function of miRNA concentration does not yield significantly different results, as reported 
 in Text S1. The use of Hill functions to model regulations by miRNAs is coherent with their established catalytic action in 
 animals \cite{Alberts}. A stoichiometric model has been studied in the context of sRNA regulation in 
bacteria \cite{Levine07,Shimoni07,Metha08}, in which each sRNA can pair with one messenger and drive its sequestration or degradation in an irreversible
 fashion. A comparison with a possible 
stoichiometric action is shown in Text S1.\\ 
 The probability of finding in our cell exactly $(w,q,s,r,p)$ molecules at time $t$ satisfies the master equation:

\bea
& &\partial_t P_{w,q,s,r,p} = ~~k_{w} (P_{w-1,q,s,r,p} - P_{w,q,s,r,p}) + k_q w (P_{w,q-1,s,r,p}-P_{w,q,s,r,p}) \nn \\
& & + k_r(q) (P_{w,q,s,r-1,p} - P_{w,q,s,r,p}) + k_s(q) (P_{w,q,s-1,r,p} - P_{w,q,s,r,p}) \nn\\
& & +k_p(s) r (P_{w,q,s,r,p-1}-P_{w,q,s,r,p}) + g_w \Bigl[ (w+1)P_{w+1,q,s,r,p} - w P_{w,q,s,r,p} \Bigr]\nn\\
& & + g_q \Bigl[ (q+1)P_{w,q+1,s,r,p} - q P_{w,q,s,r,p} \Bigr]+  g_r \Bigl[ (r+1)P_{w,q,s,r+1,p} - r P_{w,q,s,r,p} \Bigr]\nn\\
& & + g_s \Bigl[ (s+1)P_{w,q,s+1,r,p} - s P_{w,q,s,r,p} \Bigr]+ g_p \Bigl[ (p+1)P_{w,q,s,r,p+1} - p P_{w,q,s,r,p} \Bigr],
\label{maFFL}
\eea

where $k_{w}, k_r(q), k_s(q)$ are transcription rates, $k_q, k_p(s)$ are translation rates, and $g_{x_i}$ represents the degradation rate of the species 
$x_i$.\\
In order to solve the master equation for $<x_i>$ and $\s_{x_i}$ for all $x_{i} \in(w,q,s,r,p)$ at the steady state we have to linearize Hill functions.
This is by now a standard procedure \cite{Thattai01,Komorowski09}. 
The idea is that at the steady state the distributions of  
regulators (TFs or miRNAs) have a finite width and sample only  small regions of the domains of the corresponding Hill functions. 
We may therefore approximate Hill functions 
by their linearizations around the mean values of the regulators $q$ or $s$ (see Text S1 for details of the linearization), ending up with:

\bea
k_r(q) & \sim & k_{r}^0 + k_{r}^1 q \nn\\
k_s(q) & \sim & k_{s}^0 + k_{s}^1 q \nn\\
k_p(s) & \sim & k_{p}^0 - k_{p}^1 s.
\label{linearization}
\eea

We would like to emphasize that linearizing the Hill functions does not mean to linearize the model. In fact, even with a linearized dependence on the miRNA copy number, 
our model keeps a nonlinear contribution in the term encoding the target translation (due to the fact that it depends on both the number of miRNAs and mRNAs). 
As we will see later, this nonlinearity leads to non trivial consequences.\\
Despite this nonlinearity, the moment generating function approach \cite{Thattai01,Swain08,Komorowski09} can be succesfully used. By defining the 
generating function:

\be
F(z_1,z_2,z_3,z_4,z_5) = \sum_{w,q,s,r,p} z_{1}^w ~z_{2}^q ~ z_{3}^s~ z_{4}^r ~z_{5}^p ~P_{w,q,s,r,p},
\ee

and using the linearization in equation \ref{linearization} we can convert equation \ref{maFFL} into a second-order partial differential equation:

\bea
{\partial_t F } &= & k_w ( z_1 F - F) + k_q z_1(  z_2  {\partial_{z_1}} F -  {\partial_{z_1}} F ) + k_{r}^{0} (z_4 F - F)\nn\\
& & + k_{r}^{1} z_2 ( z_4 {\partial_{z_2}} F - {\partial_{z_2}} F ) + k_{s}^{0} ( z_3 F - F) + k_{s}^{1} z_2 ( z_3  {\partial_{z_2}} F -  {\partial_{z_2}} F )\nn\\
& & + k_{p}^{0} z_4 ( z_5  {\partial_{z_4}} F - {\partial_{z_4}} F ) - k_{p}^{1} z_3 z_4 (z_5 {\partial_{z_3,z_4}} F -  {\partial_{z_3,z_4}} F )\nn\\
& & + g_w ( {\partial_{z_1}} F - z_1 {\partial_{z_1}} F ) + g_q ( {\partial_{z_2}} F - z_2 {\partial_{z_2}} F ) + g_s ( {\partial_{z_3}} F - z_3 {\partial_{z_3}} F )\nn\\
& & + g_r ( {\partial_{z_4}} F - z_4 {\partial_{z_4}} F ) + g_p ( {\partial_{z_5}} F - z_5 {\partial_{z_5}} F ).
\label{maF}
\eea

We now use  the following properties of the moment generating function: $F|_1=1$; ${\partial_{z_i}} F = <x_i>$; ${\partial^{2}_{z_i}} F = <x_i^2> -<x_i>$ where $|_1$ 
means evaluation of $F$ at $x_i =1$ for all $i
$. At the steady state (${\partial_{t} }F =0$) differentiation of equation \ref{maF} generates equations for successively higher moments. In particular, we 
are interested in $<p>$ and $\sigma_p$ and differentiating up to the fourth moments leads to their analytical expressions (see Text S1 for details of the 
calculation).\\
Noise in protein expression is originated by the combination of two types of fluctuations: intrinsic and extrinsic ones. Intrinsic fluctuations are essentially due to the stochasticity of the gene expression process. Extrinsic ones, instead, are due to the environment. 
In the latter case a
prominent role is played by the noise
transmitted by upstream genes \cite{Pedraza05,Volfson06}.  As a matter of fact there is a certain degree of 
arbitrariness
in the definition of extrinsic and intrinsic noise \cite{Paulsson05}. Since we focus on the target production we define ``intrinsic" the noise derived 
from the stochastic steps of its expression 
(transcription, translation and degradation) and ``extrinsic" the noise propagating from its regulators ($s,q$) that makes the parameters ($k_r(q),k_p(s)$) 
fluctuate through the Hill functions. 
Therefore in our model we do not have to include extrinsic noise with an arbitrary distribution as it naturally arises from the stochastic steps of 
production of regulators and propagates to the target 
gene. 

\subsection*{Comparison with a TF transcriptional control}
\label{TF-gene}

To show the noise buffering role of the miRNA-mediated incoherent FFL (Figure 2A) we first compare it to a simpler process: a gene activated by a
 TF (Figure 2B), without any post-transcriptional regulation. The strategy used to model this linear network is equivalent to the one explained in the previous section for the FFL (see Text S1
 for more details) and it is presented 
schematically in Figure 2B'. Starting from a gene activated by a 
TF, in principle the gain of a new regulator implies also a new source of extrinsic noise for the target, given that the fluctuations in the number of 
regulators propagate to downstream genes and, as 
discussed 
in \cite{Shahrezaei08}, the addition of extrinsic fluctuations generally increases the noise of a system. However, the 
peculiar structure of the FFL can overcome this problem, actually reducing noise, as was previously shown in the case of negative transcriptional 
auto-regulation \cite
{Becksei00}. Given that the two circuits lead to different mean values, the comparison  of noise strengths in target protein will be done in terms of 
the coefficient of variation ($CV_p = \sigma_p /<p>
$). With the parameter choice explained in the caption of Figure 3, the predicted  $CV_p$ are 0.147 and 0.1 for the TF-gene cascade and the 
FFL respectively. Therefore the introduction of the 
miRNA pathway  not only controls the mean value but also reduces the relative fluctuations. This effect can be clearly seen looking at the shape of
 the probability distributions in Figure 3C. It is rather easy to understand the origin of this noise buffering effect: any fluctuation in the  concentration of TFs affects the rate of 
mRNA transcription, driving  consequently the target protein away from its steady 
state, but  mRNA and miRNA concentrations tend to vary in the same direction in the FFL. In this way, miRNAs can always tune the protein production against TF 
fluctuations. As can be seen in Figure 3A and B,  there is a certain degree of correlation in the time evolution of $q,r,p$ due to noise propagation, despite the overimposed higher-frequency 
intrinsic noise of each molecular species, but in the 
case of the FFL the $p$ trajectory is less sensitive to $q$ fluctuations thanks to the action of miRNAs ($s$). It is important to stress that this result is not
affected by the Hill function linearization discussed above. In fact, 
the predictions of the model are in good agreement with Gillespie simulations (which keep into account
the full nonlinear repressing and activating Hill functions). Moreover our results turn out to be  robust with respect 
to parameter choice, showing a rather stable noise reduction essentially for any choice of expression and degradation constants (see Text S1 for details).     

\subsection*{Comparison with an open regulatory circuit}
\label{open}

The same fine-tuning of the mean target concentration achieved with a FFL could be equally obtained with an open circuit like the one in Figure 2C, 
where the miRNA gene is  controlled by an 
independent TF. If the two TFs, activating 
the miRNA and target gene expression, have the same rate of transcription, translation and degradation of the single master TF in the FFL -as well as the 
other model parameters as in Figure 2A' and C'-  the stationary mean levels of the various molecular species are the same in both circuits. 
In particular, the mean concentration of the target protein can be fine-tuned to the same desired value by both circuits.
However,  while the deterministic description 
at the steady state is the same in the two cases 
(see Text S1 for details) the behaviour of fluctuations is completely different. As we shall see below, the 
open circuit leads to much larger fluctuations in the final product than the 
FFL. It is well possible that this is the reason for which FFLs have been positively selected by evolution and are presently overrepresented in the mixed
TF-miRNA regulatory network. In fact, fine-tuning can be implemented 
advantageously only together with a fluctuation control: a 
precise setting of the mean value of a target protein  is useless without a simultaneous damping of the stochastic fluctuations.  
To assess this 
result we used the same strategy discussed above: we solved analitically  for both circuits the master equation and tested our results with a set of
Gillespie simulations.
Our results are shown in Figure 4:  the lack of correlation between the miRNA and mRNA trajectories in the open circuit 
(Figure 4B) leads to much larger deviations from the mean number of proteins with respect to the FFL case. Using the same 
parameter values of Figure 3, the predicted $CV_p$ for the open circuit is $CV_p=0.175$, to be compared with the value $CV_p=0.1$ of the FFL. 
Different cell-to-cell variability can be clearly seen comparing the distributions of the number of target proteins for the two circuits (Figure 4C).
Note that a target embedded in an open circuit has an even more noisy expression than a gene simply regulated by a TF,  for which $CV_p=0.147$. 

\subsubsection*{Deviant effects}

Stochastic equations are the natural formalism to describe a set of biochemical reactions when the number of molecules involved is small and thus fluctuations are important. 
As the number of molecules increases, the stochastic description smoothly converges, at least for linear systems, toward a deterministic one and stochastic equations can be
substituited by ordinary differential equations (ODE). 
It is usually expected that even in the regime in which fluctuations cannot be neglected one could use ODE if interested only in 
the time evolution of the mean values. This approximation can be thought as a sort of ``mean field" approach (by analogy with statistical mechanics
where the mean field approximation is implemented by neglecting fluctuations). However, similarly to what happens in statistical mechanics in the proximity of a critical point, it may happen that, even at the level of mean values, the ODE description does not coincide with the (more rigorous) stochastic one. These breakdowns between the 
two descriptions are
known as ``deviant effects" \cite{Samoilov06} and are typically a consequence of nonlinear terms in the equations or of strong extrinsic fluctuations
 \cite{Shahrezaei08,Shahrezaei08b}. In some cases these deviant effects can have relevant phenomenological consequences.
This is the case of our system: although the FFL (Figure 2A,A')  and the open circuit (Figure 2C,C') have the same 
deterministic description at the steady state (see Text S1 for details), the master equation approach gives a mean value of the target protein systematically 
lower in the FFL circuit, with respect to the same quantity in the open circuit. 
This is a non trivial consequence of the correlated fluctuations in the number of mRNAs and miRNAs in the FFL. This correlation between fluctuations 
obviously cannot be 
taken into account in the deterministic description and as a consequence the ODE analysis correctly describes  
the steady state mean number of target proteins only for the open circuit. 
This result can be understood by looking at the analytical expression of the mean value of $p$:
\be
<p> = k_{p}^0 <r> - k_{p}^1 < rs >.
\ee

In a FFL, fluctuations of $r$ and $s$ are highly correlated (Figure 3A), because the transcription rates of messengers and miRNAs depend on a shared TF. 
The result 
is that in this case $<rs> ~ > ~ <r> <s>$. On the other hand, the production of $s$ and $r$ is independently regulated in an open circuit, 
implying that  $<rs> ~ \sim ~ <r> <s>$. A deterministic description does not take into account fluctuations so correctly 
describes  $<p>$ only when uncorrelated noise is averaged out without affecting  mean values. 
In conclusion, the correlation in fluctuations 
introduced by the FFL topology affects the target protein mean value, establishing  a systematic 
decrease with respect to the deterministic description. This deviant effect can be substantial and underlines the necessity of a stochastic 
nonlinear modeling. A fully 
linearized description, as for example the one proposed by \cite{Komorowski09} for post-transcriptional regulation, would not be able to show this type of effects.

\subsection*{The incoherent feedforward loop is effective in reducing extrinsic fluctuations}

In the previous sections we compared different regulatory circuits keeping the same amount of input noise, i.e. the same amount of fluctuations in
the copy number of master TFs. Since the topology of a regulatory motif can have stronger effects on extrinsic rather than intrinsic noise \cite{Shahrezaei08}, 
it would be very interesting to study how the mixed incoherent FFL behaves as a function of such extrinsic noise. As a matter of fact extrinsic and intrinsic fluctuations are 
generally coupled in a non-trivial way in biochemical networks \cite{Tanase06}, but we developed a strategy to control fluctuations in upstream TF expression, 
known to be one of the major sources of extrinsic noise in eukaryotes \cite{Volfson06}, without affecting the copy number of the molecular species in the circuit. 
From equation \ref{maF} we can calculate  $<q>$ (which denotes the mean number of TFs) and its noise strength $CV_q$:

\bea
<q> & = & \frac{k_q k_w} {g_q g_w}\nn\\
CV_{q} & = & \frac{1}{<q>} \sqrt{<q> \frac{g_q +g_w +k_q}{g_q + g_w}},
\eea

where, as above, the parameters $k_w$ and $k_q$   denote the rate of transcription and translation of the TF respectively, and 
$g_w$ and $g_q$ the corresponding degradation constants.\\ 
Setting the rates of degradation ($k_q$ and $k_w$) and the product $k_w k_q$ to constant values, we end up with: $<q>\sim constant$ and $CV_q \sim \sqrt{k_q}$. 
This is a  well known result: fluctuations in the protein level are stronger when the rate of translation is higher \cite{Raj08} and can
be tuned (while keeping the mean value $<q>$  fixed) by changing the ratio  $k_w/k_q$  with $k_w k_q = constant$. 
This represents a perfect theoretical setting to test the dependence of the target noise $CV_p$ on the input noise $CV_q$, while maintaining unchanged the 
mean value of all the molecular species involved in the circuit.\\ 
We report in Figure 5 the results of such analysis for the three  circuits discussed in the previous sections. 
While extrinsic fluctuations increase, so does the FFL's performance in filtering out noise, compared to the other circuits. 
Once again it is easy to understand the reason of this behaviour: the FFL architecture channels fluctuations of an upstream factor into parameters with opposite effect on the target protein, forcing them to combine destructively. 
Therefore the specific FFL topology seems 
effective in the maintenance of gene expression robustness despite noisy upstream regulators. The introduction of a miRNA pathway, building a FFL from a TF-gene 
cascade, really makes the difference in 
situations of strong upstream noise. This feature can explain why miRNAs can often be deleted without observable consequences \cite{Wu09}, since experiments 
usually do not measure fluctuations and are typically performed in controlled 
environments,  where noise is kept at minimal levels.

\subsection*{Noise filtering optimization}
\label{optimal}

The efficiency of the FFL in controlling the fluctuations of the target protein is a function of three main 
parameters: the number of master TFs (which in turn is a function of $k_w$ and $k_q$), the number of miRNA
copies (function of $k_s$ and $h_s$) and the strength of miRNA repression (defined as $1/h$). In this section we shall study the efficiency of the FFL in
buffering  noise as a function
of each one of these three quantities, changing a corresponding parameter while keeping fixed all others. As we shall see, in all three cases the noise reduction  with respect to 
a simple TF-target interaction (i.e. without the miRNA) shows a U-shaped profile with a well defined minimum which allows us to identify the values of the
parameters which optimize the noise reduction property of the FFL. This pattern is rather robust, and only marginally  depends on the way the variable of interest is tuned 
(for instance, by changing $k_s$ or $h_s$ in the case of
 miRNA concentration). It is important to stress that in all three cases optimal noise filtering does not imply strong repression, 
a result which well agrees with the observation that miRNAs embedded in an incoherent FFL usually have a fine-tuning effect on 
the targets instead of switching them off completely. It is exactly in the intermediate region of the parameters,
 in which fine-tuning occurs, that we also have optimal noise reduction.

\subsubsection*{Optimal repression strength}

The repression strength is defined as $1/h$ (inverse of the dissociation constant in the Hill function 
of equation \ref{repr-hill}). As it can be seen in  Figure 6A, the FFL exhibits a noise profile with a typical U-shape and reaches an optimal value of noise reduction 
(measured as the difference in the noise strength $CV_p$ with respect to the simple TF-gene circuit) for intermediate values of repression strength. The open circuit, constrained to give the
same mean value  $<p>$,  always introduces larger target fluctuations. 
As mentioned above, 
optimal noise filtering is reached for intermediate values of the repression strength and does not require strong target repression. 
For instance with the choice of parameter values of Figure 6, 
 optimal noise reduction is obtained for a mean value of the target protein which is
about $66\%$ of the value obtained without the miRNA, i.e. with a simple TF-target interaction. This prediction could  
be experimentally tested via manipulation of the repression strength, in analogy 
to the work of  \cite{Dublanche06}  on the transcriptional autoregulatory motif.  It is  instructive to notice the analogies between the behaviour of the mixed FFL and that of the negative transcriptional autoregulation loop. This purely transcriptional network motif occurs ubiquitously in transcriptional regulatory networks and was 
recently studied in great detail \cite{Shahrezaei08,Singh09}. Also in this case, optimal noise filtering is obtained for a well defined value of the 
repression stength. It is easy to understand the reason of this behaviour. In a negative transcriptional autoregulation, the protein expressed from a gene inhibits its own transcription by increasing expression when protein numbers are low, 
while decreasing expression when protein numbers are 
high. Increasing the repression strength improves the circuit potential to reduce stochasticity, but at the same time decreases the copy number of 
mRNAs and proteins, with a rise in intrinsic 
fluctuations that can overcome any attenuation. Consistently, experiments show a well defined
 range of repression strength for which noise minimization is optimal 
\cite{Dublanche06}.

\subsubsection*{Optimal miRNA concentration}

Another variable which can be tuned in order to achieve optimal noise reduction is the miRNA concentration. If we keep the number of TFs constant then
the miRNA concentration $<s>$ can 
only depend on the maximum rate of 
transcription of the miRNA gene ($k_s$) and on the
affinity of its promoter to the TF (proportional to $1/h_s$, where  $h_s$ is the dissociation constant in equation \ref{activ-hill}). 
In Figures 6B and 6C we analyze the noise strength $CV_p$ of the
target protein in the FFL for different miRNA concentrations and compare it to the $CV_p$ obtained with the simple TF-gene interaction and with the open circuit.
Changing the miRNA concentration by varying $k_s$ 
(Figure 6B) or $h_s$ (Figure 6C)  we find very similar U-shaped profiles for  $CV_p$. 
As for the previous analysis, also in this case it is possible to find  an optimal miRNA concentration, and again 
it is such that the effect on the protein target is only a modest reduction (in this case $\sim 60\%$ of the value obtained without the miRNA). 
Apart from the conserved U-shaped profile, there 
are rather deep differences in the noise behaviour
depending on the choice of the tuning parameter. 
In fact, while an increase of $k_s$ always induces an increase of $<s>$, this quantity becomes insensitive to  $h_s$ 
above a certain threshold. 
Since the number of TFs is constant in this analysis, it is clear that increasing $1/h_s$ (Figure 6C) the system 
can reach at best the value of $<s>$ consistent 
with the  maximum rate of transcription. At the same time a large value of $1/h_s$ means that very few TFs are enough to support the maximum transcription 
rate for the miRNA gene, so fluctuations in the number of 
TFs become irrelevant despite the 
topology of the circuit. As a consequence the $CV_p$ curves for the FFL and the open circuit converge to a commom value (Figure 6C). 
A refined experimental control of miRNA concentration 
through graded miRNA overexpression or silencing  would permit a test of the U-shaped profile of $CV_p$ in a FFL.       

\subsubsection*{Optimal TF concentration}

The last case that 
we consider in this section is the effect of different TF concentrations  
on the noise buffering properties of the FFL. It is expected that for high TF 
concentrations (i.e. high values of $<q>$) the  activation functions in 
equations \ref{activ-hill} reach the saturation point, making the system insensitive to variations in TF concentration. As long as the number of TFs does not 
fluctuates below the saturation point, the miRNA and the target gene are maximally 
transcribed, with no reference to the exact number of TFs. Accordingly, $CV_p$ becomes asymptotically constant for large $<q>$ for each circuit topology
(Figure 6D). The gap 
between the asymptotic values of the direct TF regulation and the two other circuits 
is due to the fact that the former does not suffer for the additional external noise due to the fluctuations in the miRNA number. On the other hand, 
for small values of $<q>$  also the number of target proteins  is very small
as its expression is hardly activated regardless of the circuit type,  
with a consequent increase  of the noise strength (Figure 6D). The central region is the  most interesting one: this is the region in  which
the system is maximally sensitive to changes in TF concentration. In this regime the FFL outperforms both the simple direct
regulation and the open circuit in buffering 
noise. 
Also in this case the  optimal TF concentration  is placed in a region corresponding to a modest reduction  of $<p>$,
despite the miRNA repression.

\subsubsection*{Exploring the parameter space}

To give a more comprehensive insight into the relation between noise control and target repression, we finally evaluate the buffering of fluctuations ( $CV_{p}/CV_{p_{0}}$) for each average number of TFs $<q>$ and each degree of target suppression ($<p>/<p_{0}>$), where  $<p_0>$ and $CV_{p_{0}}$ represent here the constitutive mean expression and fluctuations in absence of miRNA regulation.
Results of this analysis are reported in Figure 7A. As discussed above, noise reduction can be implemented successfully in the parameter region where the target activation function (in Figure 7B) is not saturated, since this is the region where the target is sensitive to changes in TF concentration and therefore also to its fluctuations, regardless of the presence or absence  of miRNA regulation. It is exactly in this region that  noise buffering can be observed (compare Figures 7A and B). In particular, for each TF concentration the best noise reduction appears for a target level around 60\% of its constitutive expression. In the optimal setting, noise can be remarkably reduced to about one half of its constitutive value, but the reduction remains  substantial also for weaker repressions, further confirming that a strong miRNA repression is not required for noise control.\\  
We consider the characterization of the optimal setting of miRNA-mediated incoherent FFLs for noise buffering,  and the resulting U-shaped profile predicted for the noise reduction factor, as one of the major results of our analysis which, we expect, should be amenable 
of direct  experimental validation. The fact that an optimal noise filtering is obtained with a rather modest reduction in the amount  of the target protein also agrees with 
the recent experimental observation that  miRNA down-regulation of targets is often modest \cite{Baek08,Selbach08} and apparently dispensable from a functional
point of view.
In this respect it is tempting to conjecture that, at least for some targets of incoherent FFLs, the  down-regulation 
could only be the side effect of an evolutionary design aiming instead to optimize noise reduction.

\subsection*{Comparison with purely transcriptional incoherent feedforward loops}

The capability of incoherent FFLs to reduce fluctuations was previously studied with simulations in the contest of purely transcriptional networks \cite{Shahrezaei08}.
In this section we present a comparison of the noise properties of  microRNA-mediated FFLs (scheme in 
Figure 1A') and purely transcriptional ones (detailed scheme of reactions in Figure 8A), where the miRNA is replaced by a protein that
inhibits transcription (rather than translation, as miRNAs do).
The transcriptional FFL can be modeled with the same strategy explained previously for the miRNA-mediated version and analogously mean values and standard deviations of the various molecular species can be calculated analytically  with the moment generating function method (see Text S1 for more details on calculations and model assumptions).
In order to make an unbiased comparison of the noise properties of these two circuits,  the corresponding models must be constrained to produce the same amount of target proteins. Although there is no  unambiguous way to put this constraint, due to the presence of more free parameters ($k_{j}$ and $g_{j}$) in the purely transcriptional case, a reasonable choice is to keep the shared parameters to same values (i.e repression/activation efficiencies and  production/degradation rates) and choose the two additional ones to make the amount of repressor proteins $j$ in the transcriptional case equal to the amount of miRNAs $s$ in the mixed circuit. With this choice we can evaluate the target noise $CV_p$ as a function of the repression strength ($1/h$) for both circuits (Figure 8B). Even though the transcriptional version can potentially reduce target fluctutions, buffering efficiency appears clearly increased by the use of  miRNAs as regulators. Furthermore, a comparison of Figure 8C and Figure 7B points out that a miRNA-mediated FFL can buffer fluctuations over a wider range of conditions as well as to a greater extent. 
This is mainly due to the additional step of translation required for the production of proteins $j$ which unavoidably adds noise to the system.
We would like to emphasize that the shown efficiency in noise reduction, achieved with the transcriptional FFL, should be  considered as an upper bound. In fact, the constraints imposed on $k_{j}$ and $g_{j}$ limit the translational burst size, i.e. the average number of proteins produced from a single mRNA, and this  parameter crucially influences the intrinsic fluctuation amplitude of proteins $j$  \cite{Kaern05} (see Text S1 for details on parameter constraints). With the parameter values used in Figure 8, the translational burst size is $\sim 0.3$, while in eukaryotes it is expected to be larger (certainly larger than one) because of the long average  half-life of messenger RNAs compared to the time required for one translation round   \cite{Larson09}. Therefore the noise added by the step of translation of proteins $j$ should realistically be more substantial than reported for this model setting, with harmful consequences on the noise buffering efficiency of the purely transcriptional circuit.\\ 
Moreover some peculiarities (not currently included in our model) of the mixed-motif can further explain why it can be better suited for noise buffering.
Firstly, fluctuations in  RNA polymerase and ribosome abundance are possible sources of extrinsic noise in gene expression \cite{Swain02}. These fluctuations are expected to influence unspecifically the rate of transcription and translation respectively of each gene. In a  miRNA-mediated FFL the correlation between target mRNA and miRNA production, which is crucial for noise reduction,  is robust to these additional sources of noise  as mRNAs and miRNAs are identically affected only by global RNA polymerase fluctuations. On the other hand, in  purely transcriptional FFLs the number of repressor proteins $j$ is exposed to the independent fluctuations in ribosome concentration, so the regulator-regulated correlation can be compromised with potentially negative consequences on the circuit's noise reduction efficiency.\\
Secondly, delays in the action of  regulators (miRNA or proteins) in the indirect pathway from the master TF to the target can damage the noise buffering function (see Text S1 for a more detailed study of the impact of time delays on noise control). However, the biogenesis of miRNAs is thougth to be faster than the one of proteins, and thus miRNAs may affect the target expression with a shorter delay with respect to factors regulating nuclear events like a TF \cite{Li09}. This feature should enable  miRNAs to produce rapid responses, as required to counteract fluctuations.\\
Finally, the presence of a nucleus makes the eukaryotic cell a two-compartment system with stochastic transport channels, with consequences on gene expression noise \cite{Xiong10}. In fact,  transcriptional regulation requires an additional transport step with respect to the post-transcriptional one. In a transcriptional FFL, the repressor protein (replacing the miRNA) must return into the nucleus to act on the target. This again potentially reduces the correlation of its fluctuations with the target ones, affecting the noise buffering ability. 

\subsection*{Cross-talk between microRNA targets}

A recent study pointed out that the action of a miRNA on a specific target gene expression is affected by the total number of miRNA targets and their mRNA abundance \cite{Arvey10}, a phenomenon called ``dilution effect". This is 
presumably a consequence of target competition for a finite intracellular pool of miRNAs.  In particular, the degree of downregulation of an individual target expression is generally reduced by the presence of other transcribed target genes. A similar cross-talk between targets has been previously shown for sRNA regulation in bacteria \cite{Levine07} both theoretically and experimentally. 
Therefore, the functionality of a genetic circuit that involves miRNA regulations, as the one studied in this paper, can be influenced by the expression level of miRNA targets not embedded in the circuit.  To address this issue we evaluate in this section the impact  of an additional miRNA target independently transcribed (a situation depicted in Figure 9A) on the circuit ability in noise buffering.    

\subsubsection*{Stoichiometric versus catalytic models of miRNA action}

The model used so far for miRNA regulation was based on the hypothesis of perfectly catalytic action. The rate of translation of target mRNAs was assumed to be a nonlinear decreasing function of miRNA concentration, neglecting the details of mRNA-miRNA physical coupling with the implicit assumption that the downregulation process does not affect the available miRNA pool. A perfectly catalytic action does not predict any competition effect among multiple targets at equilibrium, since each target can only sense the available number of miRNAs without altering it. On the other hand, a stoichiometric model has been proposed in the context of sRNA regulation in bacteria \cite{Levine07,Shimoni07,Metha08}, in which each sRNA can pair with one messenger leading to mutual degradation. In this latter case the expression of a secondary target can capture a significant portion of the sRNAs, with a resulting decrease in the average repression acting on the first target. The nature of miRNA regulation is presumably somewhere in between these two extreme possibilities, although usually generically referred to as catalytic. 
In this view, in order to address the effect of target cross-talk on miRNA-mediated FFLs, we consider a deterministic model (introduced previously in \cite{Levine07}) that explicitely takes into account the physical coupling of miRNAs and target mRNAs and the catalytic/stoichiometric nature of this coupling. While the full detailed model is presented in Text S1, the effective equations describing the dynamics of the mean number of miRNAs $<s>$, mRNAs $<r>$ of the target in the FFL and mRNAs $<r_2>$ of the secondary miRNA target are:   

\bea
\frac{d {<s>}}{{d t}} & =& k_{s}(<q>)  - g_{s} <s> - \alpha ~(\gamma_{1} <r> <s> +~ \gamma_{2} ~<r_{2}> <s>)  \nn \\
\frac{d {<r>}}{{d t}} & =& k_{r}(<q>)  - g_{r} <r> - \gamma_{1} <r> <s> \nn \\
\frac{d {<r_{2}>}}{{d t}} & =& k_{r_{2}}  - g_{r_{2}} <r_2> - \gamma_{2} <r_{2}> <s> ,
\label{levine}
\eea

where $\gamma_1$ and $\gamma_2$ describe the probability of miRNA-mRNA coupling for the target in the FFL and the secondary target respectively,  while $\alpha$ is the probability (assumed equal for both targets) that a degradation event of a mRNA, induced by a miRNA, is accompained by the degradation of the miRNA itself. The limit $\alpha=1$ describes a stoichiometric mode of action, while the opposite situation of $\alpha=0$ represents a perfectly catalytic mode in which the rate of mRNA degradation is a linear function of the number of miRNAs.\\
The corresponding stochastic model, of which equations \ref{levine} describe the mean-field limit, cannot be solved analytically starting from the master equation, therefore noise properties  will be examined in the following with simulations only.

\subsubsection*{Dilution effect}

In the first place we evaluate the dependence of the target protein downregulation on the expression rate of the secondary target, starting from the model described by Equations \ref{levine}. The dilution effect is shown in Figure 9B for different values of $\alpha$:  the downregulation exerted on the FFL target depends on the rate of expression of the secondary target, in line with the observed inverse correlation between target abundance and mean downregulation in higher eukaryotes \cite{Arvey10} and in bacteria \cite{Levine07}. Similar results can be obtained by  varying the coupling constant $\gamma_2$ with respect to $\gamma_{1}$ (as reported in \cite{Levine07}). Therefore, the noise buffering function and the optimality criteria discussed in previous sections could be compromised in the presence of many or highly transcribed independent miRNA targets. This issue will be addressed in details in the following section.\\
As expected, a perfectly catalytic mode does not feel the effect of secondary mRNA targets (red line in Figure 9B), while the stoichiometric mechanism is the most 
sensitive (green line in Figure 9B). This result suggests that a catalytic mode (at least approximately), like the miRNA one, can allow a larger proliferation of the number of targets while limiting the effects of their cross-talk. 

\subsubsection*{Consequences of dilution effect and secondary target fluctuations on noise buffering}

Since a high level of expression of secondary targets can determine a decrease of the average downregulation, it can potentially reduce the FFL ability in filtering out target fluctuations. In fact,  also the noise reduction $CV_p/CV_{p_{0}}$ (where $CV_{p_{0}}$ is the constitutive noise in absence of miRNA) is a function of the additional target expression, as shown in Figure 9C. As the expression of the out-of-circuit target increases, its messengers are able to capture more and more miRNAs and the efficiency in noise reduction is gradually compromised. Finally the FFL target fluctuations $CV_p$ approach the constitutive ones $CV_{p_{0}}$ when the messengers of the FFL target become a small fraction of the total miRNA targets. The robustness of the circuit functioning with respect to the dilution effect is again dependent on the repression mode (that changes with $\alpha$). 
Moreover, as discussed in Text S1,  different modes (stoichiometric/catalytic) of miRNA action have a different potential in reducing fluctuations: even in absence of secondary targets, where models with different $\alpha$ have been constrained to produce the same amount of target protein, the noise buffering efficiency decreases with $\alpha$ (Figure 9C). This observation highlights  that the level of miRNA ability to  avoid mutual degradation while targeting a mRNA can play a role in the optimization of fluctuation counteracting, besides conferring stability with respect to target cross-talk.\\ 
While the corruption of the noise-buffering ability seems mainly due to the increase in the mean level of secondary messengers, there is another more subtle cause that gives a contribution: the uncorrelated fluctuations of secondary messengers. Since the secondary target is independently transcribed  (not under the control of the master TF activating the miRNA gene) its fluctuations are expected to be completely uncorrelated with the miRNA ones, implying a random sequestration of miRNAs. To disentagle this contribution from the dilution effect, we studied the case of a secondary target transcribed at the same effective rate of the FFL target, but with different levels of fluctuations (see Figure 9D). 
In the case of equal transcription rates the dilution effect has a negligible impact on the noise buffering activity of the circuit (see Figure 9C), nevertheless the level of noise reduction  ($CV_p/CV_{p_{0}}$) is progressively reduced as the second target concentration becomes more and more noisy, as reported in Figure 9D. This effect seems especially relevant for a hypothetically stoichiometric miRNA repression.
Therefore, the noise level of additional targets is a variable that must be taken into account in evaluating the cross-talk effect on the noise-buffering efficiency of the circuit.   
 Although the FFLs are overrepresented in the mixed network \cite{Shalgi07,Tsang07,Yu08,Re09}, a single microRNAs can downregulate hundreds of target genes and consequently not every  target is expected to be under the control of the same TF regulating the miRNA  gene (see Text S1 for a more detailed discussion). Therefore, even though most motif function analysis are carried out looking at the motif operating in isolation, we have shown that the presence of additional miRNA targets in the network can alter the functioning of a miRNA-mediated motif. 
In fact, the efficiency of miRNA-mediated FFLs as noise controllers should be considered contest-dependent. While this circuit seems properly designed to filter out fluctuations when the miRNA-target interaction is specific or secondary targets are poorly transcribed, cell types or conditions that require a high expression of out-of-circuit miRNA targets can significantly corrupt this circuit property. 
Besides the understanding of the function of endogenous miRNA-mediated FFLs, this analysis of  target cross-talk effects can be a useful warning for the growing field of synthetic biology \cite{Mukherji09}: the implementation of  genetic circuits incorporating small RNA regulations for specific scopes must take into account the sRNA specificity and the level of expression (and fluctuations) of eventual other targets.

\section*{Discussion}

\subsection*{Experimental and bioinformatic evidences of the relevance of miRNA mediated FFLs in gene regulation.}

Few cases of incoherent miRNA-mediated FFLs have been experimentally verified until now: a case involving c-Myc/E2F1 regulation \cite{Donnell05} and  more 
recently a miR-7 mediated FFL in {\it 
Drosophila} \cite{Li09}.  As a matter of fact, miR-7 has indeed been found to be essential to buffer external fluctuations, providing robustness to the eye developmental program. 
The fact that miR-7 is interlocked in an 
incoherent FFL provides a first hint that our model can be biologically relevant.\\
On the purely computational side,  it is interesting to notice that in \cite{Re09}
 it was  shown that the typical targets of these FFLs are not randomly distributed but are instead remarkably enriched in TFs. 
These are the typical genes for which a control of stochastic
fluctuations should be expected:  the noise in a regulator expression propagates to all its targets, affecting the reliability of signal transmission in the downstream network.\\
Finally, a significant enrichment in oncogenes within the components of the FFLs was also observed \cite{Re09}. The mentioned  FFL containing c-Myc/E2F1 is just an example \cite{Kerscher06}.
In view of the emerging idea that non-genetic heterogenetity, due to stochastic noise, contributes to tumor progression \cite{Brock09} and affects apoptotic signal response \cite{Spencer09}, the role 
of miRNA-mediated FFLs in reducing fluctuations can explain why they are often  involved in cancer-related pathways.

\subsection*{Concluding remarks}

The type of regulatory action which a miRNA exerts on its targets can be rather well understood looking at the degree of coexpression with the targets 
\cite{Hornstein06,Flynt08,Bartel04,Bartel09,Bushati07}. 
In particular, an incoherent mixed-FFL implies a high level of miRNA-target coexpression, so it is suitable to implement a fine-tuning interaction. The target is not switched off 
by miRNA repression, rather its mean level is adjusted post-transcriptionally to the desired value. However, many cells can have a protein concentration far from the finely controlled mean value, if strong fluctuations are allowed. Hence, a noise buffering mechanism can be crucial at the level of single cells, and a fine-tuning interaction will be effective for 
a large part of the cell population only if coupled with a noise control. Some authors  proposed the conjecture that the incoherent mixed-FFL can actually have a role in noise buffering 
\cite{Hornstein06,Tsang07,Wu09} and biological evidences that miRNAs can effectively be used as expression-buffers have been recently found \cite{Wu09,Li09}. 
From this point of view the miRNA-target interactions classified as neutral \cite{Bartel09}, as the mean level of the target only changes inside its functional range by the presence/absence of miRNAs, actually could have been selected by evolution to prevent potentially harmful fluctuations. In this paper we demonstrated, through stochastic modeling and simulations, that the incoherent mixed-FFL has the right characteristics to reduce fluctuations, giving a proof to the previously proposed intuitive conjecture and supplying the lacking quantitative description. In particular, we showed that this circuit filters out the noise that is propagating from the master TF, giving robustness to the target gene expression in presence of noisy upstream factors. Furthermore, our theoretical description led  to the prediction that there is a value of the miRNA repression strength for which the noise filtering is optimal. A maximum of target-noise attenuation appears likewise varying the miRNA concentration or the TF concentration and this robust prediction could be tested experimentally. In all cases the implementation of the best noise filter does 
not imply  a strong suppression of the target protein expression, coherently  with a fine-tuning function and in agreement with the observation that the miRNA down-regulation of a target is often modest \cite{Baek08,Selbach08}.\\
Our paper presents the first model explicitly built on the mixed version of the FFL. From a theoretical point of view, we addressed the detailed master equation describing the system (without neglecting the dynamics of mRNA), instead of the approximate Langevin description, and we were able to apply the moment generating function approach despite the presence of nonlinear terms that can give rise to deviant effects. This approach allowed us to take into account extrinsic fluctuations as the noise propagating from upstream genes,  without an arbitrary definition of the extrinsic noise distribution. 
This strategy can be naturally extended to other circuits in the mixed network to test their potential role in the control of stochasticity.\\
Furthermore, we compared, in terms of noise buffering ability,  miRNA-mediated FFLs with their  purely transcriptional counterparts, where the miRNA is replaced by a protein that inhibits transcription rather than translation. This comparison shows that a miRNA regulator can be better suited for the noise buffering purpose.\\
Finally, we tryed to overcome the limitations  in the analysis that can arise from considering a genetic circuit as operating in isolation. In this perspective, we evaluated the impact that the recently discovered dilution effect \cite{Arvey10,Levine07} can have on the noise buffering function of miRNA-mediated incoherent FFLs. More specifically, we showed than an efficient noise control requires the minimization of the number of miRNA target sites on out-of-circuit genes, especially if highly expressed or strongly fluctuating in the mRNA level. \\
The hypothesis of a role of miRNAs in noise buffering can shed new light on peculiar characteristics of miRNA regulation. As discussed in \cite{Wu09} and \cite{Li09}, it can explain why miRNAs are 
often highly conserved, controlling key steps in development,  but in many cases they can be deleted with little phenotypic consequences. On the evolutionary side, the origin of 
vertebrate complexity seems to correspond to the huge expansion of non-coding RNA inventory (including miRNAs) \cite{Heimberg08}. This can suggest a further reasoning: the morphological complexity
 requires a high degree of signaling precision, with a strict control of stochasticity, and miRNA regulation can satisfy these requirements if embedded in an appropriate circuit, as we showed for the 
ubiquitous miRNA-mediated FFL.


\section*{Methods}

Simulations were implemented by using Gillespie's first reaction algorithm \cite{Gillespie76}. The reactions simulated were those presented in schemes  2A',B',C' and 8A. Reactions that depend on a regulator were  allowed to have as rates the corresponding full nonlinear Hill functions. All the results are at steady state, which is assumed to be reached when the deterministic evolution of the system in analysis is at a distance from the steady state (its asymptotic value) smaller than its 0.05\% (see Text S1 for details). For the parameter set used for Figures 3-9 the steady state was assumed at 5000 seconds, around 14 times the protein half-life. Each data point or histogram is the result of 100000 trials.


\section*{Acknowledgments}
We would like to thank Mariama El Baroudi and Antonio Celani for useful discussions.

\newpage

\section*{Figures}

\begin{figure}[!ht]
\begin{center}
\includegraphics[width=6in]{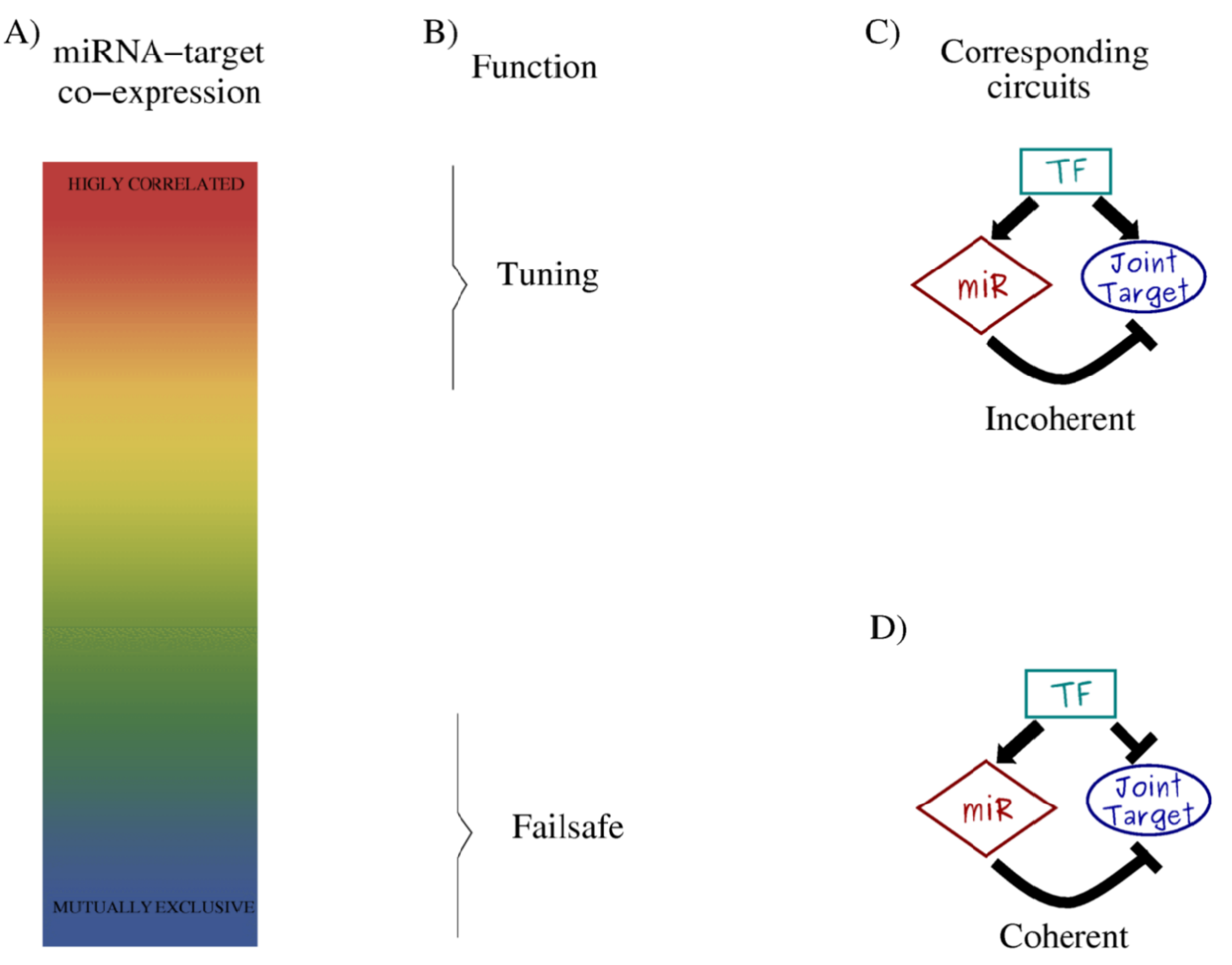}
\end{center}
\caption{
{\bf  Overview of the connections between miRNA-target expression, miRNA function and regulatory circuitry.} (A) MiRNAs and corresponding targets can present different degrees of coexpression between the two extremes of concurrent expression (high correlation) and exclusive domains (high anticorrelation). These two opposite situations are expected to correspond to different functional roles (B) for the miRNA repression. A peculiar expression pattern, evidence of a functional aim, is a consequence of the network structure in which miRNAs are embedded. A high miRNA-target correlation can be achieved through the incoherent FFL (C), where the miRNA repression is opposite to the TF action. Whereas a  failsafe control can be performed by a coherent FFL (D), in which the miRNA reinforces the TF action leading to mutually exclusive domains of miRNA-target expression.}
\end{figure}

\begin{figure}[!ht]
\begin{center}
\includegraphics[width=6in]{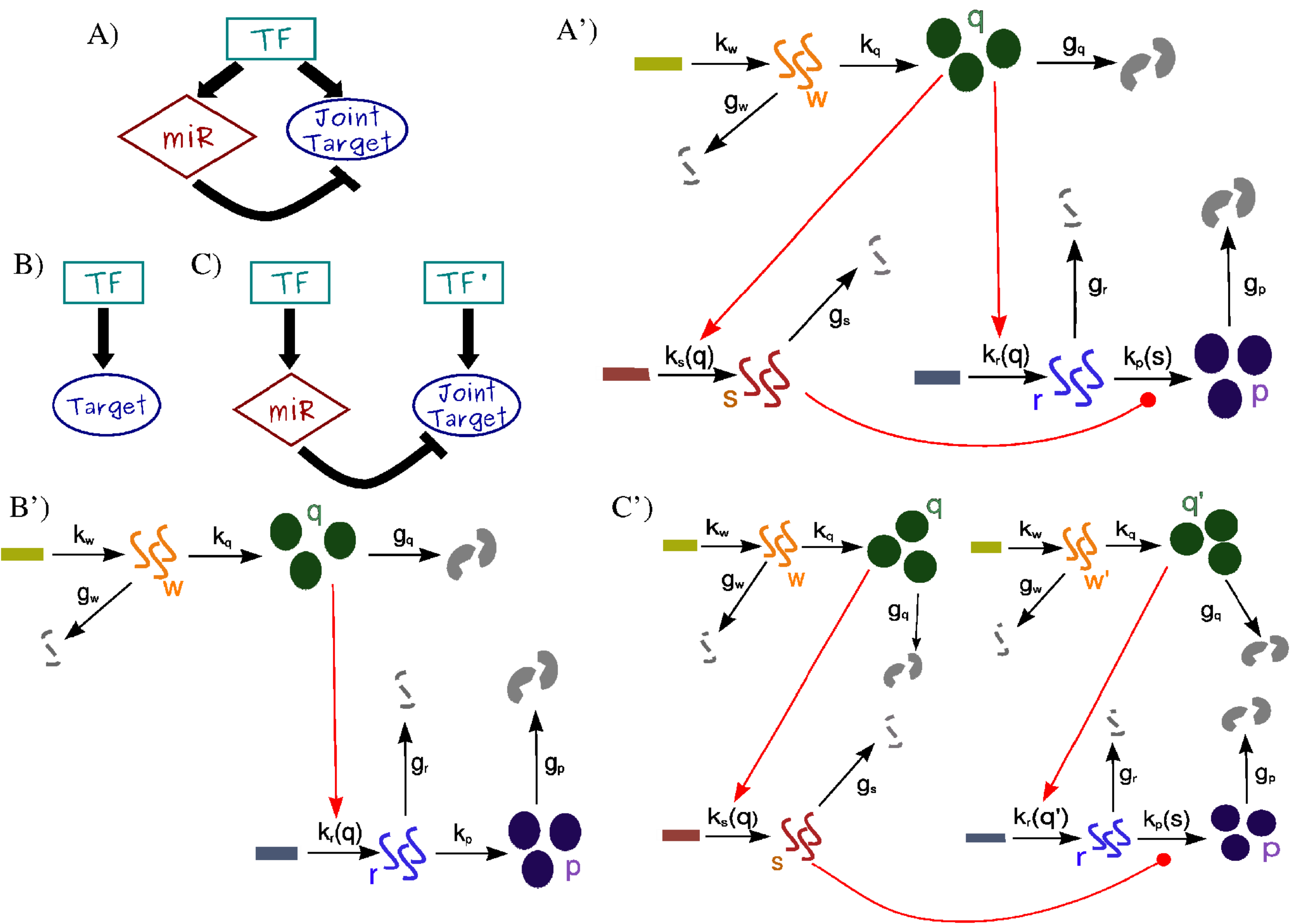}
\end{center}
\caption{
{\bf Representation of the incoherent FFL and the two circuits used for comparison.}  (A) A miRNA-mediated incoherent FFL that can be responsible for miRNA-target coexpression; (B) a gene activated by a TF; (C) an open circuit that leads to the same mean concentrations of the molecular species of the FFL in scheme A. (A')(B')(C') Detailed representation of the modelization of the corresponding circuits. Rectangles represent DNA-genes, from which RNAs ($w,s,r$) are transcribed and eventually degraded (broken lines). RNAs can be translated into proteins ($q$ is the TF while $p$ is the target protein)  symbolized by circles, and proteins can be degraded (broken circles). Rates of each process (transcription, translation or degradation) are depicted along the corresponding black arrows. Regulations are represented in red, with arrows in the case of activation by TFs  while rounded end lines in the case of miRNA repression. TF regulations act on rates of transcription that become  functions of the amount of regulators. MiRNA regulation makes the rate of translation of the target a function of miRNA concentration.}
\end{figure}

\begin{figure}[!ht]
\begin{center}
\includegraphics[width=6in]{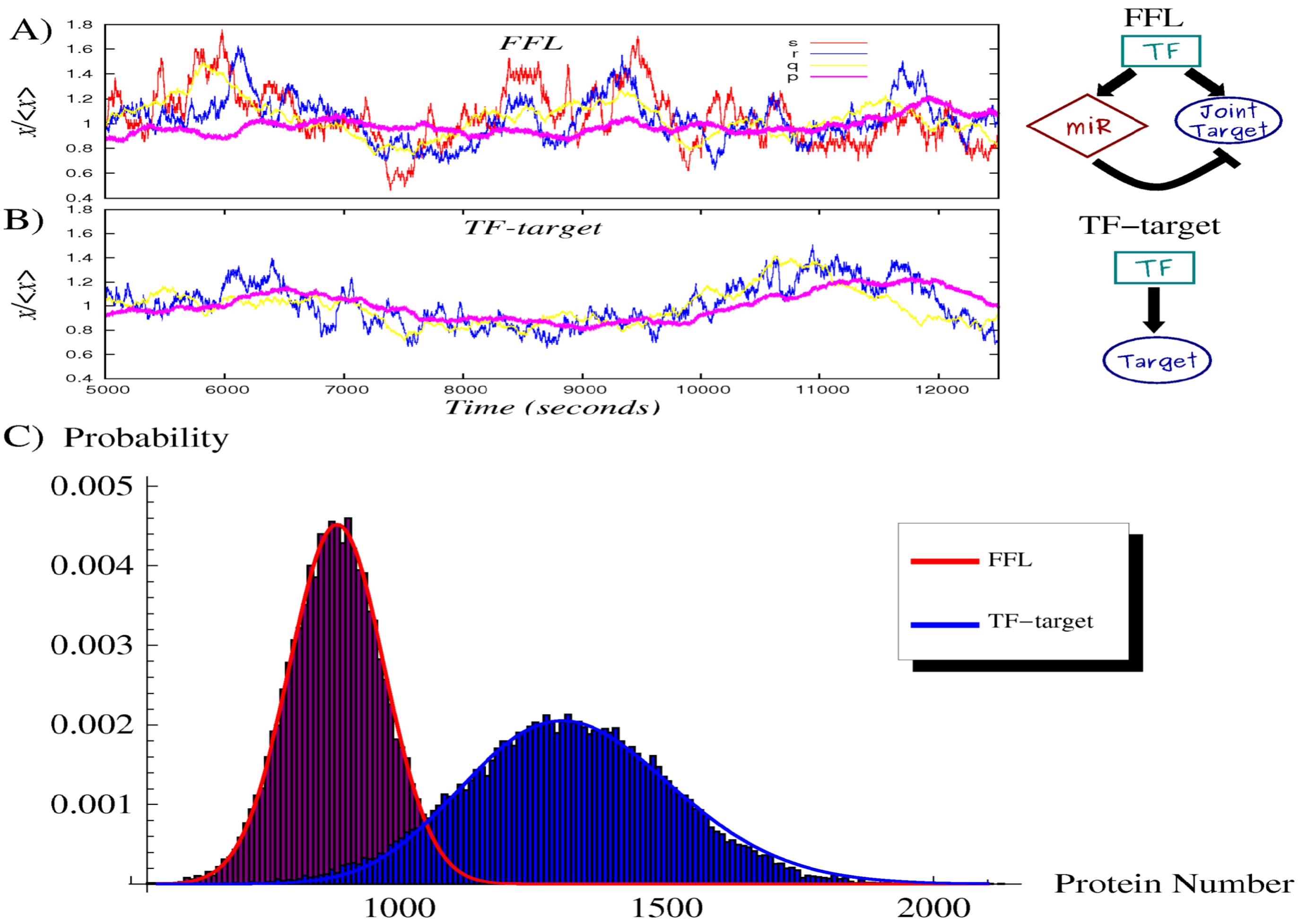}
\end{center}
\caption{
{\bf  Noise properties of the FFL compared with a TF-gene linear circuit. } (A) An example of simulation results for the FFL (scheme on the right or more detailed in Figure 2A'). The normalized trajectory of each molecular species is shown as a function of time after reaching the steady state. The rate of transcription of the TF is $k_w =0.06 ~\mathrm{s}^{-1}$ and of translation $k_q = 0.04 ~\mathrm{s}^{-1}$. Proteins degrade with a rate $g_q=g_p=0.002 ~\mathrm{s}^{-1}$, while mRNAs and miRNAs with $g_w = g_r = g_s = 0.006 ~\mathrm{s}^{-1}$. The parameters in the Hill functions of regulation (equations \ref{activ-hill},\ref{repr-hill}) are the following: the maximum rate of transcription for mRNAs is $k_r =0.8 ~\mathrm{s}^{-1}$, while for miRNAs is $k_s=0.5 ~\mathrm{s}^{-1}$; the maximum rate of translation of the target is $k_p=0.04 ~\mathrm{s}^{-1}$; dissociation constants are $h_s=200,h_r=200,h=60$; Hill coefficients are all $c=2$, as typical biological values range from 1 (hyperbolic control) to 30 (sharp switching)\cite{Thattai01}. (B) Time evolution in a simulation for the molecular players in the linear TF-gene cascade (scheme on the right or more detailed in Figure 2B'). Compared to the FFL case, the $p$ evolution is more sensitive to TF fluctuations. (C) The probability distribution of protein number for the two circuits. Histograms are the result of Gillespie simulations while continuous lines are empirical distributions (gaussian for the FFL and gamma for the TF-gene) with mean and variance predicted by the analytical model.}
\end{figure}

\begin{figure}[!ht]
\begin{center}
\includegraphics[width=6in]{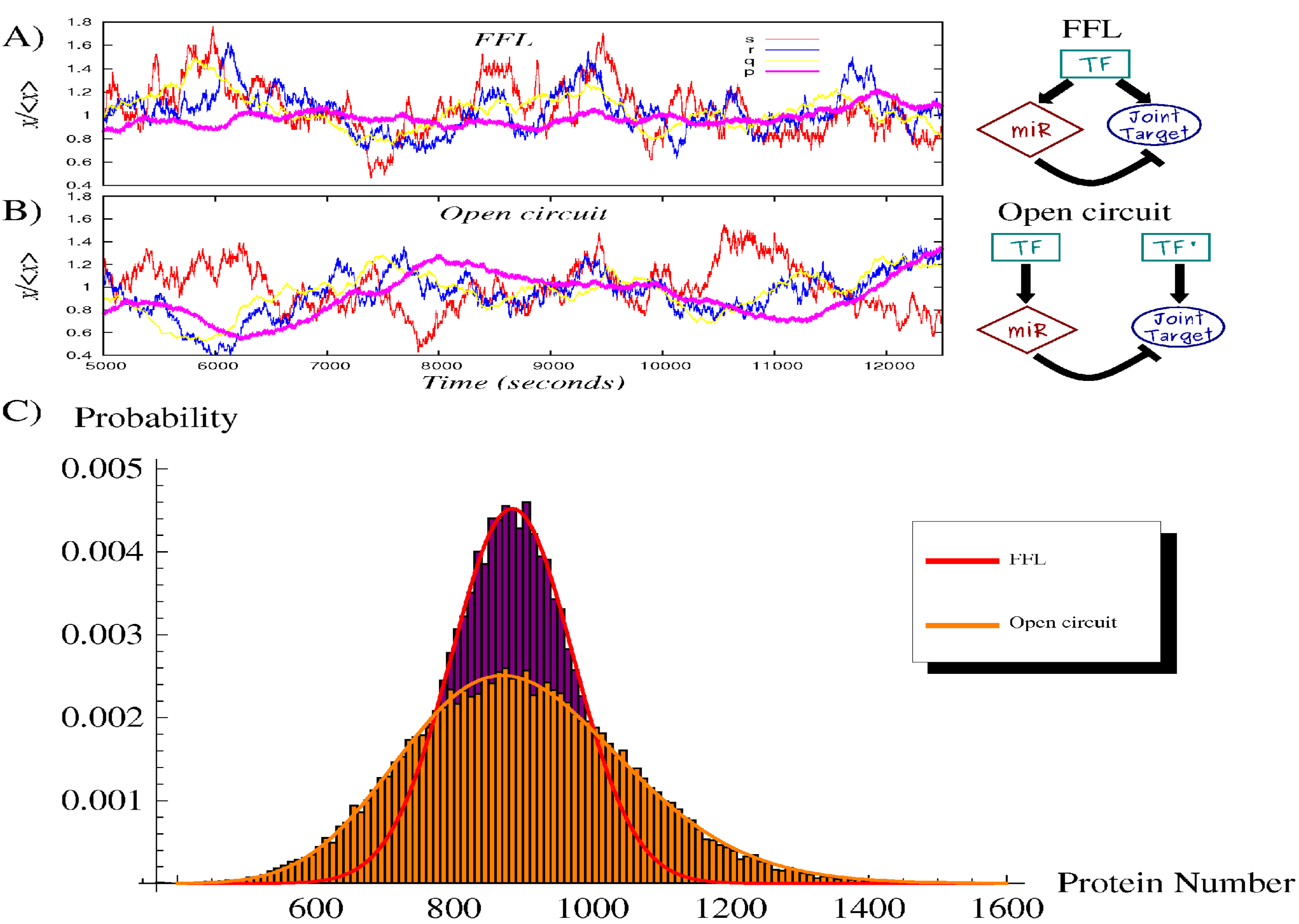}
\end{center}
\caption{
{\bf  Noise properties of the FFL compared with an analogous open circuit.} (A) An example of simulation results for the FFL (scheme on the right or more detailed in Figure 2A'). The parameter values are the same of Figure 3.(B) Time evolution in a simulation for the molecular players in the open circuit (scheme on the right or more detailed in Figure 2C'). The correlation between the $s$ and $r$ trajectories that is present in the FFL (A) is completely lost in the open circuit. As a consequence while the mean value of $p$ is approximately the same, its fluctuations are appreciably greater in the open circuit case.(C) The probability distribution of protein number for the two circuits. Histograms are the result of Gillespie simulations while continuous lines are empirical distributions (gaussian for the FFL and gamma for the open circuit) with mean and variance predicted by the analytical model.}
\end{figure}

\begin{figure}[!ht]
\begin{center}
\includegraphics[width=6in]{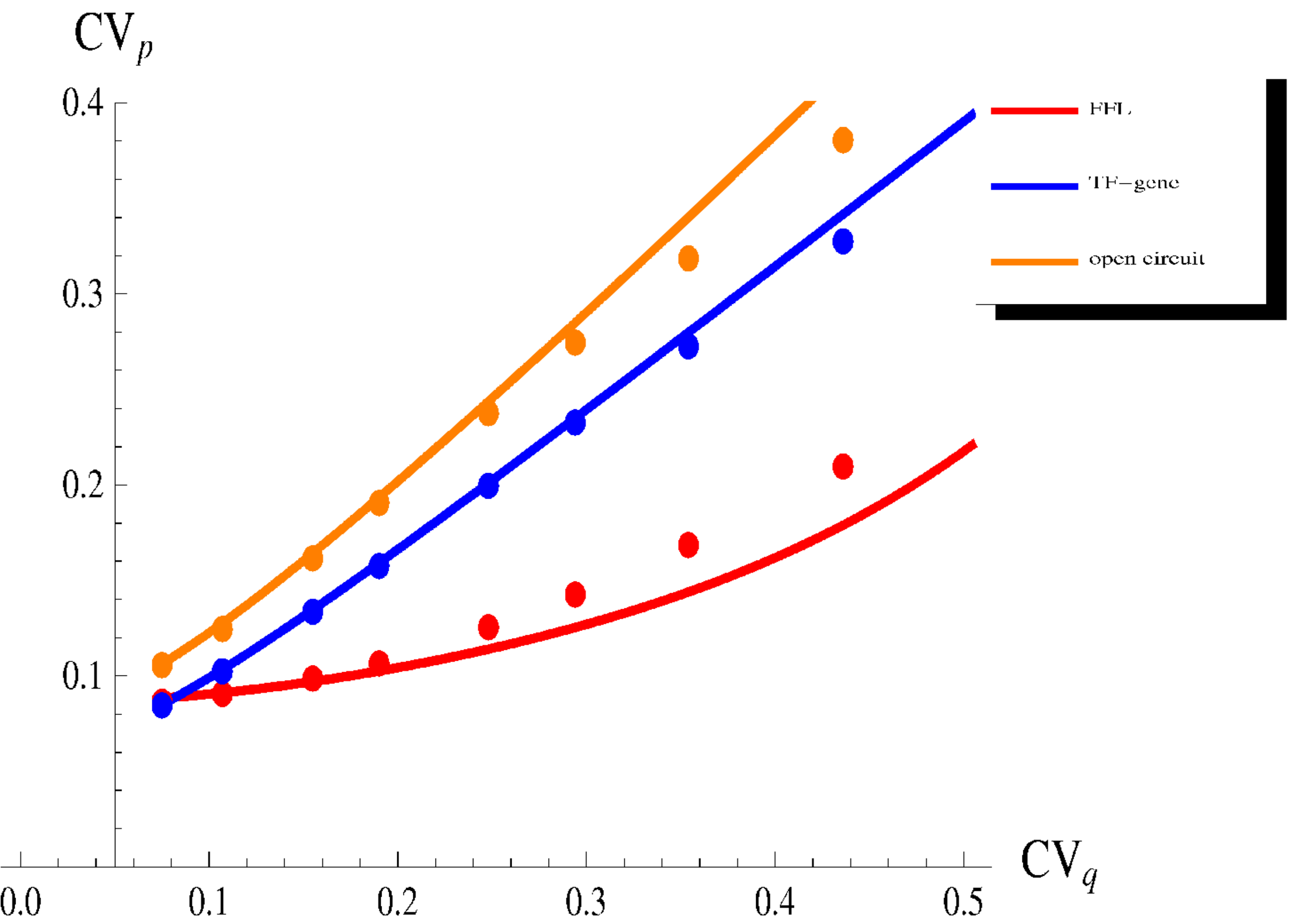}
\end{center}
\caption{
{\bf  The effect of fluctuations in an upstream TF.} We maintain constant the number of TFs $<q>$, while we vary its relative fluctuations $CV_q$, tuning the relative contribution of transcription (rate $k_w$) and translation (rate $k_q$). All the other parameters have the values reported in caption of Figure 3. The incoherent FFL makes the target less sensitive to fluctuations in the upstream TF. The extent of the noise reduction, with respect to the other circuits, depends on the magnitude of the 
TF noise, pointing out that the FFL topology is particularly effective in filtering out extrinsic fluctuations. Dots are the result of Gillespie simulations with the full nonlinear dynamics while continuous lines are analytical predictions.}
\end{figure}

\begin{figure}[!ht]
\begin{center}
\includegraphics[width=6in]{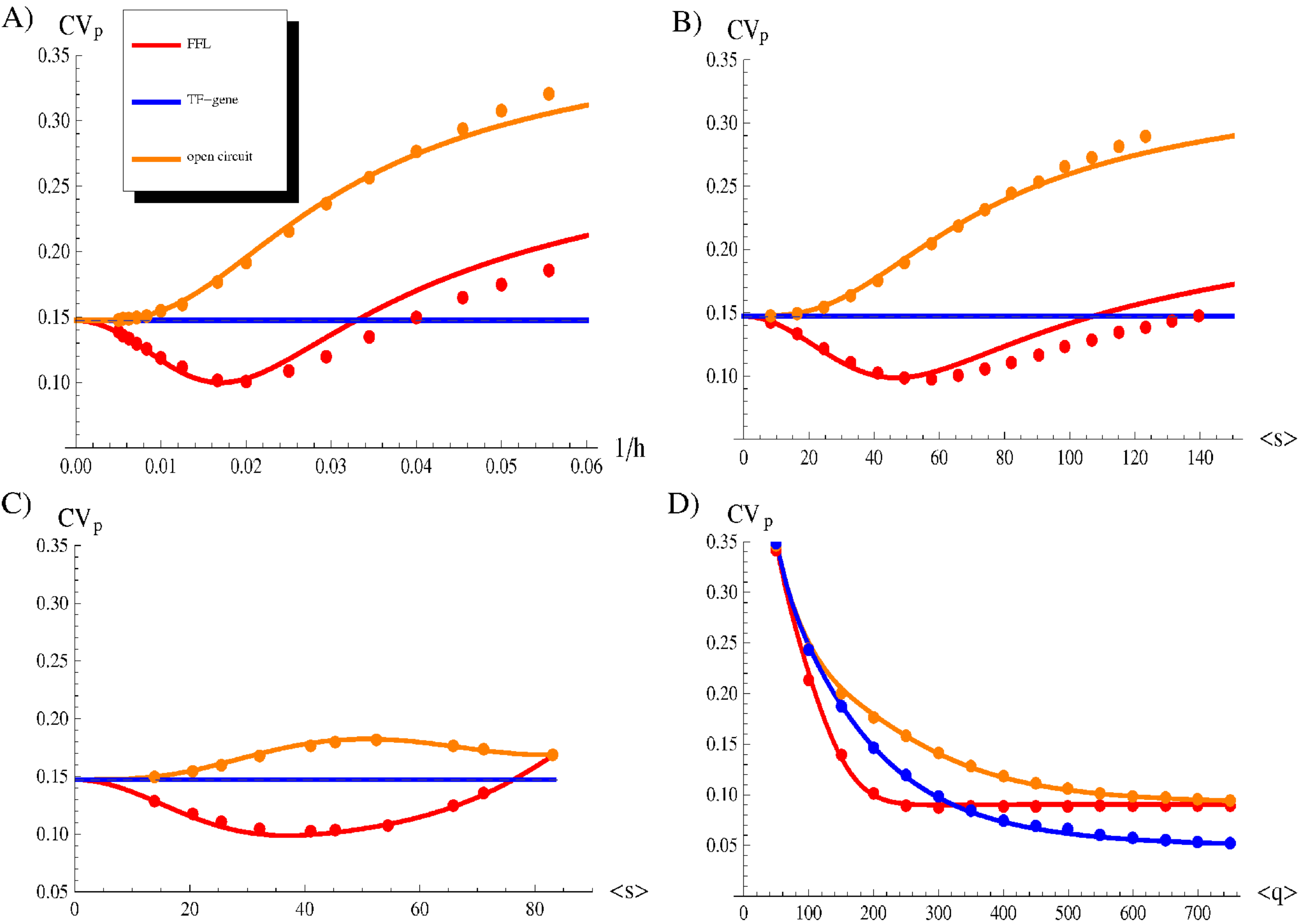}
\end{center}
\caption{
{\bf  How an optimal noise filter can be built.} (A) The coefficient of variation of the target protein $CV_p$ as a function of the repression strength $1/h$. The Figure shows the presence of an optimal repression strength for which the introduction of a miRNA regulation in a FFL minimizes noise. (B) $CV_p$ as a function of the mean number of miRNAs $<s>$. In this case $<s>$ is changed through the maximum rate of transcription $k_s$ (see equation \ref{activ-hill}). (C) $CV_p$ as a function of  $<s>$, varying the dissociation constant $h_s$. In both cases (B and C) is evident a U-shaped profile, implying an optimal noise buffering for intermediate miRNA concentrations. (D) $CV_p$ as a function of the mean number of TFs  $<q>$. The number of TFs depends on the rate of their transcription $k_w$ and of their translation $k_q$. The Figure is obtained manipulating $k_q$, but the alternative choice of $k_w$ leads to equivalent results (see Text S1). For intermediate concentration of the TF, the noise control by the FFL outperforms the one of the other circuits. In each plot, dots are the result of Gillespie simulations while continuous lines are analytical predictions. The values of parameters kept constant are the same of Figure 3, however the results are quite robust with respect to their choice (see Text S1 for details).}
\end{figure}

\begin{figure}[!ht]
\begin{center}
\includegraphics[width=6in]{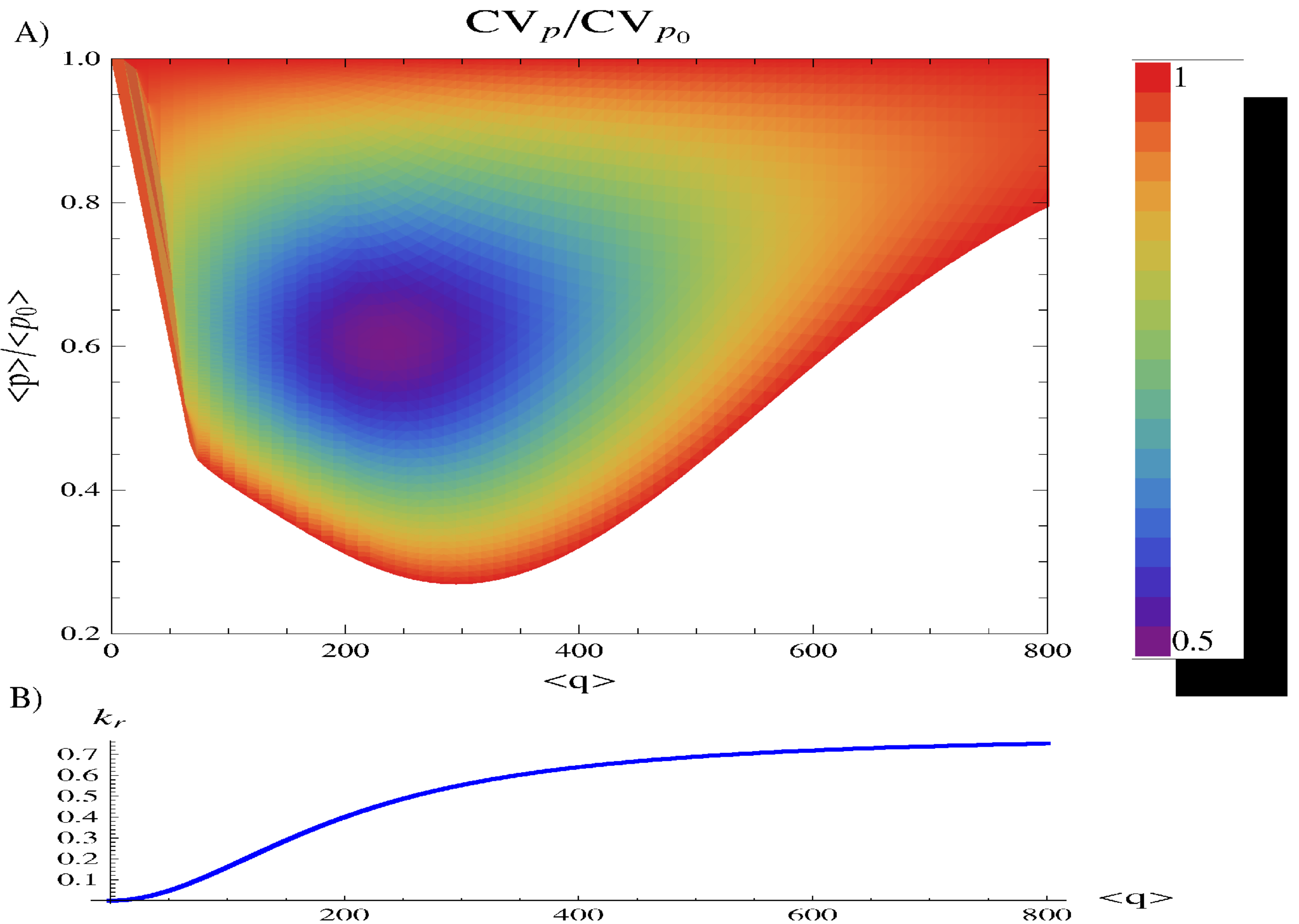}
\end{center}
\caption{
{\bf  Exploring the parameter space.} (A) The target noise  $CV_p$, achieved with the FFL, is evaluated with respect to noise deriving from constitutive expression $CV_{p_{0}}$ (i.e. in absence of miRNA regulation) for different mean levels of the TF $<q>$ and different degrees of reduction  of the target protein level $<p>/<p_{0}>$ (where $<p_{0}>$ is the mean constitutive expression).  The TF level is changed through its rate of translation $k_q$ (equivalent results can be obtained changing the rate of transcription $k_w$), while the target level is tuned varying the repression strength. All the other parameters have the values reported in caption of Figure 3  except $k_w = 0.01263$ (lower than in Figure 3 to explore a more noisy situation). The region where miRNA repression leads to larger fluctuations with respect to constitutive ones is shown in white. When a noise reduction is gained the value of $CV_{p}/CV_{p_{0}}$ is reported with the color code explained in the legend. The best noise control is achieved with a modest suppression of target expression, around the 60\% of its constitutive value. (B) The rate of transcription of the target mRNA as a function of the mean number of TFs. The noise reduction shown in the above plot can be obtained outside the saturation regime (where the slope of the activation curve tends to zero). }
\end{figure}

\begin{figure}[!ht]
\begin{center}
\includegraphics[width=6in]{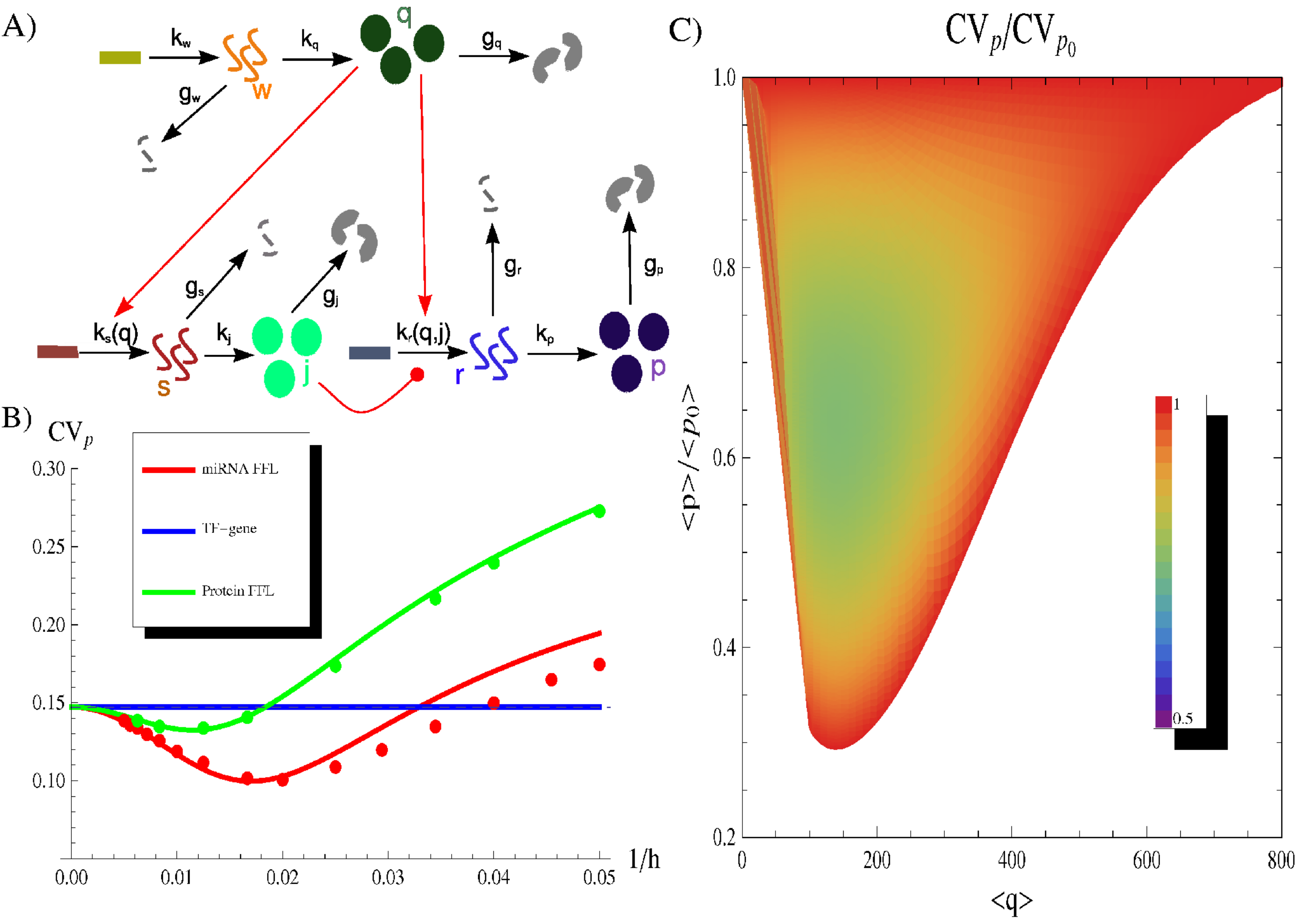}
\end{center}
\caption{
{\bf  Comparison with a purely transcriptional incoherent FFL.} (A) Detailed scheme of a purely transcriptional incoherent FFL. (B) The coefficient of variation of the target protein $CV_p$ as a function of the repression strength $1/h$ for a miRNA-mediated FFL and for its transcriptional counterpart. Thanks to the constraints imposed on parameters we can directly compare their noise-buffering efficiency with respect to a gene only activated by a TF. Both circuitries lead to a $CV_p$ curve with a minimum for an intermediate repression strength, but the miRNA-mediated circuit appears more efficient in filtering out fluctuations. The values of parameters kept constant are the same of Figure 3. Dots are the result of Gillespie simulations with the full nonlinear dynamics while continuous lines are analytical predictions. Also in this case, analytical solutions fit nicely with simulation results. (C) The noise reduction  $CV_p/CV_{p_{0}}$,  achieved with a purely transcriptional incoherent FFL, evaluated for different mean levels of the TF $<q>$ and different degrees of reduction  of the target protein level $<p>/<p_{0}>$. The parameter values and the color code are the same of Figure 7 so as to allow a direct comparison.}
\end{figure}

\begin{figure}[!ht]
\begin{center}
\includegraphics[width=6in]{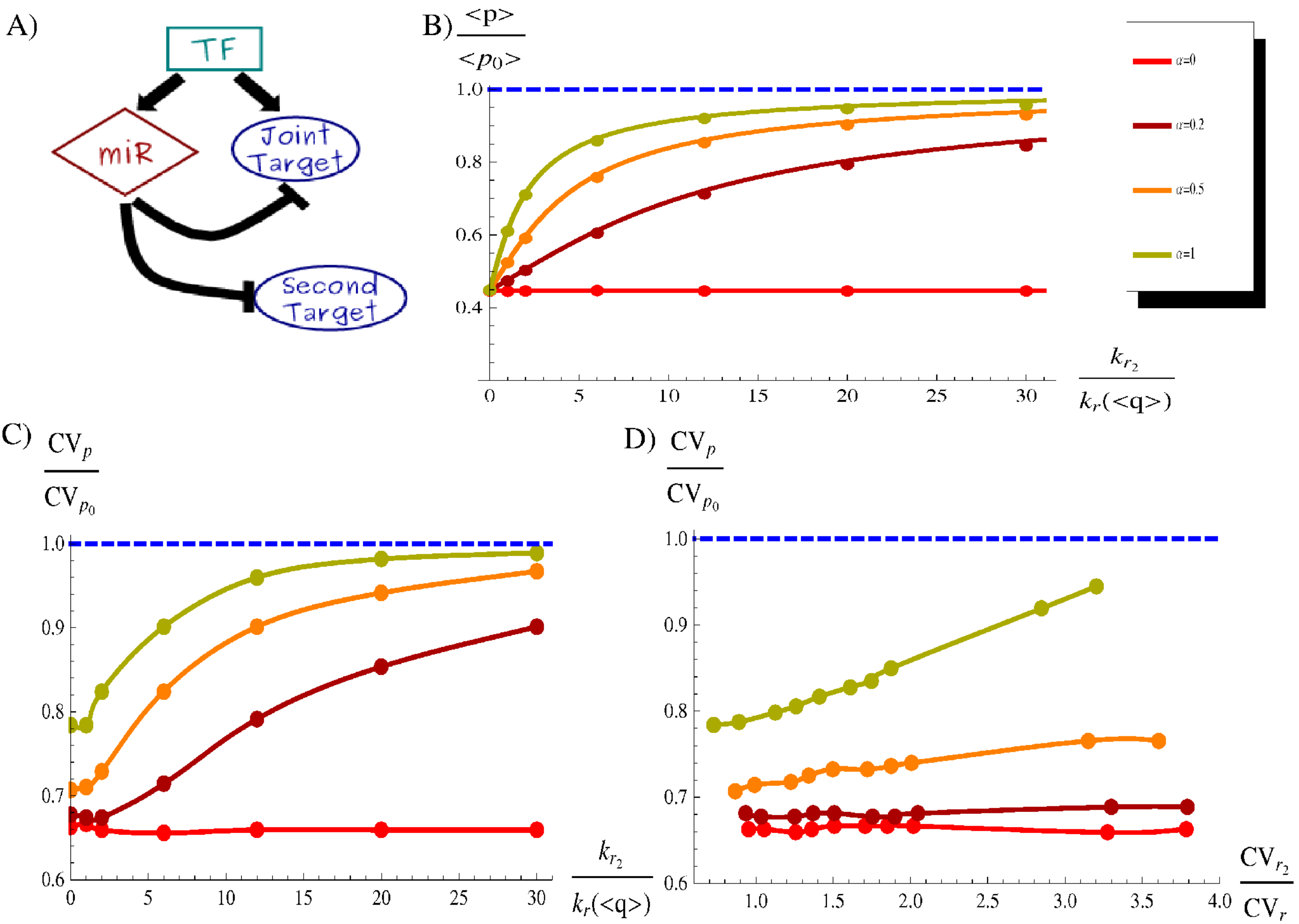}
\end{center}
\caption{
{\bf  Effects of cross-talk between miRNA targets.} (A) Scheme of a miRNA-mediated FFL with an additional independently transcribed target gene (second target). (B) The degree of protein downregulation $<p>/<p_{0}>$ is depicted as a function of the ratio of effective transcription rates of the secondary target ($k_{r_{2}}$) and of the FFL joint target ($k_r(<q>)$), for different values of $\alpha$. Since the rate of transcription of the joint target is a function of the TF concentration, we consider for this analysis the effective mean rate $k_r (<q>)$ as a reference (where $<q>$ is constant as we are not tuning the TF concentration). The transcription of the second target is modeled as an independent birth-death process with birth rate $k_{r_{2}}$. In this plot the coupling constants between targets and  miRNAs are assumed equal ($\gamma_{1}=\gamma_{2} =\gamma$) and  for  each $\alpha$ value the coupling constant $\gamma$ is chosen so as to start with the same amount of target proteins ($<p>$) in absence of secondary targets (the complete set of parameters values is presented in Text S1). In the limit of infinite out-of-circuit target expression, the joint target protein level approaches its constitutive value if $\alpha>0$, while remains constant in the ideal case of perfectly catalytic miRNA repression (red curve). Continuous lines are analytical solutions of the deterministic model (Equations \ref{levine}), while dots are the result of stochastic simulations. (C) With the parameter setting of Figure 9B, the noise reduction $CV_{p}/CV_{p_{0}}$  is evaluated in the same $k_{r_{2}}/k_{r}(<q>)$ range. Dots are the result of Gillespie simulations while continuous lines come from trivial interpolations. (D) The noise reduction is evaluated as a function of the out-of-circuit  mRNA fluctuations  $CV_{r_{2}}$, relative to the joint target fluctuations $CV_{r}$. The fluctuations of the second target are modulated considering its rate of transcription as a function of an independent TF and changing the TF noise with the same strategy used for Figure 5 (see Text S1 for more details). The concentrations of the TFs activating the two targets are constrained to be equal so as to study the situation of two independent targets with the same effective transcription rate.  Dots are the result of Gillespie simulations, simply interpolated with continuous lines.}
\end{figure}

\clearpage
\newpage

\title{{\bf Text S1:} Supporting Information for the article ``The role of incoherent microRNA-mediated feedforward loops in noise buffering''}
\author{M Osella, C Bosia, D Cor\'{a} and M Caselle \\
}

\maketitle

\tableofcontents

\newpage

\section{Deterministic models}

In this section we consider the mean field descriptions  at the steady state of the three networks in analysis: a TF-gene linear circuit without any
 post-transcriptional control, an incoherent miRNA-mediated FFL and an open circuit. As we will show the open circuit is built in order to have the same  
 mean levels of molecular species (in particular of target proteins) that are obtained with the FFL. This feature makes the open circuit a suitable null 
 model in order to disentangle the topology contribution to noise buffering. In this section the variables that describe the state of the system 
 ( $\{w,q,r,s,p\}$ for the FFL) are continuous variables.

\subsection{The TF-gene linear circuit}

\begin{figure}[h!]
\centering
\includegraphics[width=0.7\textwidth]{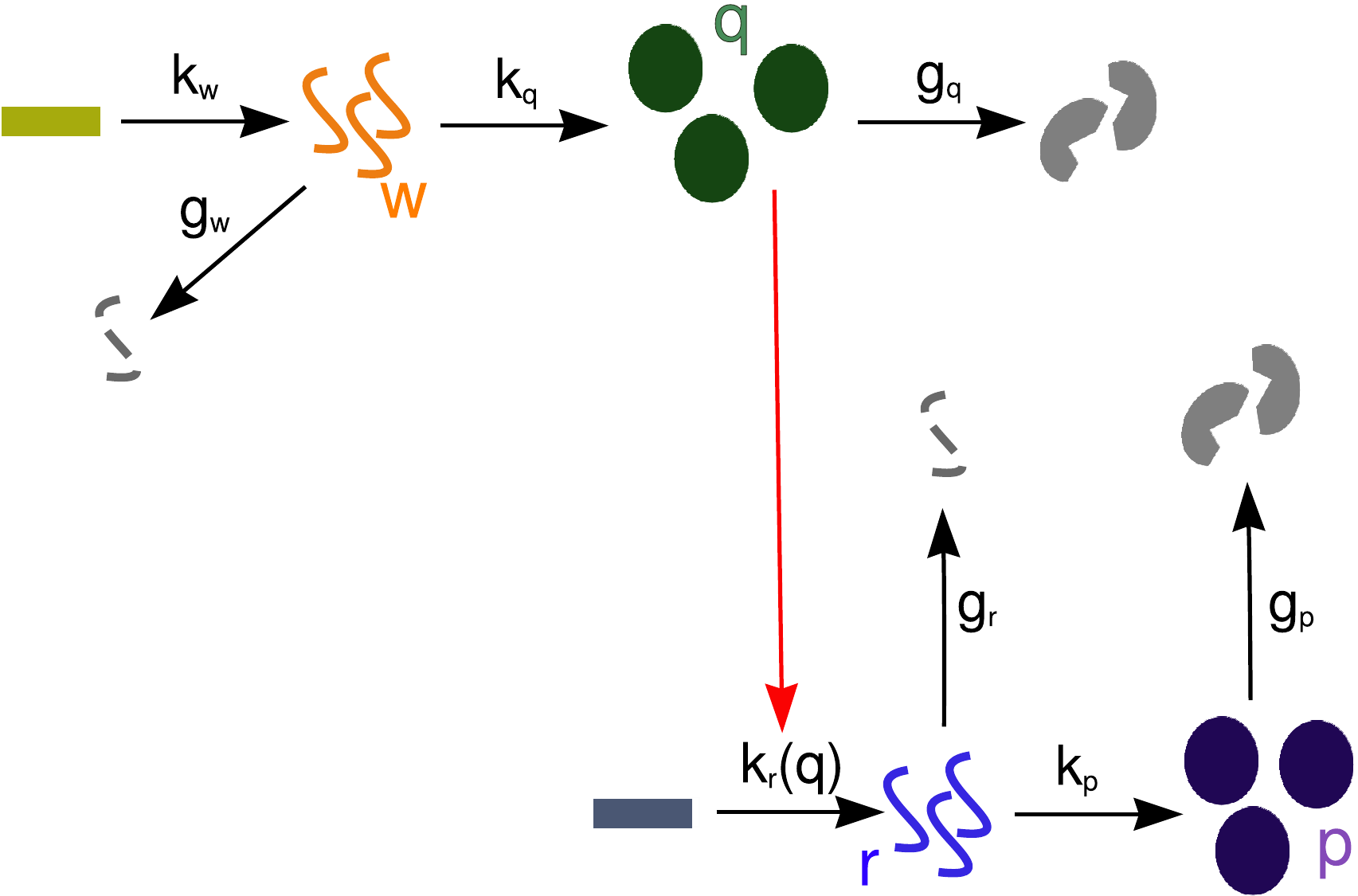}
\caption{Scheme of a TF-gene linear circuit. Rectangles represent DNA-genes, from which RNAs ($w,r$) are transcribed and eventually degraded (broken lines). 
RNAs can be translated into proteins ($q$ is the copy number of  TFs while $p$  of  target proteins)  symbolized by circles, and proteins can be degraded 
(broken circles). Rates of each process (transcription, translation or degradation) are depicted along the corresponding black arrows. 
Regulations are represented in red,  with arrows indicating activation by TFs.}
\label{fig1}
\end{figure}

A deterministic description of the scheme in Fig.\ref{fig1} is given by the equations:

\bea
\frac{d {w}}{{d t}} & =& k_{w} - g_{w} w \nn \\
 \frac{d {q}}{{d t}}& =& k_{q} w - g_{q} q \nn \\
\frac{d {r}}{{d t}} & =& k_{r}(q)  - g_{r} r \nn \\
\frac{d {p}}{{d t}} & =& k_{p} r - g_{p} p, 
\eea

where the rate of transcription of the target mRNAs ($r$) is a Hill function of the number of TFs ($q$): 

\be
k_r(q)  = \frac {k_r q^c} {h_{r}^c +q^c}.
\ee

At the steady state, where $d x_i / dt =0 ~\forall x_i\in\{w,q,r,p\}$, the system of equations can be solved, ending up with:

\bea
w_{ss} & = & \frac{k_{w}}{g_{w}}\nn\\
q_{ss} &= & \frac{k_{q} k_{w}}{g_{q} g_{w}}\nn\\
r_{ss} &= & \frac{k_{q}^2 k_{r} k_{w}^2}{g_{r} (g_{q}^2 g_{w}^2 h_{r}^2 + k_{q}^2 k_{w}^2 )}\nn\\
p_{ss} & = & \frac{k_{p} k_{q}^2 kr k_{w}^2} {g_{p} g_{r} (g_{q}^2 g_{w}^2 h_{r}^2 +k_{q}^2 k_{w}^2)},
\label{nm1}
\eea

where the subscript $ss$ indicates the evaluation at the steady state  and we assumed $c=2$.

\subsection{The FFL}
\label{modelFFL-det}

\begin{figure}[h]
\centering
\includegraphics[width=0.7\textwidth]{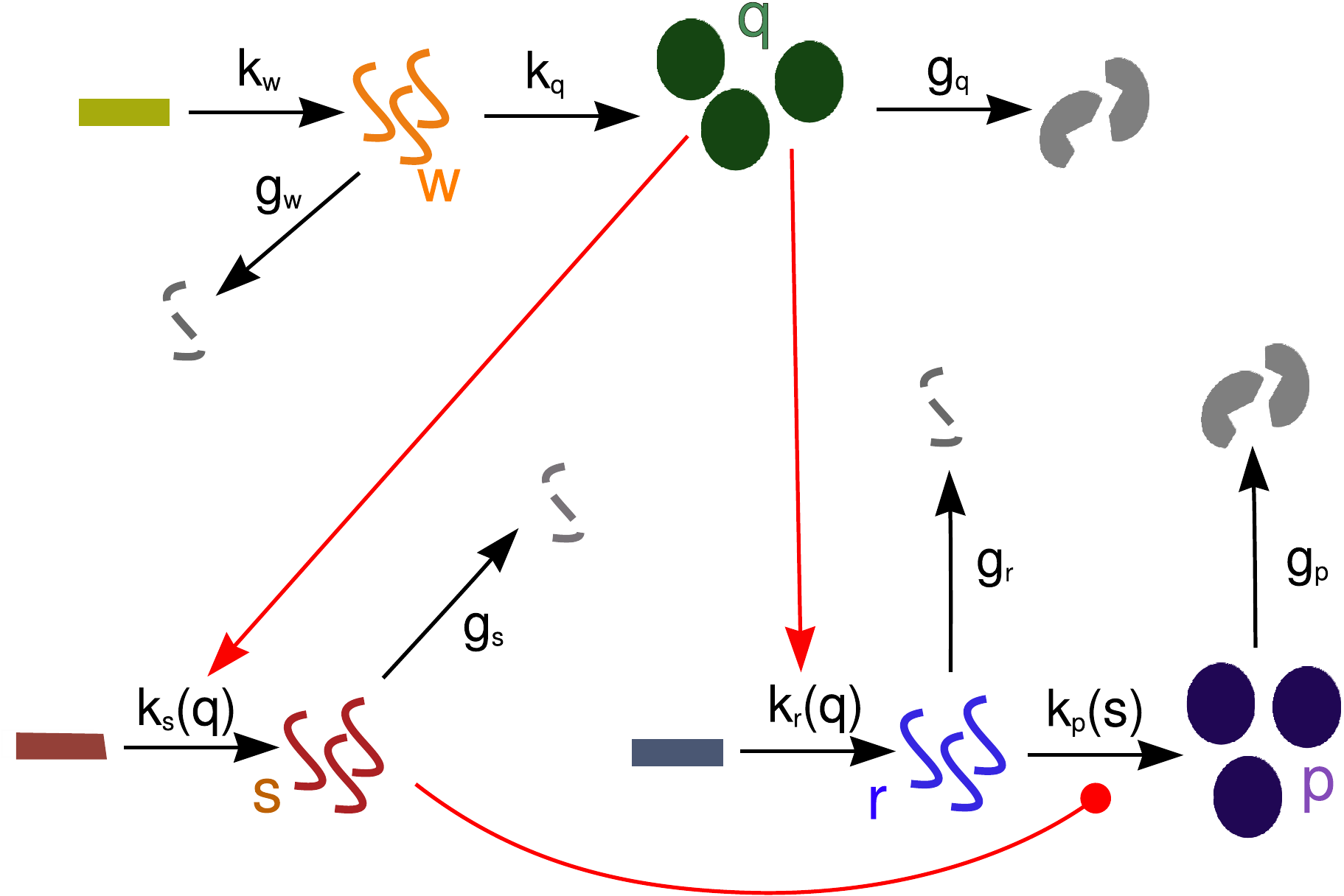}
\caption{Scheme of a miRNA-mediated incoherent FFL. Notations are the same of Fig.\ref{fig1}. The only difference with respect to Fig.\ref{fig1} is the presence of 
  the miRNA gene, 
activated by the TF (red arrow). MiRNA regulation of the target (red rounded end line) makes its  rate of translation  a function of miRNA concentration.}
\label{fig2}
\end{figure}

A deterministic description of the scheme in Fig.\ref{fig2} is given by the equations:

\bea
\frac{d {w}}{{d t}} & =& k_{w} - g_{w} w \nn\\
 \frac{d {q}}{{d t}}& =& k_{q} w - g_{q} q \nn\\
\frac{d {s}}{{d t}} & =& k_{s}(q)  - g_{s} s \nn \\
\frac{d {r}}{{d t}} & =& k_{r}(q)  - g_{r} r \nn \\
\frac{d {p}}{{d t}} & =& k_{p}(s) r - g_{p} p .
\label{detFFL}
 \eea

The transcription rates of the miRNA gene and of the target gene are Hill functions of the number of TFs ($q$), while the translation rate of the 
target gene is a repressive Hill function of the number 
of miRNAs ($s$):

\bea
k_r(q) & =& \frac {k_r q^c} {h_{r}^c +q^c}\nn \\
k_s(q) & =& \frac {k_s q^c} {h_{s}^c +q^c} \nn\\
k_p(s) & = & \frac {k_p} {1 + (\frac{s}{h})^c}.
\label{hill}
\eea

For simplicity we use the same Hill coefficient $c$ for each Hill function, but the analysis can be straigthforwardly generalized to the case of Hill functions with different steepness.\\\
At the steady state, where $d x_i /dt =0 ~\forall x_i \in\{w,q,s,r,p\}$, the system of Eqs.\ref{detFFL} can be solved (we assume again $c=2$), ending up with:

\bea
w_{ss} & = & \frac{k_{w}}{g_{w}} \nn\\
q_{ss} &=&  \frac{k_{q} k_{w}}{g_{q} g_{w}}\nn\\
s_{ss} &=&  \frac{k_{q}^2 k_{s} k_{w}^2}{g_{s} (g_{q}^2 g_{w}^2 h_{s}^2 + k_{q}^2 k_{w}^2 )}\nn\\
r_{ss} &=& \frac{k_{q}^2 k_{r} k_{w}^2} {g_{r} (g_{q}^2 g_{w}^2 h_{r}^2 + k_{q}^2 k_{w}^2 )}\nn\\
p_{ss} & = & \frac{h^2 k_{p} k_{q}^2 k_{r} k_{w}^2}  {g_{p} g_{r} (g_{q}^2 g_{w}^2 h_{r}^2 + k_{q}^2 k_{w}^2 ) \left( h^2 +\frac{ k_{q}^4 k_{s}^2 k_{w}^4}  {g_{s}^2 (g_{q}^2 g_{w}^2 h_{s}^2 +k_{q}^2 k_{w}^2 )^2 }\right) }.
\label{detFFLss}
\eea

The results for $w_{ss},q_{ss},r_{ss}$ coincide with the corresponding ones of the TF-gene linear cascade (see Eqs.\ref{nm1}). 
$s_{ss}$ and $r_{ss}$ have the same functional dependence on the input parameters (except for the obvious substitutions $k_r \leftrightarrow k_s$, $g_r \leftrightarrow g_s$ and
$h_r \leftrightarrow h_s$)  as their expression depends on the amount of TFs in the same way. 
On the contrary  $p$ has a different expression with respect to the linear circuit TF-gene, as in this case  additional terms, related to miRNA repression, appear.

\subsection{The open circuit}

\begin{figure}[h]
\centering
\includegraphics[width=0.7\textwidth]{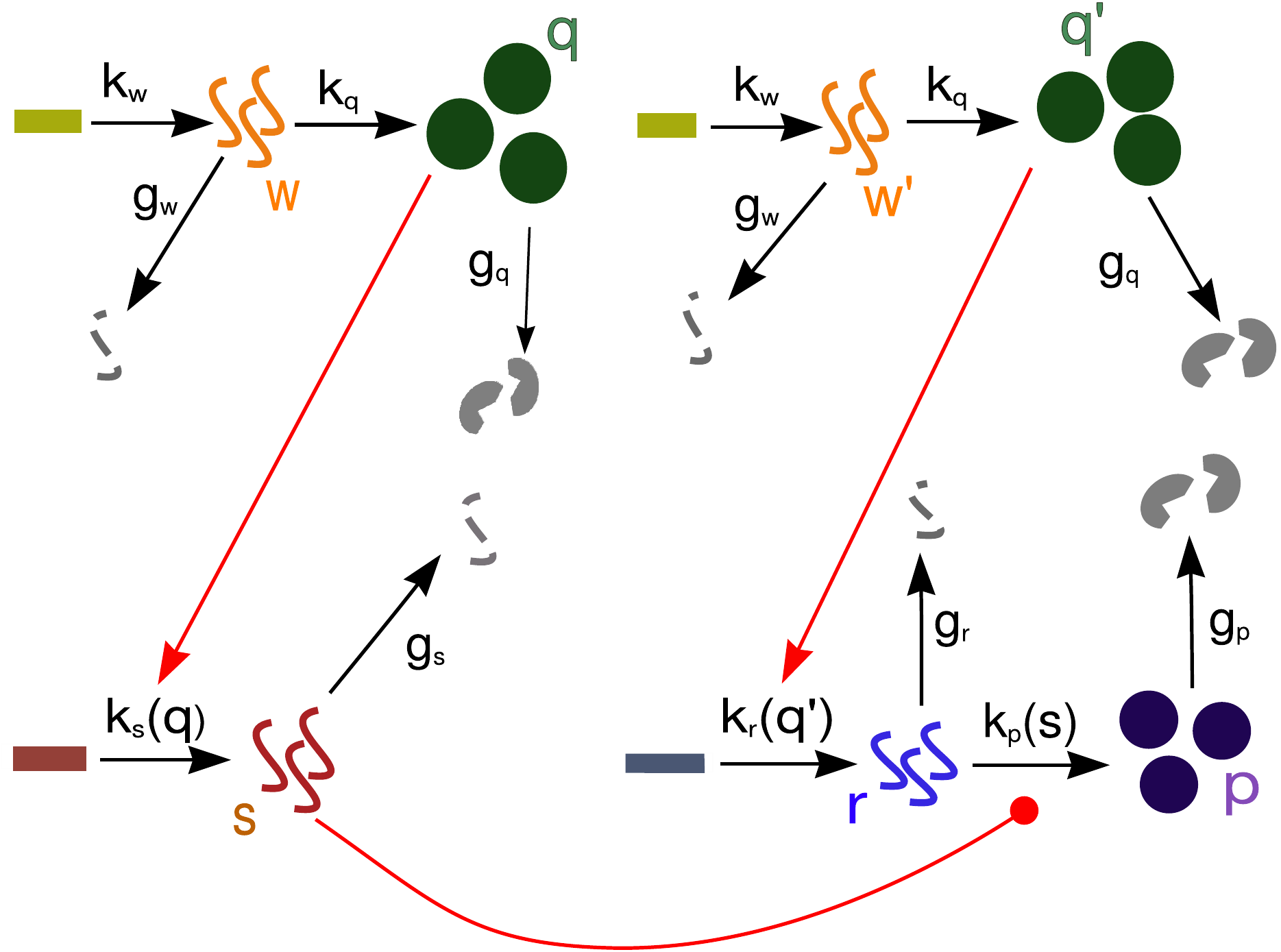}
\caption{Scheme of an open circuit that can lead to the same mean concentrations of molecular species of the FFL. 
The notation is the same of Fig.\ref{fig1}. Unlike the FFL case, here the miRNA and the target gene are activated by two indipendent TFs, 
present in copy numbers $q$ and $q'$. }
\label{fig3}
\end{figure}

A deterministic description of the scheme in Fig.\ref{fig3} is given by the equations:

\bea
\frac{d {w}}{{d t}} & =& k_{w} - g_{w} w \nn\\
\frac{d {q}}{{d t}}& =& k_{q} w - g_{q} q \nn\\
\frac{d {w'}}{{d t}} & =& k_{w} - g_{w} w' \nn\\
\frac{d {q'}}{{d t}}& =& k_{q} w' - g_{q} q' \nn\\
\frac{d {s}}{{d t}} & =& k_{s}(q)  - g_{s} s \nn \\
\frac{d {r}}{{d t}} & =& k_{r}(q')  - g_{r} r \nn \\
\frac{d {p}}{{d t}} & =& k_{p}(s) r - g_{p} p .
\label{detOpen}
\eea

The presence of two independent TFs (copy numbers $q$ and $q'$) that regulate respectively the transcription of $s$ and $r$ 
does not change the expression of $p_{ss}$ previously obtained  in the  FFL case, as long as their rate of transcription, 
translation and degradation are the same of that of the single TF in the FFL and assuming that the Hill functions of 
activation of the target gene and the miRNA gene are exactly the same. The solutions of Eqs.\ref{detOpen} at equilibrium  (with $c=2$) are:

\bea
w_{ss} & = & w'_{ss} = \frac{k_{w}}{g_{w}} \nn\\
q_{ss} &=& q'_{ss}= \frac{k_{q} k_{w}}{g_{q} g_{w}}\nn\\
s_{ss} &=&  \frac{k_{q}^2 k_{s} k_{w}^2}{g_{s} (g_{q}^2 g_{w}^2 h_{s}^2 + k_{q}^2 k_{w}^2 )}\nn\\
r_{ss} &=& \frac{k_{q}^2 k_{r} k_{w}^2} {g_{r} (g_{q}^2 g_{w}^2 h_{r}^2 + k_{q}^2 k_{w}^2 )}\nn\\
p_{ss} & = & \frac{h^2 k_{p} k_{q}^2 k_{r} k_{w}^2}  {g_{p} g_{r} (g_{q}^2 g_{w}^2 h_{r}^2 + k_{q}^2 k_{w}^2 ) \left( h^2 +\frac{ k_{q}^4 k_{s}^2 k_{w}^4}  {g_{s}^2 (g_{q}^2 g_{w}^2 h_{s}^2 +k_{q}^2 k_{w}^2 )^2 }\right) }.
\eea

Therefore, this open circuit  allows the same setting of the concentration of target proteins. As mentioned above, 
this feature makes the open circuit a good null model for comparison with the FFL: as 
the mean field description is the same, any difference between the two will be due to stochastic  fluctuations. 

\section{Stochastic Models}
\label{stoch-model}

We present the master equations for the three circuits discussed in the previous section (Fig.\ref{fig1},\ref{fig2},\ref{fig3}), keeping into 
account the discrete and stochastic nature of chemical reactions. The strategy to find the expression of $<p>$ and $CV_p$ at the steady state 
is the method of the moment generating function (as 
discussed in 
the main text). In this section the variables describing the system ($\{w,q,r,s,p\}$ for the FFL) are discrete and represent the 
actual number of molecules at a specific time. The notation $<x_i>$ 
indicates the mean value at the steady state for the variable $x_i$.

\subsection{Linearization of Hill functions}

As a first step, following \cite{Thattai01SI,Komorowski09SI} we linearize Hill functions in Eqs.\ref{hill}.
This is a commonly used approximation \cite{Thattai01SI,Komorowski09SI} and it is based on the idea that at the steady state the distributions of  
regulators (TFs or miRNAs) have a finite width and sample only small regions of the domains of the corresponding Hill functions.   
We may therefore approximate Hill functions 
by their linearizations about mean values of the regulators $q$ or $s$:

\bea
k_r(q) & \sim & k_r(q)|_{<q>} + \partial_{q} k_r(q)|_{<q>} (q-<q>)\nn \\
k_s(q) & \sim & k_{s}(q)|_{<q>} + \partial_{q} k_s(q)|_{<q>} (q-<q>)\nn\\
k_p(s) & \sim & k_{p}(s)|_{<s>} +  \partial_{s} k_p(s)|_{<s>} (s-<s>).
\label{linearizationSI}
\eea

Defining:

\bea
k_r^0 &=& k_r(q)|_{<q>} -\partial_{q} k_r(q)|_{<q>} <q>\nn\\
k_r^1 &=& \partial_{q} k_r(q)|_{<q>} \nn\\
k_s^0 &=& k_s(q)|_{<q>} -\partial_{q} k_s(q)|_{<q>} <q>\nn\\
k_s^1 &=& \partial_{q} k_s(q)|_{<q>} \nn\\
k_p^0 &=&  k_p(s)|_{<s>} -\partial_{s} k_p(s)|_{<s>} <s>\nn\\
k_p^1 &=&  -\partial_{s} k_p(s)|_{<s>} ,
\eea

and substituting in Eqs.\ref{linearizationSI} we obtain Eqs.4 of the main text:

\bea
k_r(q) & \sim & k_{r}^0 + k_{r}^1 q \nn\\
k_s(q) & \sim & k_{s}^0 + k_{s}^1 q \nn\\
k_p(s) & \sim & k_{p}^0 - k_{p}^1 s.
\label{linearization2}
\eea

\subsection{The TF-gene linear circuit}

The master equation that describes the circuit in the scheme of Fig. \ref{fig1} is:

\bea
& & \frac{\partial P_{w,q,r,p}}{\partial t} = ~~k_{w} (P_{w-1,q,r,p} - P_{w,q,r,p}) + k_q w (P_{w,q-1,r,p}-P_{w,q,r,p})\nn \\
& & + k_r(q) (P_{w,q,r-1,p} - P_{w,q,r,p})+ k_p r (P_{w,q,r,p-1}-P_{w,q,s,r,p}) \nn\\
& & + g_w \Bigl[ (w+1)P_{w+1,q,r,p} - w P_{w,q,r,p} \Bigr]+ g_q \Bigl[ (q+1)P_{w,q+1,r,p} - q P_{w,q,r,p} \Bigr]\nn\\
& & +  g_r \Bigl[ (r+1)P_{w,q,r+1,p} - r P_{w,q,r,p} \Bigr]+ g_p \Bigl[ (p+1)P_{w,q,s,r,p+1} - p P_{w,q,s,r,p} \Bigr].
\label{maSG}
\eea

Introducing the moment generating function as:

\be
F(z_1,z_2,z_3,z_4) = \sum_{w,q,r,p} z_{1}^w ~z_{2}^q ~ z_{3}^r ~z_{4}^p ~P_{w,q,r,p},
\ee

and using the linearized form of Hill functions in Eq.\ref{linearization2}, we can convert Eq.\ref{maSG} into a first order partial differential equation (PDE):

\bea
{\partial_t F } &= & k_w ( z_1 F - F) + k_q z_1(  z_2  {\partial_{z_1}} F -  {\partial_{z_1}} F ) + k_{r}^{0} (z_3 F - F)\nn\\
& & + k_{r}^{1} z_2 ( z_3 {\partial_{z_2}} F - {\partial_{z_2}} F )  + k_{p} z_3 ( z_4  {\partial_{z_3}} F - {\partial_{z_3}} F ) \nn\\
& & + g_w ( {\partial_{z_1}} F - z_1 {\partial_{z_1}} F ) + g_q ( {\partial_{z_2}} F - z_2 {\partial_{z_2}} F ) \nn\\
& & + g_r ( {\partial_{z_3}} F - z_3 {\partial_{z_3}} F ) + g_p ( {\partial_{z_4}} F - z_4 {\partial_{z_4}} F ).
\label{mgSG}
\eea

This equation cannot be solved exactly but it is not difficult to extract the first two moments of the probability distributions $P_{w,q,r,s}$
at the steady state, thus allowing
to obtain a close expression for $p$ and $CV_p = \s_p /<p>$.

\subsubsection{Moments of the distribution}

These moments can be evaluated by deriving  Eq.\ref{mgSG} 
at the steady state ($\partial_t F =0$) and using  the following properties of the moment generating function: $F|_1=1 ; F_i = <x_i> ; F_{ii} = <x_i^2> -<x_i>$ 
(with the notation $F_i=\partial_{x_i}F$), where $|_1$ means evaluation of $F$ at $x_i =1$ for all $i$. We only discuss here the derivatives which are needed to obtain
$F_3 = <p>$ and $F_{3,3} -F_3^2 +F_3 =\s_p^2$. 

\bea
F_1 &=& k_{w}/g_{w}\nn\\
F_2 & = & \frac{k_{q} F_1}{ g_{q}}\nn\\
F_3 & = & \frac{k_{r}^0 +k_{r}^1  F_2} {g_{r}}\nn\\
<p>=F_{4} & = & \frac{k_{p} F_3} {g_{p}}\nn\\
F_{1,1} &=& \frac{k_{w} F_1}{g_{w}}\nn\\
F_{1,2} &=& \frac{k_{q} F_1 + k_{w} F_2 +k_{q} F_{1,1}} {g_{q}+g_{w}}\nn\\
F_{1,3} & = & \frac{k_r^0 F_1 +k_{w} F_3 +k_r^1 F_{1,2}} {g_{r} + g_{w}}\nn\\
F_{1,4} &=&  \frac{k_{w} F_{4} + k_{p} F_{1,3}} {g_{p} +g_{w}}\nn\\
F_{2,2} &=& \frac{k_{q} F_{1,2}} {g_{q}}\nn\\
F_{2,3} &=& \frac{k_{r}^0  F_{2} + k_{r}^1 F_{2} + k_{q} F_{1,3} + k_{r}^1 F_{2,2}} {g_{q} + g_{r}}\nn\\
F_{2,4} &= & \frac{k_{q} F_{1,4} + k_{p} F_{2,3}} {g_{p} + g_{q}}\nn\\
\s_p^2 -<p> +<p>^2 = F_{3,3}& = & \frac{k_{r}^0  F_{3} +k_r^1 F_{2,3}}  {g_{r}}\nn\\
F_{3,4} &=& \frac{k_{p} F_{3} + k_r^0 F_{4} + k_{r}^1  F_{2,4} + k_{p} F_{3,3}}  {g_{p} + g_{r}}\nn\\
F_{4,4} &=& \frac{k_{p} F_{3,4}} {g_{p}}.
\eea

\subsection{The FFL}
\label{stoch-FFL}
The master equation describing the circuit in the scheme of Fig. \ref{fig2} is:

\bea
& & \frac{\partial P_{w,q,s,r,p}}{\partial t} = ~~k_{w} (P_{w-1,q,s,r,p} - P_{w,q,s,r,p}) + k_q w (P_{w,q-1,s,r,p}-P_{w,q,s,r,p}) \nn \\
& & + k_r(q) (P_{w,q,s,r-1,p} - P_{w,q,s,r,p}) + k_s(q) (P_{w,q,s-1,r,p} - P_{w,q,s,r,p}) \nn\\
& & +k_p(s) r (P_{w,q,s,r,p-1}-P_{w,q,s,r,p}) + g_w \Bigl[ (w+1)P_{w+1,q,s,r,p} - w P_{w,q,s,r,p} \Bigr]\nn\\
& & + g_q \Bigl[ (q+1)P_{w,q+1,s,r,p} - q P_{w,q,s,r,p} \Bigr]+  g_r \Bigl[ (r+1)P_{w,q,s,r+1,p} - r P_{w,q,s,r,p} \Bigr]\nn\\
& & + g_s \Bigl[ (s+1)P_{w,q,s+1,r,p} - s P_{w,q,s,r,p} \Bigr]+ g_p \Bigl[ (p+1)P_{w,q,s,r,p+1} - p P_{w,q,s,r,p} \Bigr].
\label{maFFLSI}
\eea

Introducing the moment generating function:

\be
F(z_1,z_2,z_3,z_4,z_5) = \sum_{w,q,s,r,p} z_{1}^w ~z_{2}^q ~ z_{3}^s~ z_{4}^r ~z_{5}^p ~P_{w,q,s,r,p},
\ee

and using the linearization in Eqs.\ref{linearizationSI}, we can convert Eq.\ref{maFFLSI} into a PDE that is of second order in this case:

\bea
{\partial_t F } &= & k_w ( z_1 F - F) + k_q z_1(  z_2  {\partial_{z_1}} F -  {\partial_{z_1}} F ) + k_{r}^{0} (z_4 F - F)\nn\\
& & + k_{r}^{1} z_2 ( z_4 {\partial_{z_2}} F - {\partial_{z_2}} F ) + k_{s}^{0} ( z_3 F - F) + k_{s}^{1} z_2 ( z_3  {\partial_{z_2}} F -  {\partial_{z_2}} F )\nn\\
& & + k_{p}^{0} z_4 ( z_5  {\partial_{z_4}} F - {\partial_{z_4}} F ) - k_{p}^{1} z_3 z_4 (z_5 {\partial_{z_3,z_4}} F -  {\partial_{z_3,z_4}} F )\nn\\
& & + g_w ( {\partial_{z_1}} F - z_1 {\partial_{z_1}} F ) + g_q ( {\partial_{z_2}} F - z_2 {\partial_{z_2}} F ) + g_s ( {\partial_{z_3}} F - z_3 {\partial_{z_3}} F )\nn\\
& & + g_r ( {\partial_{z_4}} F - z_4 {\partial_{z_4}} F ) + g_p ( {\partial_{z_5}} F - z_5 {\partial_{z_5}} F ).
\label{maFSI}
\eea

As mentioned in the main text, even if we linearized the 
Hill functions in the master equation (Eq.\ref{maFFLSI}), the term related to the translation of the regulated target keeps a  nonlinear contribution 
due to the product $k_{p}^{1}sr$. This has the effect of making Eq.\ref{maFSI} a second order PDE.\\
Remarkably enough one can nevertheless obtain  closed analytical expressions for $<p>$ and $CV_p$. The only additional complication, with respect to TF-gene case
discussed in the previous section is that the calculation of some fourth moments is required. 
We do not report here the expression of all the moments for the sake of shortness but they can be easily derived with tedious but straightforward algebra.
  
\subsection{The open circuit}

The master equation describing the circuit in the scheme of Fig. \ref{fig3} is:

\bea
& & \frac{\partial P_{w,q,w',q',s,r,p}}{\partial t} = k_{w} \Bigl[(P_{w-1,q,w',q',s,r,p} - P_{w,q,w',q',s,r,p})\nn\\
& & + (P_{w,q,w'-1,q',s,r,p} - P_{w,q,w',q',s,r,p}) \Bigr] \nn\\
& & + k_q \Bigl[ w (P_{w,q-1,w',q',s,r,p}-P_{w,q,w',q',s,r,p}) \nn\\
& & + w' (P_{w,q,w',q'-1,s,r,p}-P_{w,q,w',q',s,r,p})\Bigr]\nn\\
& & + k_r(q) (P_{w,q,w',q',s,r-1,p} - P_{w,q,w',q',s,r,p})\nn\\
& & + k_p(s) r (P_{w,q,w',q',s,r,p-1}  -P_{w,q,w',q',s,r,p})\nn\\
& & + k_s(q') (P_{w,q,w',q',s-1,r,p} - P_{w,q,w',q',s,r,p})  \nn\\
& & + g_s \Bigl[ (s+1)P_{w,q,w',q',s+1,r,p} - s P_{w,q,w',q',s,r,p}\Bigr] \nn\\
& & + g_w \Bigl[ (w+1)P_{w+1,q,w',q',s,r,p}  - w P_{w,q,w',q',s,r,p} \Bigr] \nn\\
& & +g_q \Bigl[ (q+1)P_{w,q+1,w',q',s,r,p} - q P_{w,q,w',q',s,r,p} \Bigr]\nn\\
& & + g_w \Bigl[ (w'+1)P_{w,q,w'+1,q',s,r,p} - w' P_{w,q,w',q',s,r,p} \Bigr] \nn\\
& & + g_q \Bigl[ (q'+1)P_{w,q,w',q'+1,s,r,p} - q' P_{w,q,w',q',s,r,p} \Bigr]\nn\\
& & + g_r \Bigl[ (r+1)P_{w,q,w',q',s,r+1,p} - r P_{w,q,w',q',s,r,p} \Bigr]\nn\\
& & + g_p \Bigl[ (p+1)P_{w,q,w',q',s,r,p+1} - p P_{w,q,w',q',s,r,p} \Bigr].
\label{maFopen}
\eea

Introducing the moment generating function:

\be
F(z_1,z_2,z_3,z_4,z_5,z_6,z_7) = \sum_{w,q,w',q',s,r,p} z_{1}^w ~z_{2}^q ~ z_{3}^{w'}~ z_{4}^{q'} ~ z_{5}^s ~ z_{6}^r ~ z_{7}^p ~ P_{w,q,w',q',s,r,p},
\ee

and using the linearization in Eqs.\ref{linearizationSI} we can convert Eq.\ref{maFopen} into a second order PDE analogous to Eq.\ref{maFSI}. 
The expression of $<p>$ and $CV_p$ can be obtained as in the FFL case differentiating up to fourth moments.

\subsection{Numerical definition of the steady-state time-threshold} 

To perform Gillespie simulations in order to check the analytical results, we must define the time $t_c$ at which the system can be considered at the steady state. 
In previous papers \cite{Thattai01SI,Komorowski09SI} the steady state was assumed to be reached at a time equal to ten times the protein half-life. 
We tried to slightly improve  this definition.\\
For each circuit in analysis, we evaluated numerically the deterministic dynamics for the set of parameter values chosen for the simulations, assuming 
the  initial conditions $x_i=0~
\forall i$. Then we defined $t_c$ as the value above which  the difference between $p(t_c)$ and its asymptotic value $p_{ss}$ (in units of $p_{ss}$)
becomes smaller than a given threshold $\e$.
\be
\frac{p_{ss}-p(t_c)}{p_{ss}} = \e .
\label{tc}
\ee

\begin{figure}[h!]
\centering
\includegraphics[width=0.7\textwidth]{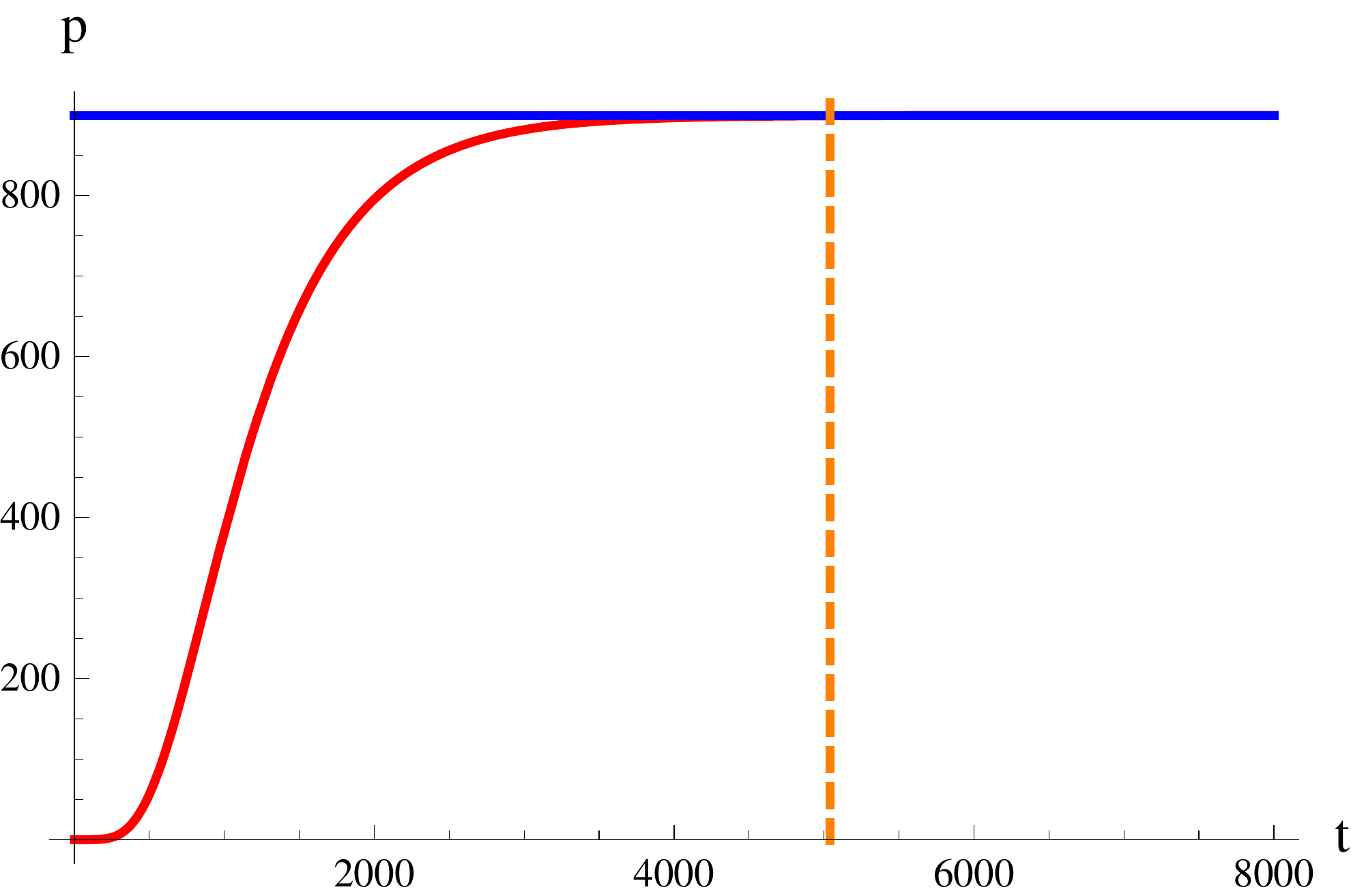}
\caption{Deterministic evolution toward the steady state. We report in blue the numerical solution of Eqs.\ref{detFFL} for $p$ with initial conditions 
$x_i=0~\forall i$. The red line represents the number of protein at equilibrium given by Eq.\ref{detFFLss}. The dashed orange line represents the time 
$t_c$ that satisfies Eq.\ref{tc}. At  time $t_c=5000 s$, resulting from  Eq.\ref{tc}  with $\e=0.05$,  the curves are almost indistinguishable. 
The rate of transcription of the TF is $k_w =0.06 s^{-1}$ and of translation $k_q = 0.04 s^{-1}$. Proteins degrade with a probability $g_q=g_p=0.002 s^{-1}$ 
while mRNAs and miRNAs with probability $g_w = g_r = g_s = 0.006 s^{-1}$. The parameters in the Hill functions (Eqs.\ref{hill}) are as follows. 
Maximum rate of transcription for mRNAs: $k_r =0.8 s^{-1}$, while for miRNAs: $k_s=0.5 s^{-1}$; maximum rate of translation of the target: $k_p=0.04 s^{-1}$; 
dissociations constants: $h_s=200,h_r=200,h=60$. The Hill coefficients are always $c=2$.}
\label{fig4}
\end{figure}

We report in fig.\ref{fig4} an example of this calculation in the case of the FFL circuit (details on the parameters and initial conditions are reported in the caption).
Setting as threshold $\e=0.05$ (which, given the size of the fluctuations which we are interested in, turns out to be a  rather conservative value) we found for this
particular circuit $t_c \sim 5000 s$, which corresponds to about 14 times the protein half-life.

This procedure allows us to treat the different circuits on the same ground and eliminates a possible source of numerical bias.
    
\subsection{Robustness of our results with respect to changes in the values of the input parameters}

\subsubsection{Constraints imposed on the FFLs by the requirement of sensitivity to changes in the master TF concentration. }

Functional FFLs can be  defined as those in which a change in the master TF concentration can cause a change in the concentration of target proteins and miRNAs \cite{Mangan03SI}. While the issue of the trade-off between sensitivity to signals and noise control is discussed in detail in Section \ref{sensitivity}, in the following we shall define more simply the conditions on parameters that ensure a sufficient dependence of miRNA and target mRNA levels on the TF concentration. Noise propagation requires a target dependence on TF concentration, therefore only in this case noise buffering can be functional.
In our contest this dependence implies that the Hill functions of activation by the TF 
and of repression by the miRNA should not be 
saturated at the steady state. Indeed, in conditions of complete saturation, signals and fluctuations cannot propagate from the master TF to the target (even in absence of miRNA regulation), therefore a noise control lose any functionality. On the other hand, in the unsaturated regime a change in the number of TFs can alter in a significant way the number of target proteins in the cell together with the number of miRNAs,  generating the correlated fluctuations needed for noise buffering.  If the TF concentration is too high (with respect to $h_r$ and $h_s$), the expression rates
 of the target and miRNA genes become insensitive to variations in TF 
concentration (unless they are so large that can escape 
from the region of saturation) limiting the sensitivity of the FFL to upstream signals. 
The same considerations hold for the target repression. If there are too many 
miRNAs (with respect to $h$), the target 
expression is drastically shut down and again the system becomes insensitive  
to changes in the number of TFs. Accordingly
we excluded from our analysis the parameter 
sets for which: $<q>\ll h_r(h_s)$ or $<q>\gg h_r(h_s)$ and $<s>\ll h$ or $<s>\gg h$. 
In other words, the circuit functionality imposes that concentrations of regulators 
 must be placed not far from the linear 
region of the corresponding Hill functions.
A high sensitivity corresponds also to an overexposure to noise, in fact noise amplification and sensitivity are correlated quantities \cite{Shibata05SI,Hornung08SI} (see also Section \ref{sensitivity}).  
Since the aim of our study is to prove the noise 
buffering role of miRNA-mediated incoherent FFL, considering the parameter space that strongly exposes to noise makes clearly sense and it seems not a limitation.\\
 With the conditon of unsaturated regulations satisfied, the qualitative results in the article apply for virtually all parameter choices. As a partial proof, in the next 
 two sections we shall discuss a few different combinations of parameters. As we shall see our results turn out to be remarkably 
 robust with respect to changes in the allowed
 (unsaturated) region of parameters.

\subsubsection{Target and miRNA genes differentially expressed}

In this section we present the target noise strength for the three circuits as a function of the ratio between the maximum rate of transcription of miRNA gene ($k_s$) 
and target gene ($k_r$), keeping fixed the TF concentration ($<q>$) and miRNA repression strength ($1/h$).
The aim is to show that the noise buffering role of the mixed FFL shows only a weak dependence on the characteristics  of miRNA and target promoters. 
In the upper part of Fig.\ref{fig5b} we plot the
target noise strength as a 
function of $k_r/k_s$. One can see that in the whole range of values the mixed FFL shows the largest noise reduction effect and in particular that
the noise buffering role of the FFL  does not require an equal rate of transcription of miRNAs and mRNAs. Indeed, as discussed in the main text, the 
noise attenuation is due to 
the correlation 
of fluctuations in the number of 
mRNAs and miRNAs and not to their absolute values. Different  maximum rates of transcription ($k_r$ and $k_s$) only change the height of peaks in mRNA 
and miRNA trajectories, without 
affecting their correlation.

\subsubsection{mRNAs and miRNAs with different stability} 

\begin{figure}[h!]
\centering
\includegraphics[width=0.9\textwidth]{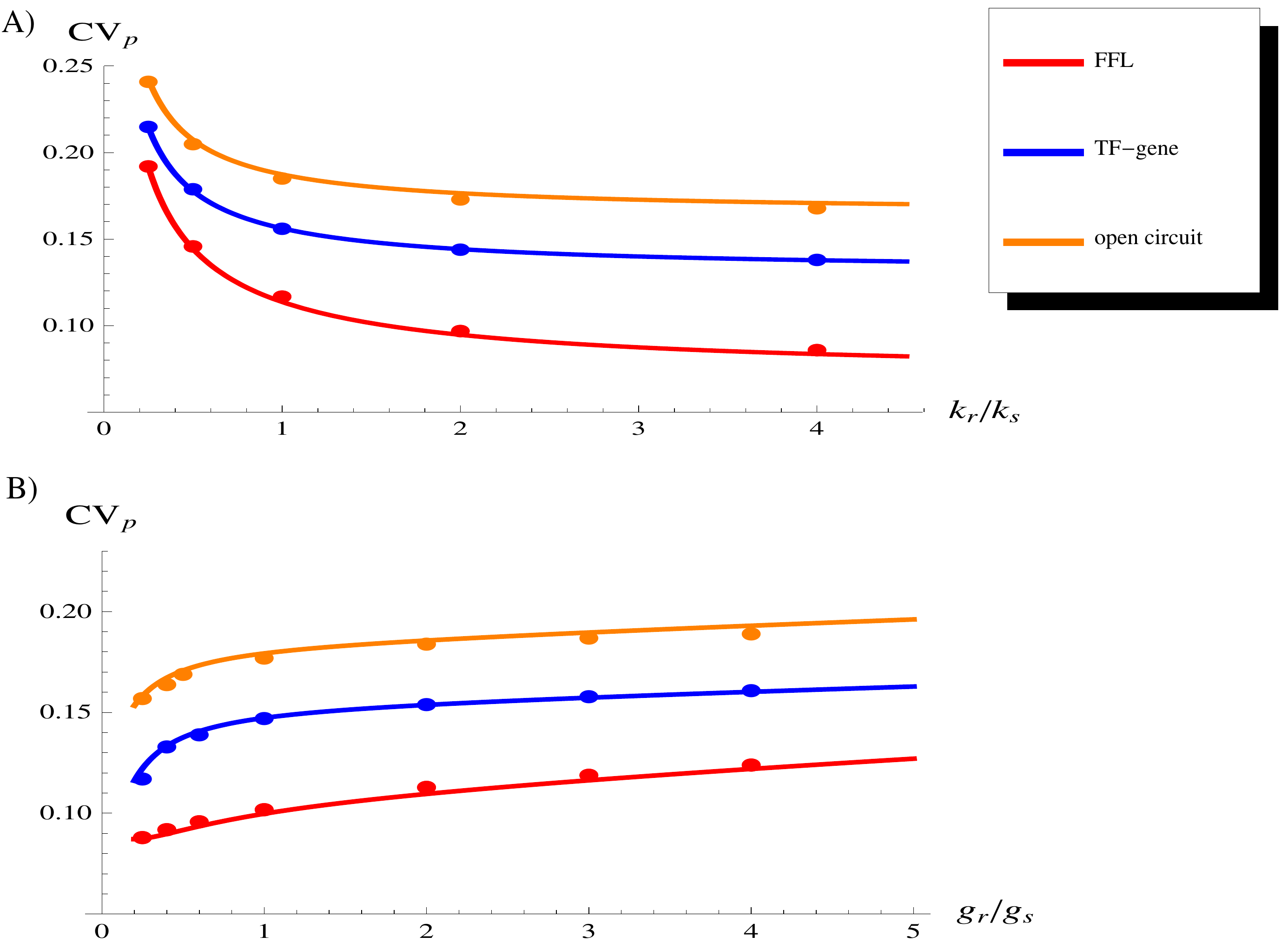}
\caption{(A) $CV_p$ as a function of the transcription rate ratio $k_r/k_s$. (B)  $CV_p$ as a function of $g_r/g_s$.}
\label{fig5b}
\end{figure}

Another important robustness test is the dependence of the FFL noise buffering efficiency on the ratio of decay constants $g_r/g_s$. In principle one could
expect a reduction in the FFL efficiency when $g_r\not= g_s$ due to the fact that with different values of $g_r$ and $g_s$ the mRNA and miRNA trajectories could 
start to fluctuate out of phase due to different 
relaxation times. 
To answer this question  we calculated the $CV_p$ for the three circuits as a function of the ratio $g_r/g_s$. The results are reported in the lower part of 
Fig.\ref{fig5b}.  

As in the previous case, we find that in the whole range of $g_r/g_s$ that we studied the mixed FFL gives the largest noise reduction effect.

These two tests together show that noise buffering is a generic feature of mixed FFLs and that 
there is no need to fine tune the half-life and/or the transcription rate of miRNAs and mRNAs to 
obtain a mixed FFL that efficiently reduces fluctuations. 

\subsubsection{Optimal TF concentration tuning $k_{w}$ instead of $k_{q}$}

In the main text we discussed the dependence of the noise strength $CV_p$ on the copy number of TFs present at the steady state (Fig.6C of the main text). 
The parameter chosen to tune $<q>$ was the rate of translation $k_q$. For the sake of completeness we report here the 
same plot obtained by varying $k_w$ instead of $k_q$. Also with this alternative protocol the FFL outperforms the other circuits  in noise control for 
intermediate concentration of the TF. This is a further proof of the robustness of our results. 

\begin{figure}[h!]
\centering
\includegraphics[width=0.7\textwidth]{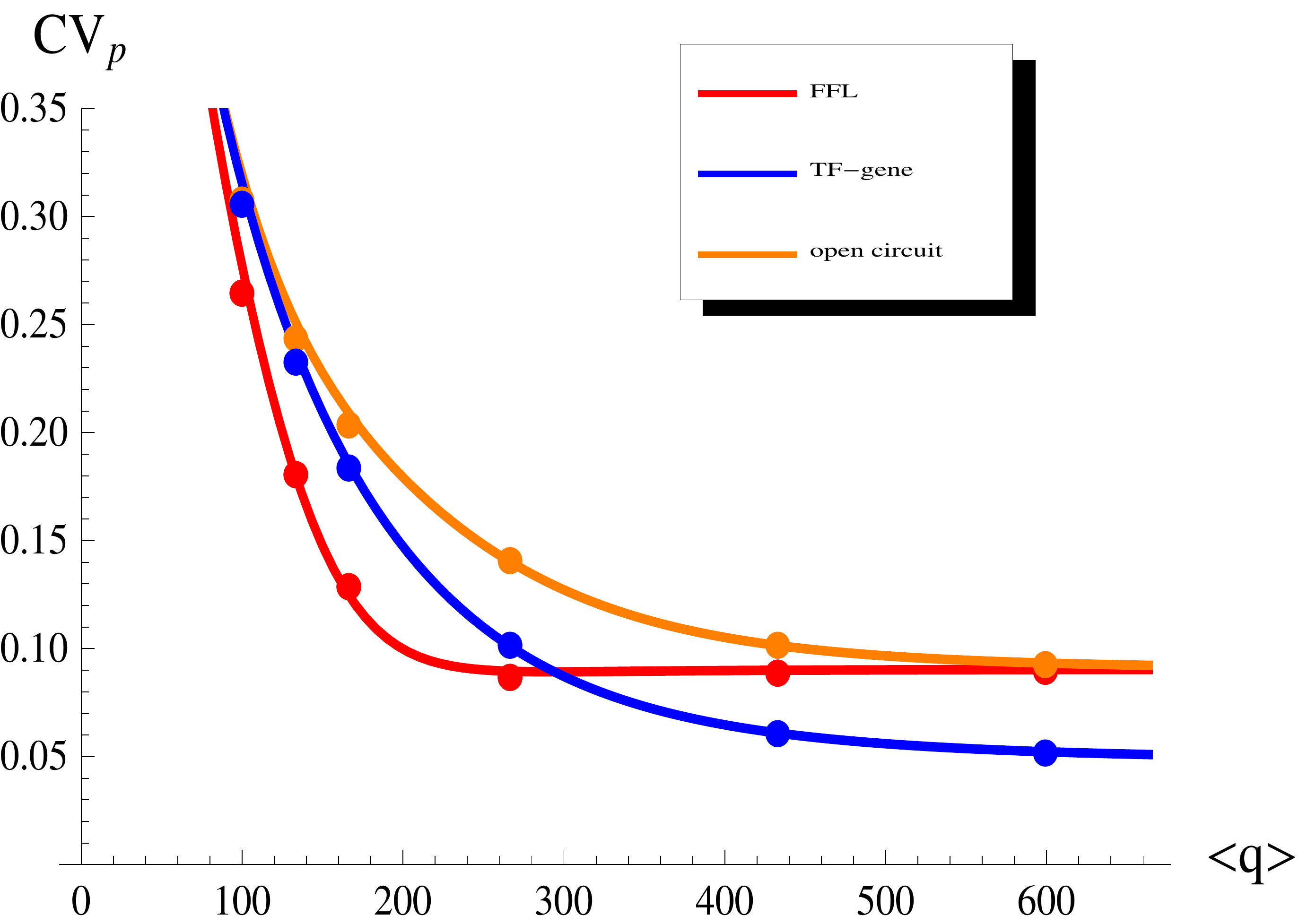}
\caption{$CV_p$ as a function of the mean number of TFs  $<q>$. Dots are the result of Gillespie simulations with the 
full dynamics while continuous lines are the analytical predictions. The parameter values are the same of Fig.\ref{fig4}}
\label{fig6}
\end{figure}

\subsubsection{Results for another set of parameters}

\begin{figure}[h!]
\centering
\includegraphics[width=0.9\textwidth]{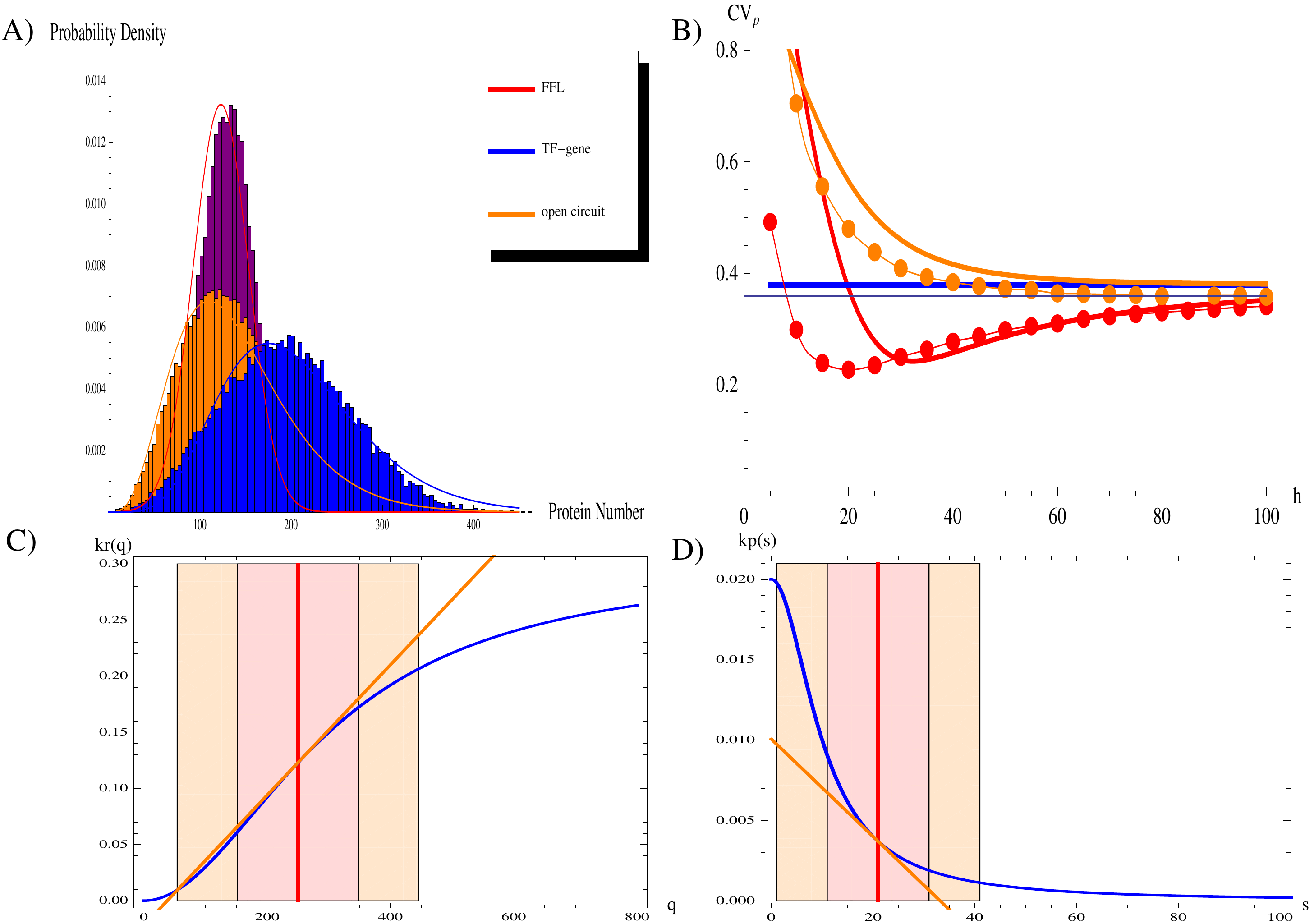}
\caption{(A)The probability distribution of target protein number for the three circuits in analysis. The parameter values are: $k_{w} = 0.01 s^{-1}$; $k_{q}=0.3 s^{-1}$;
 $g_{w}=g_{r}=g_{s}= 0.006 s^{-1}$; $ 
g_{q}=g_{p}= 0.002 s^{-1}$; $k_{r} = k_{s} =0.3 s^{-1}$; ; $h_{s}=h_{r}= 200$; $c= 2$; $h=30$. Histograms are the result of Gillespie simulations with the full nonlinear 
dynamics, while continuous lines 
are empirical distributions (gaussian for the FFL and gamma for the TF-gene and the open circuit) with mean and variance predicted by the analytical model.
(B) The coefficient of variation of the target 
protein $CV_p$ as a function of the inverse of repression strength $h$ for the three circuits. (C) The Hill function of transcriptional activation of the target gene 
(blue line). The red line represents 
the mean number of TFs $<q>$ at equilibrium, while the shaded region corresponds to intervals $[q-\s_{q},q+\s_{q}]$ and $[q-2\s_{q},q+2\s_q]$. 
The orange line represents the linearized function used for 
the analytical solution. (D) The Hill function of translational repression of the target gene (blue line) by the miRNA, in the strong repression region ($h=10$). 
The red line represents the mean number of miRNAs 
$<s>$ at equilibrium, while the shaded region corresponds to intervals $[s-\s_{s},s+\s_{s}]$ and $[s-2\s_{s},s+2\s_s]$. The orange line represents the linearized 
function used for the analytical 
solution.}
\label{fig5}
\end{figure}

As a final test of robustness we solved the master equations for the three circuits with a choice of input parameters (reported in the caption of Fig.\ref{fig5}A) leading to sizeable fluctuations in 
the number of master TFs ($CV_q\sim0.4$). 
This should be compared with the values  of the case dicussed in the main text (whose parameter set is reported in the caption of Fig.\ref{fig4}) for which the 
noise in the number of TFs was only $CV_q\sim0.17$. Also in this case  the TF fluctuations are efficiently attenuated by the 
FFL, leading  to a final value of the noise strength in the target protein of $CV_p=0.25$ to be compared with $CV_p=0.38$ for the direct TF-gene 
regulation and $CV_p=0.46$ for the open circuit (see the histograms in Fig.\ref{fig5}A). These values agree with the observation reported in the main text that the 
noise attenuation effect due to the FFL circuit becomes larger and larger as the size of TF fluctuations increases.  
The U-shaped profile of $CV_p$ for the FFL steps out also for this parameter set,  further supporting the idea that this property  does not depend 
on their particular choice but is a generic feature of the model (see Fig.\ref{fig5}B).

\subsection{Testing the effect of Hill function linearization}
\label{alt-set}

Besides the robustness against the choice of input parameters another important issue which one would like to address is the effect of the linearization of the Hill
functions. This can be easily achieved by comparing analytical versus numerical (Gillespie) results for the noise reduction.
Since this is the only approximation that we made in our analysis it is important to understand which is the range of parameters in which we can trust our analytical
results not only qualitatively but also quantitatively. It is easy to guess that the linear approximation
 should give sizeable errors only when the  fluctuations in the variables become large enough to cover a wide portion of the Hill function thus
 exploring also its non-linear part. A good example to discuss this issue is given by the set of input parameters discussed in the previous section. In this case, even if the analytical solution still captures qualitatively the main features of the systems, it is less precise in its quantitative predictions. 
This is clearly visible in Fig.\ref{fig5}B where analytical predictions are compared with the results of Gillespie simulations (which keep into account the full non linear
dynamics of the FFL) as a function of the inverse of repression strength $h$. 
While for the value of $h$  discussed above ($h=30$) the agreement is very good, as $h$ decreases the 
gap between the two curves becomes larger and larger. 
 This is a consequence of the linearization of  Hill functions and shows that 
if fluctuations are too large, as it happens in the strong repression regime, the linear approximation may become too crude. It is interesting to study how the approximation breaks
down since it is a typical example of the subtle effects which the two step nature of gene expression may have on noise propagation.
 With the choice of parameters of the figure, the $q$ 
fluctuations cover a wide region of the domain of $k_r(q)$ and $k_s(q)$ (Fig.\ref{fig5}C), but the line tangent in $<q>$ still captures quite well the Hill function trend, 
with only a slight 
overestimation ($<r>=<s> =20$ from simulations, compared to the predicted value of $21$). On the other hand, the large fluctuations in $s$ ($CV_s=0.48$) make the 
linearization of $k_p(s)$ a poor 
approximation (Fig.\ref{fig5}D). The $s$ distribution spreads on a domain region where the Hill function widely changes its curvature,  therefore the tangent line 
introduces in many trajectories a 
sizeable underestimation of the rate of target translation. As a result we have $<p> =43$ from simulations while only $<p>=28$ from the analytical model. 
In a similar way also the 
standard deviation turns out to be uncorrectly estimated by the analytical solution. These disagreements explain the displacement of analytical curves 
in Fig.\ref{fig5}B with respect to simulations. This example shows however that, despite its quantitative failure, 
the analytical model describes fairly well the qualitative behaviour of the system even in presence of large fluctuations and, as mentioned above,
it becomes more and more precise when 
fluctuations around steady state values cover a domain where the Hill functions are approximately linear (which is the usual assumption in literature).

\section{MiRNA-mediated promotion of mRNA degradation}
\label{degradation-catalytic}

\begin{figure}[h!]
\centering
\includegraphics[width=0.7\textwidth]{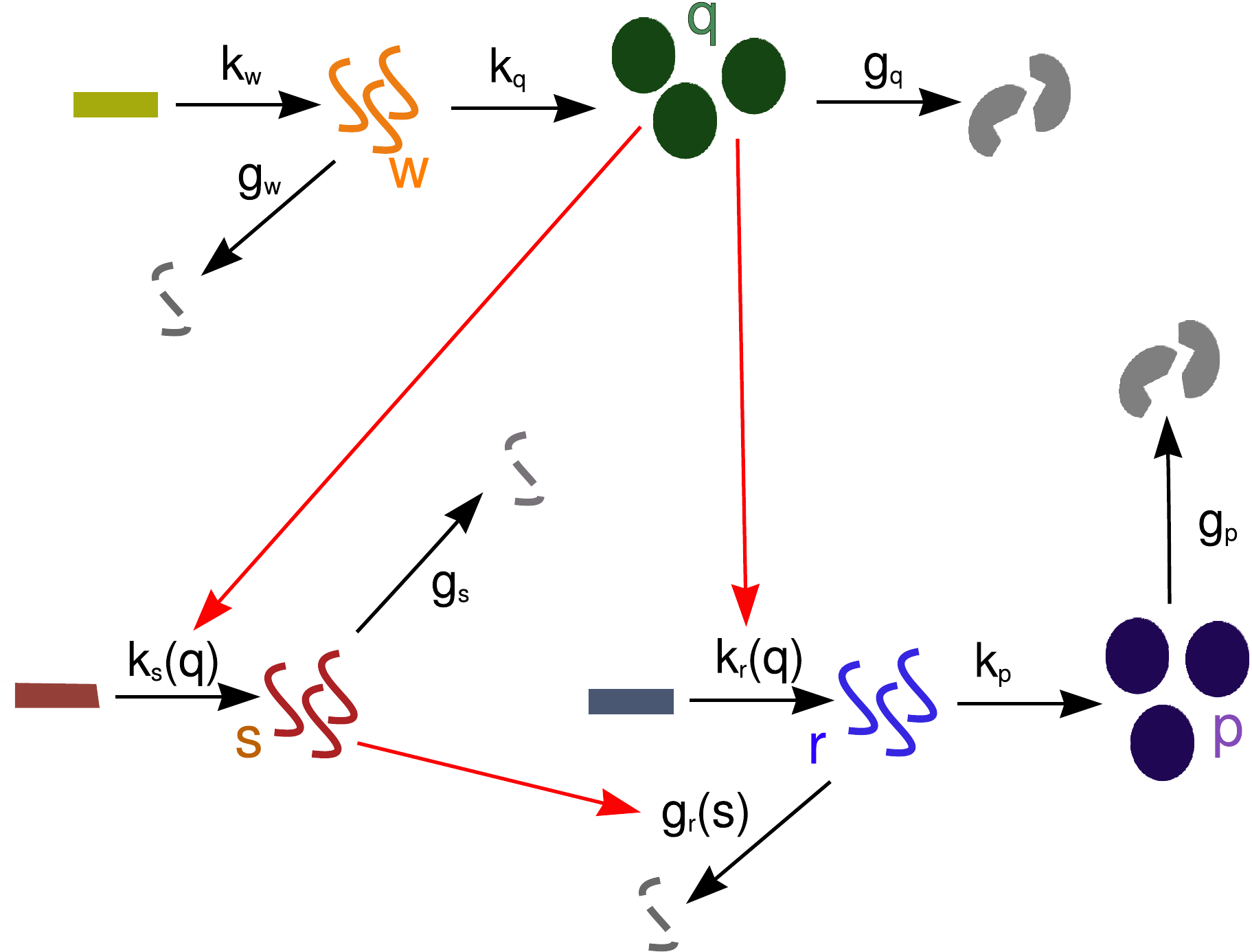}
\caption{Scheme of a miRNA-mediated incoherent FFL, where the miRNA performs its repressive function by promoting mRNA degradation. The notation is the 
same of Fig.\ref{fig1}. The red arrow 
starting from $s$ represents the regulation of the rate of degradation $g_{r}(s)$, which in this case is a non-linear increasing function of miRNA concentration.}
\label{fig7}
\end{figure}

By base pairing to mRNAs, miRNAs can mediate translational repression or mRNA degradation \cite{Sanchez06SI,Pillai07SI,Filipowicz08SI}. As discussed in the main 
text, we developed our model considering the miRNA action as repressing the target translation, making the 
rate of mRNA translation a nonlinear decreasing function of the number of miRNAs.
In this section we will prove the validity of our results even in the case of miRNA repression based on promotion of target mRNA degradation.
In this case we can introduce the miRNA action adding to the basal rate of mRNA degradation $g_r$ (in absence of miRNAs) an 
increasing Hill function of the copy number of 
miRNAs:   

\be
g_{r}(s) = g_{r} + \frac{g_{max} s^c} {h_{deg}^c +s^c},
\label{deg-hill}
\ee
 
where  $g_{max}$  represents  the  maximum 
possible increase of the degradation rate in case of high miRNA concentration (if $s\to\infty , g_{r}(s) \to g_{r} + g_{max}$); 
$h_{deg}$ is the dissociation 
constant of miRNA-mRNA interaction;  $c$ is the Hill coefficient.\\
 The stochastic models built on this assumptions cannot be solved with the same strategy explained in section \ref{stoch-model}. The closure of equations for $<p>$ and $\s_p$ would require further linearizations.
However, we  ran simulations for the alternative mechanism of miRNA-mediated 
 promotion of  target mRNA degradation to check 
the robustness of our results. Strikingly enough these simulations can be fit quite well with the analytical predictions based on the assumption of a miRNA-mediated repression of translation.

\subsection{Deterministic model}

The TF-gene linear circuit  is modelled as previously shown. We present here only the deterministic equations for the FFL, since the open circuit case 
can be easily obtained from the FFL description following the same steps discussed above for the translational repression case.
The mean field description of the system in Fig.\ref{fig7} is:

\bea
\frac{d {w}}{{d t}} & =& k_{w} - g_{w} w \nn\\
 \frac{d {q}}{{d t}}& =& k_{q} w - g_{q} q \nn\\
\frac{d {s}}{{d t}} & =& k_{s}(q)  - g_{s} s \nn \\
\frac{d {r}}{{d t}} & =& k_{r}(q)  - g_{r}(s) r \nn \\
\frac{d {p}}{{d t}} & =& k_{p} r  - g_{p} p ,
 \eea

where $k_{s}(q)$ and $k_{r}(q)$ are the Hill functions of activation shown in Eqs.\ref{hill}, while the form of $g_{r}(s)$ is shown in Eq.\ref{deg-hill}.\\
Assuming $c=2$ the expressions at steady state of $w_{ss},q_{ss},s_{ss}$ are the same of Eqs.\ref{detFFLss}, as nothing is changed in their dynamics, while 
the expressions of $r_{ss}$ and $p_{ss}$ become:

\bea
r_{ss}& = &  \frac{ k_{q}^2 k_{r} k_{w}^2 (g_{q}^4 g_{s}^2 g_{w}^4 h_{deg}^2 h_{s}^4 + 2 g_{q}^2 g_{s}^2 g_{w}^2 h_{deg}^2 h_{s}^2 k_{q}^2 k_{w}^2} { (g_{q}^2 g_{w}^2 h_{r}^2 + k_{q}^2 k_{w}^2 ) (g_{q}^4 g_{r} g_{s}^2 g_{w}^4 h_{deg}^2 h_{s}^4 + 2 g_{q}^2 g_{r} g_{s}^2 g_{w}^2 h_{deg}^2 h_{s}^2 k_{q}^2 k_{w}^2 } ..\nn\\
& & \frac{+ k_{q}^4  ( g_{s}^2 h_{deg}^2 + k_{s}^2) k_{w}^4 )} {+ k_{q}^4 (g_{r} g_{s}^2 h_{deg}^2 + (g_{r}+ g_{max}) k_{s}^2) k_{w}^4)}\nn\\
p_{ss} &=& r_{ss} k_{p} / g_{p}.
\label{degSS}
\eea

\subsection{Comparison with miRNA-mediated repression of mRNA translation}
\label{comp-deg}

\begin{figure}[h!]
\includegraphics[width=0.9\textwidth]{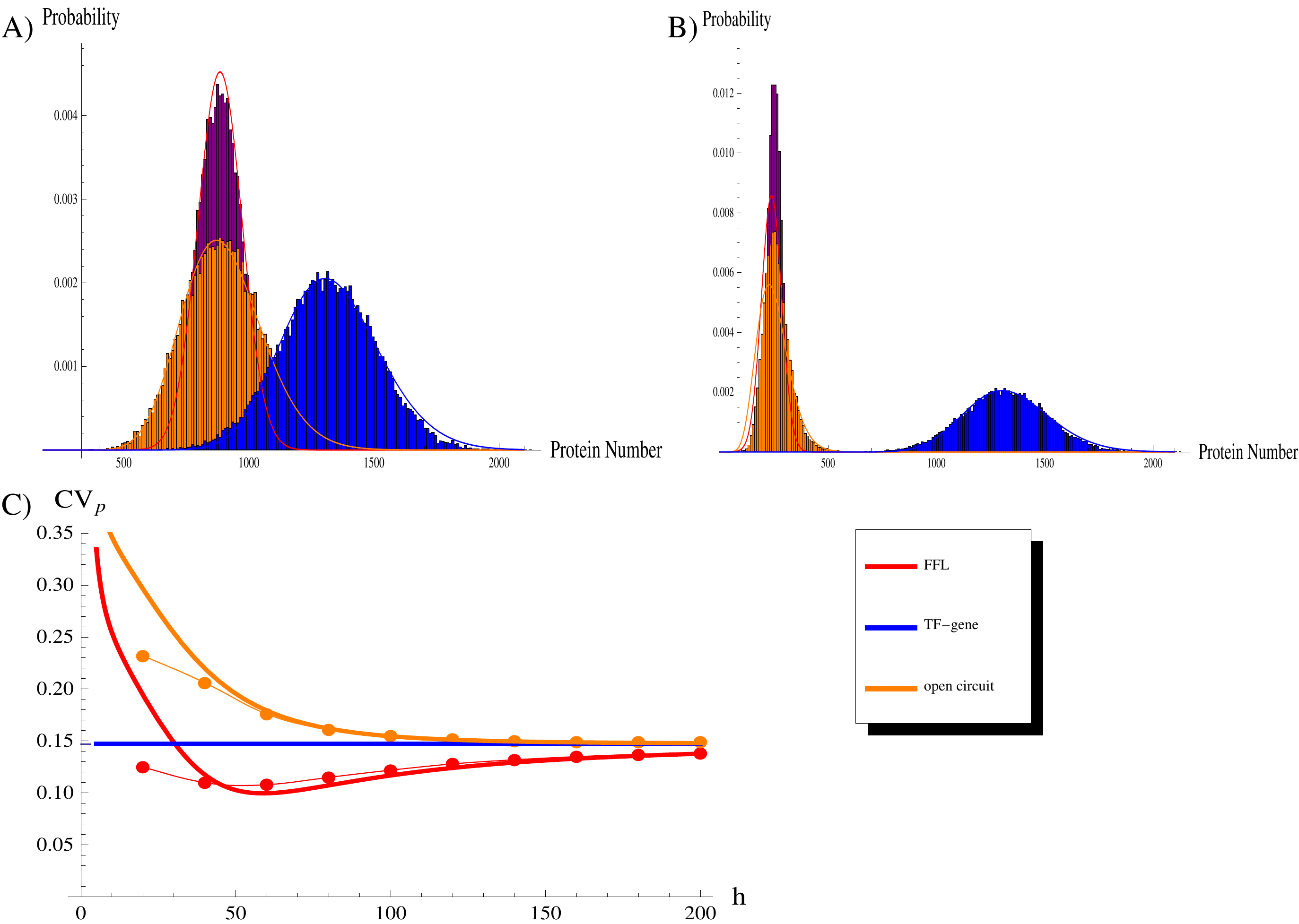}
\centering
\caption{Attenuation of noise by a FFL with a miRNA promoting degradation. (A) The probability distribution of target protein number for the three circuits in analysis.
 Histograms are the result of 
Gillespie simulations with the nonlinear dynamics depicted in Fig.\ref{fig7}. Continuous lines are empirical distributions (gaussian for the FFL and gamma for the 
TF-gene and the open circuit) with mean and 
variance predicted by the analytical stochastic model shown in section \ref{stoch-FFL}. The parameter values are those explained in caption of Fig.\ref{fig4}. 
Even if the analytical model is built on 
the hypothesis of repression of mRNA-translations, it fits equally well the distributions resulting from simulations based on miRNA-mediated  promotion of mRNA degradation. 
(B) Same histograms of A with a stronger repression ($h=20$, 
all other parameters as stated before). In the regime of strong repression the analytical model  tends to overestimate the variance $\s_p$. 
(C) The coefficient of variation of the target protein $CV_p$ 
as a function of the inverse of repression strength $h$ for the three circuits. The figure shows the presence of an optimal repression strength 
even in the case of a degradation-based miRNA repression. 
Dots are the results of Gillespie simulations with the hypothesis of a miRNA-mediated promotion of  mRNA degradation, while thick lines are analytical predictions. 
Apart from the mentioned overestimation in the strong repression region the model fits quite well Gillespie simulations.}
\label{fig8}
\end{figure}

In order to compare in an unbiased way the noise properties of the mixed FFL with different mechanisms of miRNA action, 
we set up the parameters of the two alternative models (Fig.\ref{fig2} and \ref{fig7}) so as to achieve the same final levels of the target protein $p_{ss}$. 
This can be obtained by choosing the same parameters for the two models except those involved in the miRNA regulation. These last  may then be fixed by equating
the values of $p_{ss}$ in Eq.\ref{detFFLss} and \ref{degSS}.
As show in Fig.\ref{fig8} the result of this comparison 
is that a mixed FFL with a degradation-based repression gives essentially the same results of the corresponding circuit with a 
translation-based repression.

In particular, we show in Fig.\ref{fig8}A the analogous (for the present repression scheme) 
of the histograms of Fig.3C and 4C of the main text. 
As in the translational repression case also in this model the noise buffering effect of the FFL is clearly visible thus suggesting that the inchoerent
FFL loop performs equally well its noise buffering function with either type of repression mechanism. 
Superimposing the distributions with mean and variance calculated analytically for the miRNA-mediated repression of translation  we find again a very good agreement, apart from a slight disagreement  
in the strong repression regime (small $h$). In conclusion, all the results presented in the main paper hold despite the mechanism of miRNA repression and even if the analytical predictions are based on the assumption of a miRNA-mediated repression on mRNA translation, they can be applied also to this case.

\section{Stoichiometric mechanism of repression}

Regulatory small noncoding RNAs (sRNAs) play a crucial role also in prokaryotes gene regulation. In particular, the class of {\it trans}-acting sRNAs has many features 
in common with miRNAs  in 
eukaryotes: most of them bind to the UTR of the target mRNAs through base-pairing (often imperfect) recognition
to prevent their translation or to promote their degradation. 
However, as discussed in \cite{Metha08SI}, 
unlike their eukaryotic counterpart they usually act stoichiometrically on their targets, since a given sRNA molecule is often degraded along with its target,
 instead of being used to regulate other targets. 
Different authors \cite{Metha08SI,Levine07SI,Shimoni07SI} studied the peculiar features of this noncatalytic sRNA-mediated regulation, developing a simple kinetic
 model for sRNA gene silencing.\\
In this section  we shall study the noise buffering properties of incoherent FFL motifs  assuming a stoichiometric modality of repression, and compare our results
with the previously discussed catalytic case. 

\begin{figure}[h!]
\centering
\includegraphics[width=0.7\textwidth]{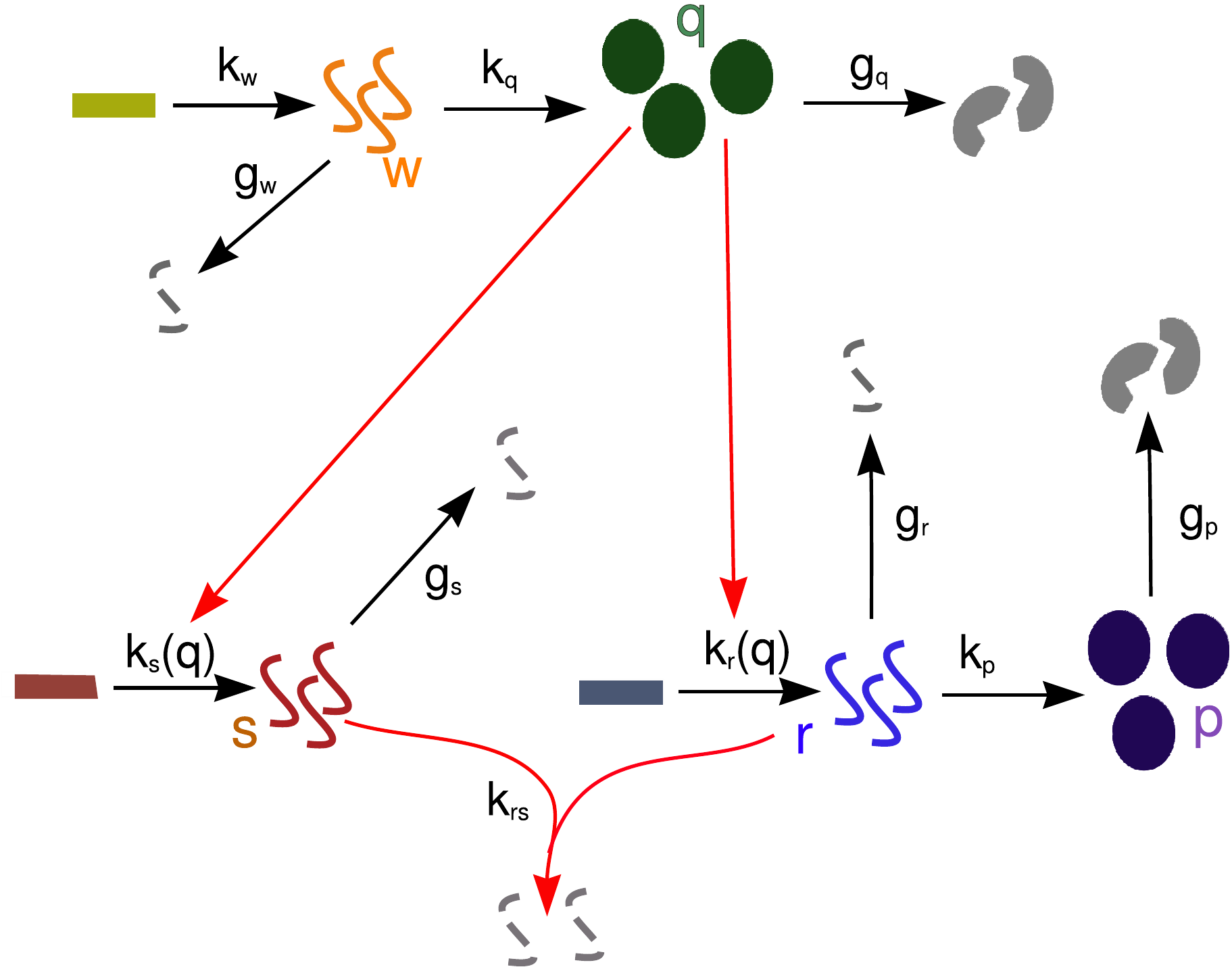}
\caption{Scheme of a miRNA-mediated incoherent FFL, where pairing of miRNA and mRNA exposes both molecules to co-degradation. The coupled degradation of the miRNA-mRNA 
 pair is described through a second-order kinetic constant $k_{rs}$ .}
\label{fig9}
\end{figure}

\subsection{Deterministic model}

The scheme of a mixed FFL in which the coupling between  sRNAs ($s$) and mRNAs ($r$) is stoichiometric is depicted in Fig.\ref{fig9}. Following 
\cite{Metha08SI,Levine07SI,Shimoni07SI}, we assume that both sRNA and mRNA are co-degraded when paired with a rate that depends on the sRNA-mRNA 
interaction strength $k_{rs}$. The mean field kinetic of our system can be described by the equations: 

\bea
\frac{d {w}}{{d t}} & =& k_{w} - g_{w} w \nn\\
\frac{d {q}}{{d t}}& =& k_{q} w - g_{q} q \nn\\
\frac{d {s}}{{d t}} & =& k_{s}(q)  - g_{s} s -k_{rs} r s \nn \\
\frac{d {r}}{{d t}} & =& k_{r}(q)  - g_{r} r -k_{rs} r s \nn \\
\frac{d {p}}{{d t}} & =& k_{p} r - g_{p} p .
\label{det-stoich}
\eea

The stationary solutions ($d_t x_i =0 ~\forall i\in\{w,q,s,r,p\}$) can be easily calculated (not reported).  

\subsection{Comparison with catalytic repression}

In this section we will explore the consequences of the nature of sRNA-mRNA interaction (stoichiometric or catalytic) on the noise properties of the mixed FFL. 
In analogy to section \ref{comp-deg}, we shall compare the two models 
choosing the parameters so as to obtain the same $p_{ss}$  with both types of sRNA action. 
As can be seen comparing the schemes in Fig.\ref{fig3} and 
Fig.\ref{fig9}, in order to have the same number of target proteins at equilibrium we set equal rates of production and degradation of each molecular
 species and then find the relation between $k_{rs}$ in the stoichiometric model and  $h$ in the catalytic model by equating the expression for $p_{ss}$
in Eqs.\ref{detFFLss} and in the solution of Eqs.\ref{det-stoich}. 

\begin{figure}[h!]
\centering
\includegraphics[width=1\textwidth]{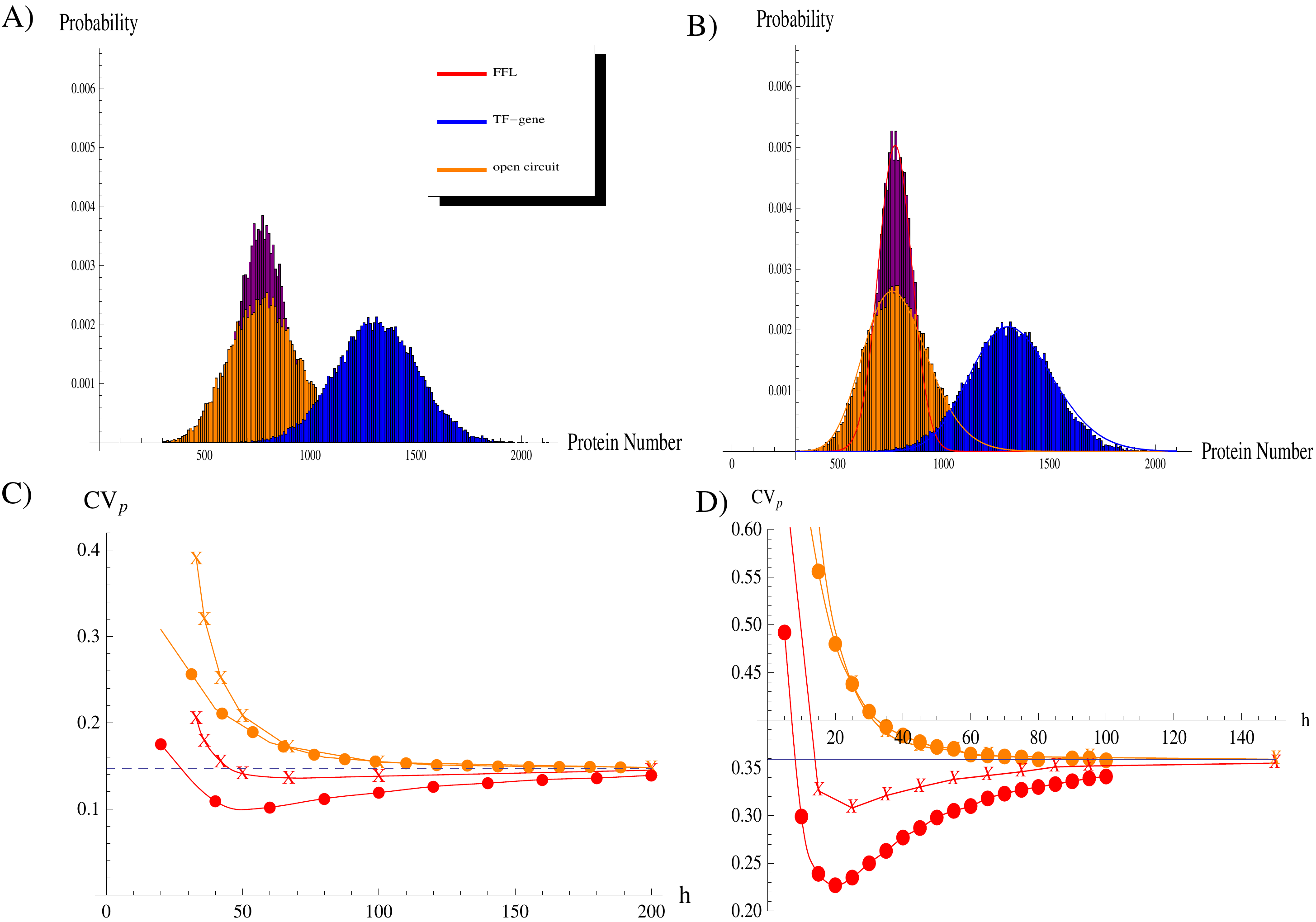}
\caption{Attenuation of noise by a FFL mediated by a sRNA that acts stoichiometrically on its mRNA target. 
We chose the same parameter set described in the caption of Fig.\ref{fig4}.
In the upper part of the figure we show the probability 
distribution of the target protein number for the three circuits in the case of stoichiometric action (A) and catalytic action (B). 
Although in both cases the FFL  reduces relative fluctuations with respect to the direct TF regulation and the open circuit, the catalytic modality turns out to be
 more efficient than the stoichiometric one. 
For the same set 
of parameters we report in (C) the $CV_p$ as a function of the inverse of the repression strength $h$. In the stoichiometric case $CV_p$ is actually a function of
 $k_{rs}$ which however can be expressed as function of $h$ (see the text). 
 To allow a simpler comparison of the various plots we plotted the stoichiometric results directly as function of $h$.
Dots are the result of simulations based on the hypothesis of a catalytic sRNA action while the x-shaped 
points derive from simulations with a stoichiometric action. For each regulatory modality we report the FFL and the open circuit data (which can be recognized because are
always higher than the FFL ones). Even if the qualitative behaviour 
is the same (in both cases a 
maximum of noise attenuation appears) the figure clearly shows that the catalytic modality is
 more efficient than the stoichiometric one in reducing the noise.
 (D) Same as (C) but for the alternative set of parameters discussed in the  caption of Fig.\ref{fig5}.}
\label{fig10}
\end{figure}

We ran simulations for the FFL, TF-gene direct regulation and the open circuit for catalytic and stoichiometric action (Fig.\ref{fig10}). The noise filtering effect 
is robust with respect to the mechanism 
of miRNA-mRNA interaction, but  a catalytic interaction makes the FFL  more efficient  in buffering fluctuations (compare the histograms in Fig.\ref{fig10}A and B).
 The U-shaped profile of the noise 
strength  $CV_p$ of a target controlled by a FFL (discussed in the main text for catalytic repression) is recovered also in the stoichiometric case. We report in 
Fig.\ref{fig10}C and D the $CV_p$ as a 
function of the inverse of repression strength for two different sets of parameter values. The maximum of attenuation is achieved for approximately the same 
value of $h$ of the catalytic case but the size of noise reduction is smaller with a stoichiometric repression. 

\begin{figure}[h!]
\centering
\includegraphics[width=0.8\textwidth]{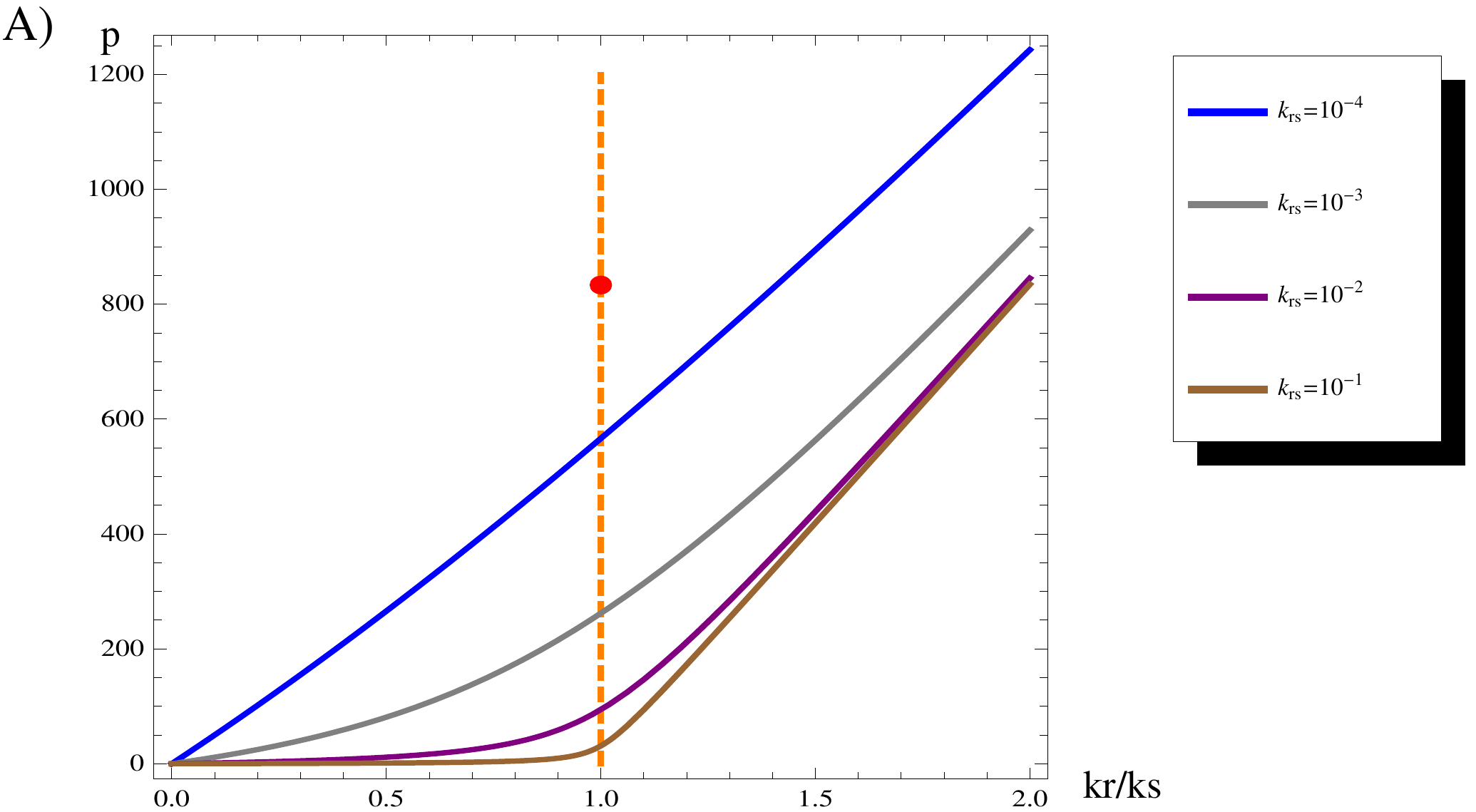}
\caption{We report the response ($p$) of the FFL as a function of $k_r/k_s$ for different values of miRNA-mRNA interaction strength. The red dot represents the protein production in absence of miRNA regulation. The threshold linear response is evident only for strong repression, while for $k_{rs}=10^{-4}$, compatible with a fine tuning regulation, the response is almost purely linear. }
\label{fig11}
\end{figure}

As reported in \cite{Levine07SI} and \cite{Elf05SI}, in the stoichiometric model described by Eqs.\ref{det-stoich} the mean protein number exhibits a threshold 
linear behaviour as a function of the ratio 
$k_r/k_s$ with the threshold in $1$ \cite{Levine07SI}. 
Following \cite{Metha08SI}, protein 
expression can be classified into three regimes: repressed ($k_r/k_s \ll 1$), crossover ($k_r/k_s\approx 1$) and expressing ($k_r/k_s \gg 1$). A threshold 
linear behaviour implies ultrasensitivity in 
the crossover regime and as a consequence the noise is enhanced near the threshold due to critical fluctuations (\cite{Metha08SI} and references 
therein). However, this threshold-linear response is expected if the 
mRNA-sRNA interaction is  strong, while for a weaker repression the threshold smoothly disappears (see Fig.\ref{fig11}) and the three regimes become 
indistinguishable. The analysis presented in Fig.\ref{fig10}  shows that the attenuation of fluctuations by a mixed FFL is observable in a regime of weak 
repression, corresponding to $k_{rs}\sim 10^{-4}$ in Fig.\ref{fig11}, where the crossover regime is vanishing and the raise in fluctuation in $k_r/k_s=1$ is negligible. \\

\newpage

\section{Purely transcriptional incoherent FFLs}

\subsection{Stochastic model}

\begin{figure}[h!]
\centering
\includegraphics[width=0.8\textwidth]{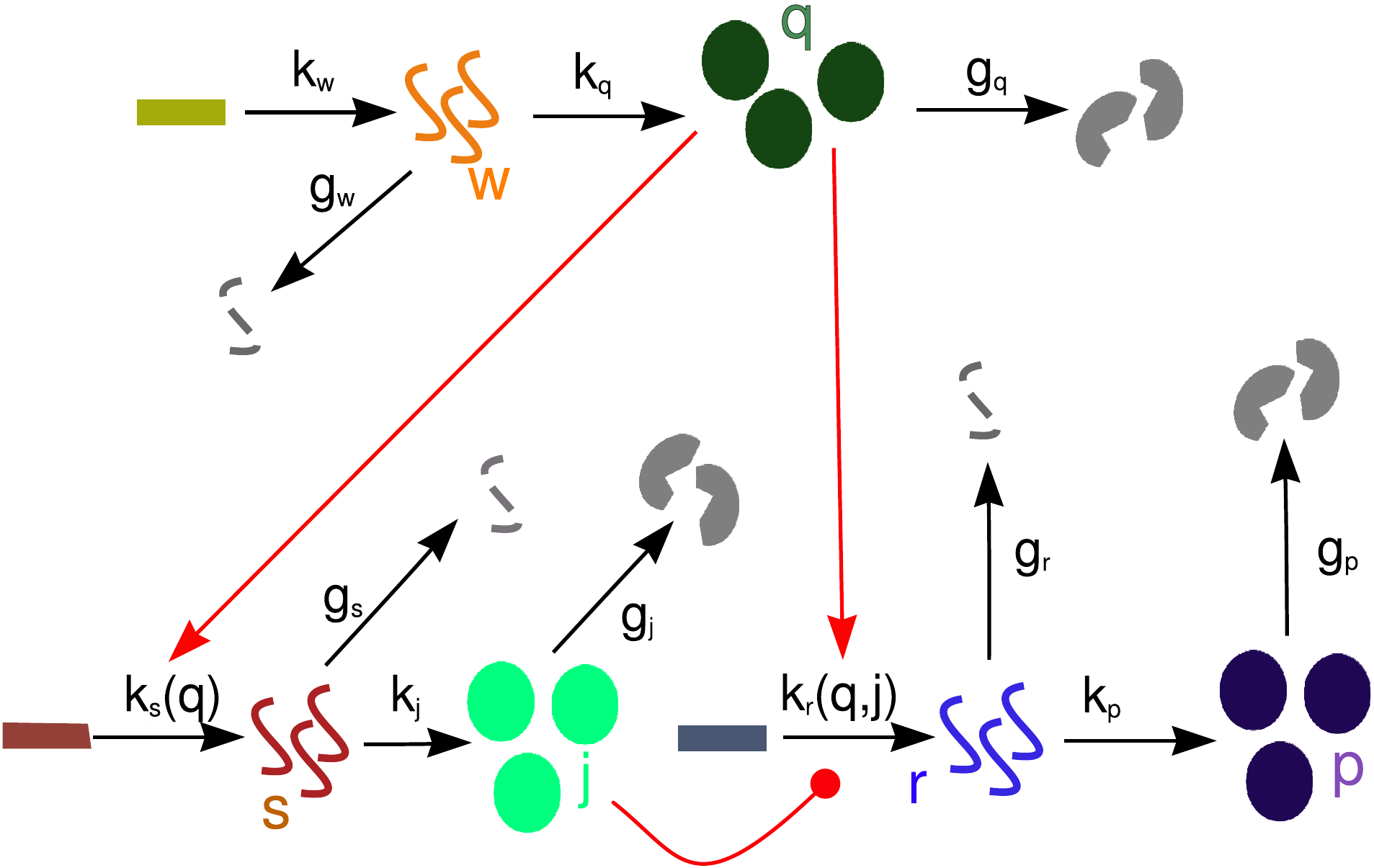}
\caption{Scheme of a purely  transcriptional incoherent FFL. Notations are the same of Fig.\ref{fig2}. The only difference with respect to Fig.\ref{fig2} is the presence of 
  a protein in the indirect pathway from the TF to target gene, therefore there are the additional reactions: translation of the protein $j$ from its mRNAs $s$ with rate $k_j$, and their degradation with rate $g_j$. The repressive action on the target is at the level of transcription in this case  (represented by the red rounded end line), resulting in a rate of target transcription $k_{r}(q,j)$ which is a function of the number of proteins $j$  and master TFs $q$.}
\label{fig12}
\end{figure}

The master equation describing a purely transcriptional incoherent FFL (depicted in the scheme of Fig. \ref{fig12})  is:

\bea
& & \frac{\partial P_{w,q,s,j,r,p}}{\partial t} = ~~k_{w} (P_{w-1,q,s,j,r,p} - P_{w,q,s,j,r,p}) + k_q w (P_{w,q-1,s,j,r,p}-P_{w,q,s,j,r,p}) \nn \\
& &  + k_s(q) (P_{w,q,s-1,j,r,p} - P_{w,q,s,j,r,p}) +k_j s (P_{w,q,s,j-1,r,p}-P_{w,q,s,j,r,p})\nn\\
& & + k_r(q,j) (P_{w,q,s,j,r-1,p} - P_{w,q,s,j,r,p})+k_p r (P_{w,q,s,j,r,p-1}-P_{w,q,s,j,r,p})\nn\\
& & + g_w \Bigl[ (w+1)P_{w+1,q,s,j,r,p} - w P_{w,q,s,j,r,p} \Bigr]+ g_q \Bigl[ (q+1)P_{w,q+1,s,j,r,p} - q P_{w,q,s,j,r,p} \Bigr]\nn\\
& & + g_s \Bigl[ (s+1)P_{w,q,s+1,j,r,p} - s P_{w,q,s,j,r,p} \Bigr]+ g_j \Bigl[ (j+1)P_{w,q,s,j+1,r,p} - j P_{w,q,s,j,r,p} \Bigr]\nn\\
& &+  g_r \Bigl[ (r+1)P_{w,q,s,j,r+1,p} - r P_{w,q,s,j,r,p} \Bigr] + g_p \Bigl[ (p+1)P_{w,q,s,j,r,p+1} - p P_{w,q,s,j,r,p} \Bigr].
\label{maFFLSI2}
\eea

The protein $j$ represses target transcription, which is also activated by the master TF $q$; consequently  the rate $k_r (q,j)$ is represented as a function of the concentration of both regulators.  In particular, we model the rate of target transcription as a product of Hill functions:

\bea
k_r(q,j) =  k_{r} \frac {q^c} {h_{r}^c +q^c}  \frac {1} {1 + (\frac{j} {h_j})^c}.
\label{Hill-protein}
\eea

While the linearization of the Hill function $k_{s}(q)$  is the one presented in Eq.\ref{linearizationSI}, we have to introduce the linearization of $k_{r}(q,j)$:

\bea
k_r(q,j) & \sim & k_r(q,j)|_{<q>,<j>} + \partial_{q} k_r(q,j)|_{<q>,<j>} (q-<q>) \nn\\
	&+ & \partial_{j} k_r(q,j)|_{<q>,<j>} (j-<j>).
\label{linearization-protein}
\eea

Therefore we can define:

\bea
k_r^0 &=& k_r(q,j)|_{<q>,<j>}-\partial_{j} k_r(q,j)|_{<q>,<j>} <j> -\partial_{q} k_r(q)|_{<q>,<j>} <q>\nn\\
k_r^1 &=& \partial_{q} k_r(q,j)|_{<q>,<j>} \nn\\
k_r^2 &=& \partial_{j} k_r(q,j)|_{<q>,<j>} .
\eea

Using the linearization just defined and  the method of moment generating function described in section \ref{stoch-model},  the analytical expression of $<p>$ and $CV_p$ can be obtained. 

\subsection{Constraints on parameters for a comparison with miRNA-mediated FFLs}

As stated in the main text, in order to make an unbiased comparison of the noise properties of these two circuits,  the corresponding models must be constrained to produce the same amount of target proteins. There are several possible ways of putting this constraint, due to the fact that there are two additional parameters  in the transcriptional FFL ($k_j$ and $g_j$) and therefore two supplementary degrees of freedom. In fact, a constraint can be inserted for example in the Hill function of target activation (tuning $h_j$) or in the rate of $s$ transcription $k_{s}(q)$, choosing accordingly the values of $k_j$ and $g_j$. This variety of options introduces some arbitrariness in the comparison. Our criterion is to keep the shared parameters to the same values (i.e. repression/activation efficiencies and  production/degradation rates) and choose the two additional ones to make equal the average amount of repressor proteins $<j>$ in the transcriptional case to the average amount of miRNAs $<s>$ in the mixed circuit. With this choice we end up with the same average amount of repressors in the two circuit versions ($<j>=<s>$), with the same efficiency of repression ($h_j=h_p$), and therefore with same impact on the target expression (making equal the  amount of target proteins produced by the two circuits). To implement this constraint, in the transcriptional FFL  the rate of translation $k_j$ must simply equal the rate of degradation $g_j$ (that we assume equivalent to the other protein degradation rates $g_{j}=g_{q}=g_{p}$). As a result, the average number of proteins $j$ which is produced from a single mRNA  is forced to $b=k_{j}/g_{s}=g_{j}/g_{s}$.  $b$ represents the translational burst size and, as discussed in the main text, it is a critical quantity in determining the noise level. As reported in \cite{Thattai01SI}, the fluctuations in the concentration of a single gene product can be expressed as:

\be
CV^{2}=\frac{1}{<p>} \left(\frac{b}{1+\eta}+1 \right).
\ee

Therefore, the noise level is dependent on the translational burst size (where $\eta$ is the ratio of protein to mRNA lifetime). \\
 We report the parameter values used in the analysis summarized in Figure 8 of the main text: $k_{j}=g_{j}=g_{q}=g_{p}=0.002; g_{w}=g_{s}=g_{r}=0.006, k_r=0.8;k_s=0.5;c=2 ;h_r= 200;h_s = 200;k_p=0.04$. For these values, the translational burst size $b$,  compatible with 
constraints, is $b= 0.33$, which is  considerably smaller than expected in eukaryotes. In conclusion, to satisfy the constraints we are probably underestimating the noise introduced by the supplementary translation step required in a purely transcriptional FFL.  This is why we expect that a miRNA-mediated FFL can overcome in noise-buffering efficiency its purely transcriptional counterpart  even more than  reported in Figure 8 of the main text.

\section{Cross-talk between miRNA targets}

\subsection{Stoichiometric vs catalytic model of miRNA action}

We start from a mass-action model for miRNA-mediated FFLs where we introduce explicitily a paramenter $\alpha$ representing the degree of catalyticity of miRNA action on targets. This type of description was introduced by Levine et al \cite{Levine07SI} and it will be  straightforwardly  applied to the FFL case in the following:

\bea
\frac{d {w}}{{d t}} & =& k_{w} - g_{w} w \nn\\
\frac{d {q}}{{d t}}& =& k_{q} w - g_{q} q \nn\\
\frac{d {s}}{{d t}} & =& k_{s}(q)  - g_{s} s - (k_{+} r s - k_{-} c)  + (1- \alpha) \beta c  \nn \\
\frac{d {r}}{{d t}} & =& k_{r}(q)  - g_{r} r - (k_{+} r s - k_{-} c)\nn \\
\frac{d {c}}{{d t}} & =&  (k_{+} r s - k_{-} c) - \beta c\nn \\
\frac{d {p}}{{d t}} & =& k_{p} r - g_{p} p ,
\label{levine-full}
\eea

where $c$ is the concentration of  miRNA-mRNA complexes, $k_{+}$  is the probability of miRNA-mRNA association, $k_{-}$ the probability of  dissociation of the complex $c$, which can degrade with rate $\beta$. The parameter  $\alpha$ represents the probability that degradation of the mRNA in the complex is accompained by degradation of the miRNA. As discussed in \cite{Levine07SI}, it is a measure of how much the miRNA action is catalytic. In this section, the variables that describe the state of the system ( $\{w,q,r,s,c,p\}$) are continuous variables, representing the average number of the various molecular species (we are omitting the notation $<..>$ for averages).
Since we are interested in steady state properties, we can simplify the model equilibrating the $c$ complex dynamics:

\bea
\frac{d {w}}{{d t}} & =& k_{w} - g_{w} w \nn\\
\frac{d {q}}{{d t}}& =& k_{q} w - g_{q} q \nn\\
\frac{d {s}}{{d t}} & =& k_{s}(q)  - g_{s} s - \alpha \gamma r s   \nn \\
\frac{d {r}}{{d t}} & =& k_{r}(q)  - g_{r} r - \gamma r s \nn \\
\frac{d {p}}{{d t}} & =& k_{p} r - g_{p} p ,
\label{levineSI}
\eea

where $\gamma= \beta k_{+} / (k_{-} + \beta)$.
The limit of $\alpha=0$ implies that for each degradation event  of  $c$ complexes,  none of the miRNAs is lost. This corresponds to a simplification of the model presented in section \ref{degradation-catalytic} of Text S1: the rate of mRNA degradation is supposed to be  a  linear function of miRNA concentration, instead of a nonlinear  Hill function. The opposite situation of $p=1$ reproduces the  stoichiometric model presented in Eq.\ref{det-stoich} (apart from the sostitution $\gamma \rightarrow k_{rs}$).\\
It is straightforward to generalize this description to the case of two miRNA targets, adding an equation describing the dynamics of a second target which is independently transcribed:

\bea
\frac{d {w}}{{d t}} & =& k_{w} - g_{w} w \nn\\
\frac{d {q}}{{d t}}& =& k_{q} w - g_{q} q \nn\\
\frac{d {s}}{{d t}} & =& k_{s}(q)  - g_{s} s - \alpha (\gamma_{1} r s + \gamma_{2} r_{2} s)  \nn \\
\frac{d {r}}{{d t}} & =& k_{r}(q)  - g_{r} r - \gamma_{1} r s \nn \\
\frac{d {r_{2}}}{{d t}} & =& k_{r_{2}}  - g_{r_{2}} r_2 - \gamma_{2} r_{2} s \nn \\
\frac{d {p}}{{d t}} & =& k_{p} r - g_{p} p .
\label{levine2}
\eea

 The analytical solutions can be found easily at the steady state.  This is the complete effective model presented partially in the main text. In the following the coupling constants between miRNAs and the mRNAs transcribed from the two target genes will be assumed equal ($\gamma_{1}=\gamma_{2}=\gamma$).

\subsection{Details on the model setting}
In this section we present a detailed view of the model setting and parameter values used for the analysis regarding the target cross-talk presented in the main text. We focus specifically on the minimal assumptions and on the parameter values used to achieve the results in Figure 9.

\subsubsection{Setting for Figure 9 B}

The solution for $<p>$ of Eqs. \ref{levine2} at the steady state depends on $\alpha$. Therefore, in order to evaluate the impact of the dilution effect for different mechanisms (stoichiometric/catalytic) of miRNA repression, we choose for each $\alpha$ the corresponding $\gamma$ value that leads to the same mean amount of target proteins $<p>$. Qualitatively, in a catalytic model ($\alpha=0$) the miRNAs are more efficient since they can affect several target mRNAs without being degraded. Consequently, as $\alpha$ decreases the $\gamma$ value must be decreased so as to mantain the same target level expression. This is the constraint that makes unique the starting point (for $k_{r_2}=0$) of the curves corresponding to different $\alpha$ values in Fig. 9 of the main text. Concerning the other parameter values, in Fig. 9 they are fixed to: $k_q=0.19 s^{-1};g_q=g_p=0.002s^{-1};g_w=g_r=g_s=g_{r_2}=0.006s^{-1};k_w=0.0126s^{-1};k_r=k_s=0.8s^{-1};k_p=0.04s^{-1};c=2; h_r=h_s=200$, while the value of $\gamma=0.00011 s^{-1}$ is assigned to the catalytic model ($\alpha=0$) - and it corresponds approximately to the optimal buffering value-,  while the $\gamma$ values for the other $\alpha$ models can be calculated as described above.

\subsubsection{Setting for Figure 9 C}

As a first approximation we assume, for the sake of simplicity, that the expression of the second target is not regulated by any TF.   Therefore, it  is a simple birth-death process, with transcription rate $k_{r_{2}}$ and degradation rate $g_{r_{2}}$ (assumed unique for both targets $g_{r_{2}}=g_r$). Since the target embedded in the FFL is regulated by the TFs $q$, $k_{r}(<q>)$ can be used as the effective rate of transcription, to be compared with  $k_{r_{2}}$  of the second target. Indeed, $k_{r}(<q>)$ represents the average rate at which the joint target is transcribed.

\subsubsection{Setting for Figure 9 D}

In this subsection we shall introduce a simple strategy to tune the second target fluctuations and analyze their impact on noise buffering efficiency. The proposed strategy is in perfect analogy with the one explained in section ``The incoherent feedforward loop is effective in reducing extrinsic fluctuations" of the main text. In brief, we add an independent TF $q'$  which activates the transcrition of the second target. Its rates of transcription $k_{w}'$ and translation $k_{q}'$ are chosen so as to produce the same mean amount of protein of the other activator ($<q>=<q'>$). Therefore, the effective mean rates of transcription of both miRNA targets turn out to be equal. Changing the ratio  $k_{w}'/k_{q}'$  while keeping costant the product $k_{w}'k_{q}'$ allows us to vary the second target fluctuations without altering its mean level.

\section{Noise reduction and signaling sensitivity}
\label{sensitivity}

Biological systems present the apparently contraddictory need for both high sensitivity to external signals both homeostatic controls,  depending on the specific function in analysis. Indeed, while one essential feature of signal transduction systems is the amplification of small changes in input signals \cite{Shibata05SI}, the reliable celullar functioning in a fluctuating environment lays on multiple homeostatic controls (the most evident is temperature control in mammals). Similarly, at the level of genetic networks there is an interplay between sensitivity to changes in the input signal and the ability to buffer stochastic fluctuations.  An increase in sensitivity to a signal results in an elevated exposure to its fluctuations, as shown for linear cascades of regulations \cite{Shibata05SI,Hooshangi05SI}. More recently the sensitivity/noise-buffering analysis has been extended to small genetic circuits, including feedback and feedforward loops \cite{Hornung08SI}. The working hypothesis of the authors is that the main function of a genetic circuit is to maximize the amplification of input signals. We argue that while this can be often the case, some circuits can have evolved to mantain reliably a functional steady state, even at the expense of a loss of sensitivity (and even thanks to that loss), to implement in other words a homeostatic control.

\begin{figure}[h!]
\centering
\includegraphics[width=1\textwidth]{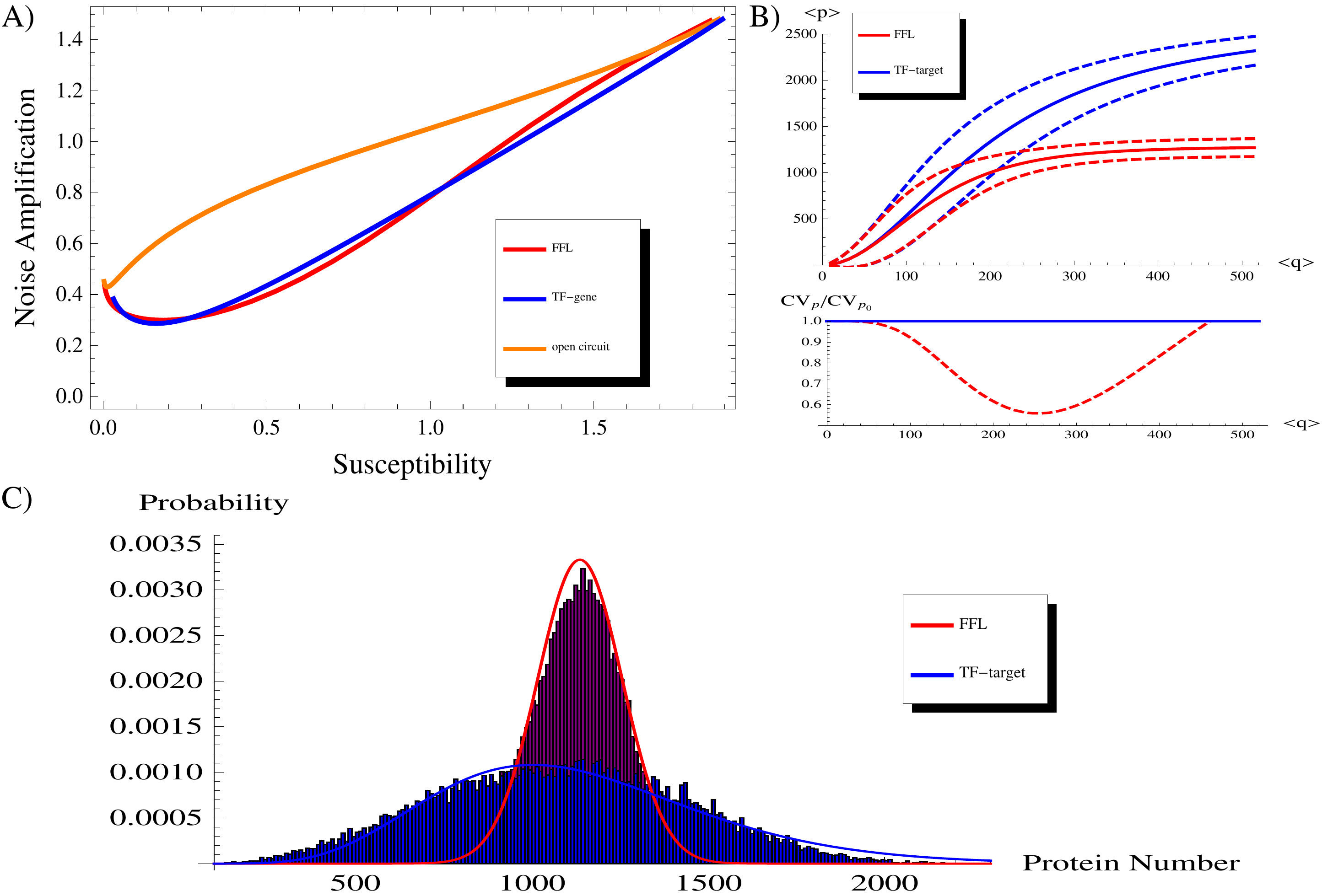}
\caption{(A) Noise amplification versus susceptibility for the three circuits: a miRNA-mediated incoherent FFL, a TF-target regulation and an open circuit. The parameter values that are fixed are those reported in caption of Fig.3 of the main text (unless $k_q = 0.19s^{-1}$ higher than in Fig.3 to increase the TF fluctuations). (B) The upper panel shows the fold change in the target level in response to a fold change in the TF level for the miRNA-mediated incoherent FFL and the TF-target linear circuit. Continuous lines represent the behaviour of mean values while dashed lines are depicted at a distance from the mean equal to one standard deviation. In the lower panel the noise reduction $CV_p/CV_{p_0}$ is depicted in same range of $<q>$.(C)  The probability distribution of protein number for the two circuits (miRNA-mediated FFL and TF-gene). In this case the two regulative circuits are constrained to produce an equal mean amount of target proteins. The same steady state is achieved with a strikingly different control of fluctuations by the two circuitries.  Histograms are the result of Gillespie simulations while continuous lines are empirical distributions (gaussian for the FFL and gamma for the TF-gene) with mean and variance predicted by the analytical model.  }
\label{fig13} 
\end{figure}

Following \cite{Hornung08SI}, the  steady state sensitivity can be defined as the relative response in output that follows a change in the input. In the contest of incoherent FFLs (scheme in Fig. \ref{fig2}) we can consider as input the mean number of TFs $<q>$ and as output the consequent level of target proteins $<p>$. Following these definitions, the susceptibility takes the form:

\be
susceptibility= \frac{<q>}{<p>} \frac{d~<p>}{d~<q>}=\frac{d~ln(<p>)}{d~ln(<q>)}.
\ee

As a measure of the quantity of noise propagating through the circuit, the noise amplification measure $\eta$ can be introduced \cite{Hornung08SI}:

\be 
\eta=\frac{CV_{p}}{CV_{q}},
\ee

defined as ratio between output and input noise. As shown in Fig. \ref{fig13} A, the incoherent miRNA-mediated FFL presents an interplay between noise amplification and susceptibility very similar to that of a gene only activated by a TF, while the same fine-tuning implemented using  an independent miRNA would imply a more severe interplay. Therefore, the noise buffering function demonstrated in this paper is achieved at the expense of  steady-state sensitivity: given a fixed value of susceptibility, the FFL and the TF-gene linear circuit lead to a  similar degree of noise amplification, while when the noise is buffered by the FFL there must be a loss of target susceptibility. Indeed, the fold change in target expression, that  follows a change in the  TF mean level $<q>$, is reduced precisely in the region where the noise control is implemented (see Fig.\ref{fig13} B). However, we propose that this is precisely the behaviour needed for a homeostatic control. The output  is highly sensitive to changes in the input concentration until a finely tuned steady state is reached, then this functional steady state is kept robust to input fluctuations even if at the expense of a sensitivity loss. The same steady state could be reached more simply without any miRNA regulation,  tuning the TF concentration in a TF-gene circuit, so as to  conserve a high sensitivity. However, in this case the equilibrium level would be affected by strong fluctuations propagating from the upstream factor, as clearly shown in Fig. \ref{fig13} B.\\ 
In conclusion, if the sensitivity is the function that have to be maximized, as it is probably the case in signaling systems, incoherent FFLs (and miRNA mediated ones) are outperformed by other circuits (like those making use of positive feedbacks loops \cite{Hornung08SI}) that support less noise amplification at a fixed susceptibility. However, in different biological contests a high  sensitivity could be  important only until a functional steady state is reached. Then a homeostatic control can be required for keeping the reached level constant in presence of noisy upstream regulators and miRNA-mediated FFLs seems properly designed for this aim. 
The proposed functioning is also in agreement with the idea of fine-tuning: when the target expression is switched on by a rise in TF concentration, the maintenance of its level into a narrow functional range can be more important than a reliable transmission of further incoming small signals.
A role of miRNA regulation in homeostasis is in line with the observation that miRNAs are often involved in signaling networks to ensure homeostatic controls (see for example \cite{Piccolo10SI}).

\section{Effects of possible delays in miRNA production.}

\begin{figure}[h!]
\centering
\includegraphics[width=0.8\textwidth]{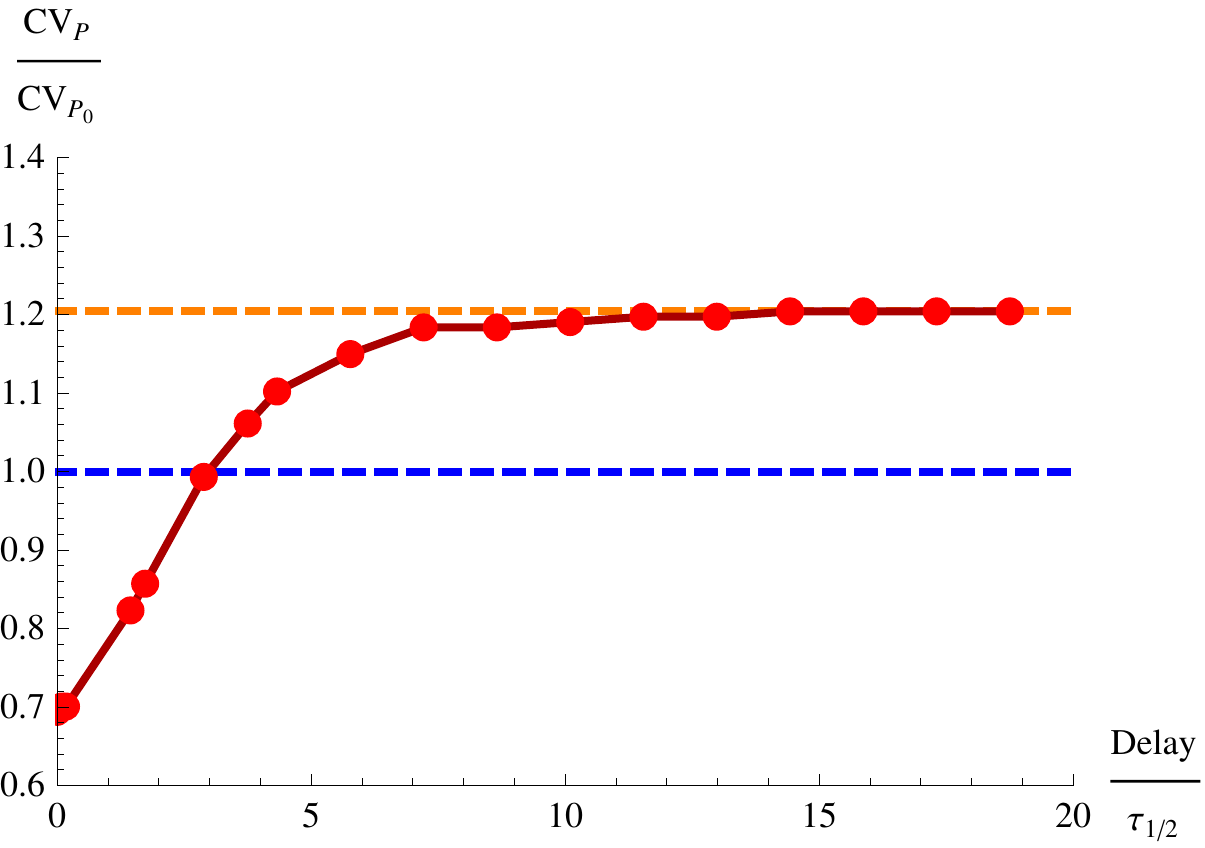}
\caption{The effect on noise-buffering efficiency  of a time-delay between miRNA transcription and  miRNA repressive action. The values of parameters are the ones in caption of Figure 3 in the main text. 
The target noise reduction $CV_p/CV_{p_{0}}$  is measured as a function of the length of the time-delay expressed in unit of protein  half-life $\tau_{1/2}$. $CV_{p_{0}}$ is the constitutive noise of a TF-gene circuit without the miRNA regulation. Until the delay length approaches approximately $3$ protein half-lives, the FFL is still able to filter out fluctuations. After that the noise level tends to the value achieved with an open circuit (dashed orange line) in which  miRNAs and target mRNAs have uncorrelated fluctuations in their level. Dots are the result of Gillespie's simulations with the full nonlinear dynamics.
} 
\label{delay}
\end{figure}

The common lore is that a RNA based post-transcriptional regulation can have  a faster action on a target gene expression with respect to TF regulation \cite{Li09SI,Mukherji09SI}. Indeed, a TF  must be transported back to the nucleus and find its target promoter to exert its regulative role. However  there is a lack of data to support quantitatively this assumption and the biogenesis of miRNA actually requires several processing steps. 
The time needed for the miRNA to be processed, loaded in RISC and in general to become active can introduce a delay between its transcription and its effect on targets. Therefore it could be interesting to consider possible effects of this time delay on the noise buffering function of mixed FFLs. 
While in  the model  presented in the main text the miRNA is supposed to act on its target instantaneously, in this section we present results of simulations performed  taking into account the time-delay that can arise from miRNA processing. More specifically the time-delay has been inserted in the Hill function of regulation of the target translation, mimicking the time required for miRNA activation. With this simulation procedure, for each chosen set of parameter values  it is possible to establish the threshold in delay time below which the circuit is able to reduce target fluctuations.\\
In Figure {\ref{delay}} the noise reduction achieved with a FFL  ($CV_{p}/CV_{p_{0}}$) is reported as a function of the time required for miRNA activation. The time-delay is expressed in unit of protein half-life, chosen as a reference since it represents the longest time-scale in the system.
The ability of the circuit in filtering out fluctuations relies on the correlation between miRNA and target mRNA fluctuations, therefore an eventual time-delay in miRNA action can negatively affect the noise buffering. More specifically, with the parameter values of Fig.3-4 of the main text,  the incoherent FFL is no more able to reduce target fluctuations if the delay is longer than approximately $3$ protein half-life (see Fig \ref{delay}).   As the processing time becomes longer and longer,  miRNA fluctuations  lose any correlation with the target ones and the target noise approches the value corresponding to the open circuit case (dashed orange line in Fig. \ref{delay}).\\
In conclusion we showed that a significant time-delay between miRNA transcription and target repression can compromise the noise-buffering function.  
When  quantitative measures of the time required for transport and processing of miRNAs and proteins will be available, it will be possible to precisely evaluate the degree of reduction of target fluctuations inserting the appropriate delays in the Hill function of regulations of our theoretical model (even if at the expense of its analytical tractability).

\section{Bioinformatical analysis of miRNAs involved in FFLs in the human mixed network.}

Although miRNA mediated FFLs have been shown to be overrepresented in real mixed networks with respect to randomized networks \cite{Shalgi07SI,Tsang07SI,Yu08SI,Re09SI}, it is equally important to establish the numerical fraction of miRNAs and miRNA targets that are actually involved in these circuits, to better highlight the effective biological relevance of miRNA-mediated FFLs. To this aim we take advantage of the genome-wide survey of human miRNA-mediated FFLs previously developed by our group \cite{Re09SI}, based on a search for overrepresented motif in human and mouse promoters and 3'-UTRs. Of the 464 miRNAs annotated as KNOWN-KNOWN in the Ensembl database (release 46) \cite{Hubbard07SI}, using the filters and the software setup of \cite{Re09SI} 193 were selected to form the post-transcriptional network (miRNA-target interactions).  Integrating this network with the transcriptional one (with TF-target and TF-miRNA interactions),  133 miRNA have been significantly associated to at least one  (usually more than one) of the 5030 mixed FFLs found in the human regulatory network (see \cite{Re09SI} for more details). Therefore  miRNAs, at least in the database considered, seem often involved in FFL circuits. Since each miRNA can regulate hundreds of targets it is also interesting to evaluate what fraction of its targets are part of FFLs. The results of this analysis are reported in the following Table where the total number of targets and the number of targets in a FFL is presented for each miRNA embedded in a FFL . While some miRNAs preferentially regulate genes through a FFL topology  this is clearly not a general trend, further confirming the importance of considering the possible cross-talk between miRNA targets (as discussed in the main text).\\
However it is important to notice that the proposed results suffer some limitations.
Firstly we cannot distinguish incoherent from coherent FFLs since sequence analysis allows the identification of putative interactions but cannot
 establish if they are positive or negative. 
Secondly the proposed regulations should be considered as potential interactions because they represent purely  bioinformatic predictions and furthermore the miRNA and its targets could  be expressed preferentially in different tissues or at different times. In this case the eventual cross-talk would be limited among co-expressed targets. 
In spite of the reported limitations, the data presented here point out that  miRNA-mediated FFLs can actually represent  an often exploited regulative circuitry, further suggesting their importance in real networks of gene regulations.\\

\begin{tabular}{||p{2.5cm}||*{4}{c|}|}
\hline
			&		  &			    &		  		\\
	miRNA gene      & Num. of targets  & Num. of targets in FFLs   &	Percentage     		\\
			&		  &			    &		  		\\
\hline
\hline	
	hsa-miR-129	&	44	&	36 		&	81.8 \%			\\
	hsa-miR-148b	&	127	&	84		&	66.1 \%			\\
	hsa-miR-149	&	55	&	36		&       65.5 \%			\\
	hsa-miR-449b	&	55	&       34		&	61.8 \%			\\
	hsa-let-7a	&	83	&	51 		&	61.4 \%			\\
	hsa-miR-199a*	&	138	&	84 		&	60.9 \%			\\
	hsa-miR-125b	&	150	&	90		&       60.0 \%			\\
	hsa-miR-199a	&	41	&	24 		&       58.5 \%			\\
	hsa-miR-101	&	105	&	61  		&       58.1 \%			\\
	hsa-miR-205	&	38	&	22 		&       57.9 \%			\\
	hsa-miR-31	&	35	&	20		&       57.1 \%			\\
	hsa-miR-203	&	51	&	29		&       56.9 \%			\\
	hsa-miR-30c	&	155	&	87 		&       56.1 \%			\\
	hsa-miR-425-3p	&	50	&	28		&       56.0 \%			\\
	hsa-miR-9	&	106	&	59		&       55.7 \%			\\
	hsa-miR-296	&	69	&	38 		&       55.1 \%			\\
	hsa-miR-194	&	90	&	49  		&       54.4 \%			\\
	hsa-miR-181d	&	120	&	64  		&	53.3 \%			\\
	hsa-miR-219	&	123	&	65  		&	52.8 \%			\\	
	hsa-miR-32	&	148	&	78           	&	52.7 \%			\\
	hsa-miR-9*	&	100	&	52  		&       52.0 \%			\\
	hsa-miR-148a	&	91	&	47  		&	51.6 \%			\\
	hsa-miR-24	&	107	&	54  		&	50.5 \%			\\
	hsa-miR-133b	&	40	&	20  		&	50.0 \%			\\
	hsa-miR-499	&	40	&	20  		&	50.0 \%			\\
	hsa-miR-30a-3p	&	48	&	23  		&	47.9 \%			\\
	hsa-miR-218	&	83	&	39  		&	47.0 \%			\\
	hsa-miR-375	&	113	&	53  		&	46.9 \%			\\
	hsa-miR-223	&	145	&	67  		&	46.2 \%			\\
	hsa-miR-100	&	46	&	21  		&	45.7 \%			\\
	hsa-miR-214	&	62	&	28  		&	45.2 \%			\\
	hsa-miR-10a	&	39	&	17  		&	43.6 \%			\\
	hsa-miR-1	&	46	&	20  		&	43.5 \%			\\
	hsa-miR-130a	&	127	&	55  		&	43.3 \%			\\
	hsa-miR-30a-5p	&	155	&	67  		&	43.2 \%			\\
	hsa-miR-802	&	76	&	31  		&	40.8 \%			\\
	hsa-miR-26a	&	129	&	52  		&	40.3 \%			\\
	hsa-miR-23a	&	152	&	60 		&	39.5 \%			\\
	hsa-miR-99a	&	46	&	18 		&	39.1 \%			\\
	hsa-miR-126*	&	181	&	70 		&	38.7 \%			\\
	hsa-miR-330	&	50	&	19 		&	38.0 \%			\\
	hsa-miR-135b	&	103	&	39 		&	37.9 \%			\\
\hline
\hline
\end{tabular}
\newpage
\begin{tabular}{||p{2.5cm}||*{4}{c|}|}
\hline
			&		  &			    &		  		\\
	miRNA gene      & Num. of targets  & Num. of targets in FFLs   &	Percentage     		\\
			&		  &			    &		  		\\
\hline
\hline	
	hsa-miR-133a	&	40	&	15 		&	37.5 \%			\\
	hsa-miR-155	&	100	&	37 		&	37.0 \%			\\
	hsa-miR-126	&	109	&	40 		&	36.7 \%			\\
	hsa-miR-140	&	106	&	38 		&	35.8 \%			\\
	hsa-miR-506	&	127	&	45 		&	35.4 \%			\\
	hsa-miR-99b	&	46	&	16 		&	34.8 \%			\\
	hsa-miR-202	&	88	&	30 		&	34.1 \%			\\
	hsa-miR-135a	&	103	&	35 		&	34.0 \%			\\
	hsa-let-7f	&	83	&	28		&	33.7 \%			\\
	hsa-miR-16	&	57	&	19		&	33.3 \%			\\
	hsa-let-7d	&	90	&	29 		&	32.2 \%			\\
	hsa-let-7e	&	127	&	40 		&	31.5 \%			\\
	hsa-miR-542-3p	&	39	&	12 		&	30.8 \%			\\
	hsa-miR-206	&	46	&	14 		&	30.4 \%			\\
	hsa-miR-34b	&	55	&	16 		&	29.1 \%			\\
	hsa-miR-34c	&	55	&	16 		&	29.1 \%			\\
	hsa-miR-342	&	49	&	14 		&	28.6 \%			\\
	hsa-miR-363	&	84	&	24  		&	28.6 \%			\\
	hsa-miR-365	&	46	&	13  		&	28.3 \%			\\
	hsa-miR-27a	&	104	&	29  		&	27.9 \%			\\
	hsa-miR-29a	&	115	&	32  		&	27.8 \%			\\
	hsa-miR-19a	&	145	&	39  		&	26.9 \%			\\
	hsa-miR-152	&	127	&	34  		&	26.8 \%			\\
	hsa-miR-199b	&	41	&	11  		&	26.8 \%			\\
	hsa-miR-141	&	146	&	38  		&	26.0 \%			\\				
	hsa-miR-212	&	58	&	15 		&	25.9 \%			\\
	hsa-miR-302c*	&	93	&	24    		&	25.8 \%			\\	
	hsa-miR-106a	&	126	&	32 		&	25.4 \%			\\
	hsa-miR-17-5p	&	126	&	32		&	25.4 \%			\\
	hsa-miR-30e-5p	&	155	&	39  		&	25.2 \%			\\
	hsa-miR-495	&	123	&	31  		&	25.2 \%			\\
	hsa-miR-144	&	146	&	36  		&	24.7 \%			\\
	hsa-miR-7	&	89	&	22  		&	24.7 \%			\\
	hsa-miR-20b	&	126	&	31  		&	24.6 \%			\\
	hsa-miR-20a	&	132	&	32  		&	24.2 \%			\\
	hsa-miR-103	&	97	&	23  		&	23.7 \%			\\
	hsa-miR-106b	&	132	&	31  		&	23.5 \%			\\	
	hsa-miR-367	&	111	&	26  		&	23.4 \%			\\
	hsa-miR-34a	&	43	&	10 		&	23.3 \%			\\
	hsa-miR-193a	&	112	&	26  		&	23.2 \%			\\
	hsa-miR-200c	&	143	&	33  		&	23.1 \%			\\
	hsa-miR-189	&	35	&	8   		&	22.9 \%			\\
\hline
\hline
\end{tabular}
\newpage
\begin{tabular}{||p{2.5cm}||*{4}{c|}|}
\hline
			&		  &			    &		  		\\
	miRNA gene      & Num. of targets  & Num. of targets in FFLs   &	Percentage     		\\
			&		  &			    &		  		\\
\hline
\hline	
	hsa-miR-93	&	83	&	19  		&	22.9 \%			\\
	hsa-miR-202*	&	49	&	11  		&	22.4 \%			\\
	hsa-miR-451	&	45	&	10  		&	22.2 \%			\\
	hsa-miR-221	&	50	&	11  		&	22.0 \%			\\
	hsa-miR-222	&	50	&	11  		&	22.0 \%			\\
	hsa-miR-138	&	60	&	13  		&	21.7 \%			\\
	hsa-miR-302b	&	134	&	29  		&	21.6 \%			\\
	hsa-miR-302c	&	134	&	29  		&	21.6 \%			\\
	hsa-miR-302d	&	134	&	29  		&	21.6 \%			\\
	hsa-miR-299-5p	&	108	&	23  		&	21.3 \%			\\
	hsa-miR-182	&	80	&	17  		&	21.2 \%			\\
	hsa-miR-142-5p	&	57	&	12 		&	21.1 \%			\\
	hsa-miR-369-3p	&	101	&	21  		&	20.8 \%			\\
	hsa-let-7b	&	83	&	17  		&	20.5 \%			\\	
	hsa-miR-494	&	122	&	24  		&	19.7 \%			\\
	hsa-miR-183	&	92	&	18  		&	19.6 \%			\\
	hsa-miR-505	&	51	&	10  		&	19.6 \%			\\
	hsa-miR-377	&	82	&	16  		&	19.5 \%			\\
	hsa-miR-96	&	133	&	26  		&	19.5 \%			\\
	hsa-miR-195	&	57	&	11  		&	19.3 \%			\\
	hsa-miR-497	&	57	&	11  		&	19.3 \%			\\
	hsa-miR-30e-3p	&	48	&	9   		&	18.8 \%			\\
	hsa-miR-381	&	165	&	31  		&	18.8 \%			\\
	hsa-miR-142-3p	&	127	&	23  		&	18.1 \%			\\
	hsa-miR-139	&	34	&	6   		&	17.6 \%			\\
	hsa-miR-30b	&	155	&	27  		&	17.4 \%			\\
	hsa-miR-30d	&	155	&	27  		&	17.4 \%			\\
	hsa-miR-302b*	&	76	&	13  		&	17.1 \%			\\
	hsa-miR-487b	&	83	&	14  		&	16.9 \%			\\
	hsa-miR-369-5p	&	90	&	15  		&	16.7 \%			\\
	hsa-miR-409-5p	&	80	&	13  		&	16.2 \%			\\
	hsa-miR-410	&	133	&	21  		&	15.8 \%			\\
	hsa-miR-329	&	93	&	14  		&	15.1 \%			\\
	hsa-miR-151	&	70	&	10  		&	14.3 \%			\\
	hsa-miR-412	&	42	&	6   		&	14.3 \%			\\
	hsa-miR-25	&	74	&	10  		&	13.5 \%			\\
	hsa-miR-192	&	45	&	6   		&	13.3 \%			\\
	hsa-miR-496	&	113	&	14  		&	12.4 \%			\\
	hsa-miR-153	&	100	&	9   		&	9.0 \%			\\
	hsa-miR-15a	&	57	&	5   		&	8.8 \%			\\
	hsa-miR-217	&	102	&	9   		&	8.8 \%			\\
	hsa-miR-323	&	57	&	5   		&	8.8 \%			\\	
\hline
\hline
\end{tabular}
\newpage
\begin{tabular}{||p{2.5cm}||*{4}{c|}|}
\hline
			&		  &			    &		  		\\
	miRNA gene      & Num. of targets  & Num. of targets in FFLs   &	Percentage     		\\
			&		  &			    &		  		\\
\hline
\hline	
	hsa-miR-484	&	100	&	6   		&	6.0 \%			\\
	hsa-miR-26b	&	129	&	7   		&	5.4 \%			\\
	hsa-miR-146b	&	40	&	2   		&	5.0 \%			\\
	hsa-miR-200a*	&	90	&	3   		&	3.3 \%			\\
	hsa-miR-200a	&	146	&	3   		&	2.1 \%			\\
	hsa-miR-200b	&	143	&	2   		&	1.4 \%			\\
	hsa-miR-429	&	108	&	1  		&	0.9 \%			\\
\hline
\hline
\end{tabular}

\end{document}